\documentclass{aa}
\usepackage[varg]{txfonts}
\usepackage{natbib}
\bibpunct{(}{)}{;}{a}{}{,} 

\usepackage{amsmath}
\usepackage{graphicx}
\usepackage{graphics}
\usepackage{epsfig}
\usepackage{pdfpages}
\usepackage{blindtext}
\usepackage{supertabular}
\usepackage{afterpage}
\usepackage{lscape}
\usepackage{tablefootnote}
\usepackage{subcaption}
\usepackage[flushleft]{threeparttable}
\usepackage{color, colortbl}
\usepackage[normalem]{ulem}
\usepackage{multirow}
\definecolor{lightgray}{gray}{0.9}

\usepackage{txfonts}

\usepackage{natbib,twoopt}
\usepackage{hyperref}
\hypersetup{colorlinks=true,linkcolor=blue,citecolor=blue,urlcolor=blue,breaklinks=true} 
\bibpunct{(}{)}{;}{a}{}{,}             
\makeatletter
  \newcommandtwoopt{\citeads}[3][][]{\href{http://adsabs.harvard.edu/abs/#3}%
    {\def\hyper@linkstart##1##2{}%
     \let\hyper@linkend\@empty\citealp[#1][#2]{#3}}}
  \newcommandtwoopt{\citepads}[3][][]{\href{http://adsabs.harvard.edu/abs/#3}%
    {\def\hyper@linkstart##1##2{}%
     \let\hyper@linkend\@empty\citep[#1][#2]{#3}}}
  \newcommandtwoopt{\citetads}[3][][]{\href{http://adsabs.harvard.edu/abs/#3}%
    {\def\hyper@linkstart##1##2{}%
     \let\hyper@linkend\@empty\citet[#1][#2]{#3}}}
  \newcommandtwoopt{\citeyearads}[3][][]%
    {\href{http://adsabs.harvard.edu/abs/#3}
    {\def\hyper@linkstart##1##2{}%
     \let\hyper@linkend\@empty\citeyear[#1][#2]{#3}}}
\makeatother

\begin{document}

\title{Dust depletion of metals from local to distant galaxies I:}
\subtitle{Peculiar nucleosynthesis effects and grain growth in the ISM\thanks{This paper is based in part on observations carried out at the European Organisation for Astronomical Research in the Southern Hemisphere under ESO programmes 065.P-0038, 065.O-0063, 066.A-0624, 067.A-0078, and 068.A-0600.}}

\author{Christina~Konstantopoulou\inst{1}\thanks{\email{christina.konstantopoulou@unige.ch}},
        Annalisa~De~Cia\inst{1},
        Jens-Kristian~Krogager\inst{2},
        C\'{e}dric~Ledoux\inst{3},
        Pasquier~Noterdaeme\inst{4},
        Johan~P.~U.~Fynbo\inst{5,6},
        Kasper~E.~Heintz\inst{5,6},
        Darach~Watson\inst{5,6},
        Anja~C.~Andersen\inst{5},
        Tanita~Ramburuth-Hurt\inst{1},
        and Iris~Jermann\inst{1}}

\institute{Department of Astronomy, University of Geneva, Chemin Pegasi 51, 1290 Versoix, Switzerland
\and Centre de Recherche Astrophysique de Lyon, Univ. Claude Bernard Lyon 1, 9 Av. Charles Andre, 69230 St Genis Laval, France
\and European Southern Observatory, Alonso de C\'{o}rdova 3107, Vitacura, Casilla 19001, Santiago, Chile
\and Institut d'Astrophysique de Paris, CNRS-SU, UMR 7095, 98bis bd Arago, 75014, Paris, France
\and Niels Bohr Institute, University of Copenhagen, Jagtvej 128, 2200 Copenhagen N, Denmark
\and Cosmic Dawn Center (DAWN), Copenhagen, Denmark}

\date{Received xxx; Accepted xxx}

 \abstract{Large fractions of metals are missing from the observable gas-phase in the interstellar medium (ISM) because they are incorporated into dust grains. This phenomenon is called dust depletion. It is important to study the depletion of metals into dust grains in the ISM to investigate the origin and evolution of metals and cosmic dust. We characterize the dust depletion of several metals from the Milky Way to distant galaxies. We collected measurements of ISM metal column densities from absorption-line spectroscopy in the literature, and in addition, we determined Ti and Ni column densities from a sample of 70 damped Lyman-$\alpha$ absorbers (DLAs) toward quasars that were observed at high spectral resolution with the Very Large Telescope (VLT) Ultraviolet and Visual Echelle Spectrograph (UVES). We used relative ISM abundances to estimate the dust depletion of 18 metals (C, P, O, Cl, Kr, S, Ge, Mg, Si, Cu, Co, Mn, Cr, Ni, Al, Ti, Zn, and Fe) for different environments (the Milky Way, the Magellanic Clouds, and DLAs toward quasars and towards gamma-ray bursts). We observed overall linear relations between the depletion of each metal and the overall strength of the dust depletion, which we traced with the observed [Zn/Fe]. The slope of these dust depletion sequences correlates with the condensation temperature of the various elements, that is, the more refractory elements show steeper depletion sequences. In the neutral ISM of the Magellanic Clouds, small deviations from linearity are observed as an overabundance of the $\alpha$-elements Ti, Mg, S, and an underabundance of Mn, including for metal-rich systems. The Ti, Mg, and Mn deviations completely disappear when we assume that all systems in our sample of OB stars observed toward the Magellanic Clouds have an $\alpha$-element enhancement and Mn underabundance, regardless of their metallicity. This may imply that the Magellanic Clouds have recently been enriched in $\alpha$-elements, potentially through recent bursts of star formation. We also observe an S overabundance in all local galaxies, which is an effect of ionization due to the contribution of their \ion{H}{ii} regions to the measured \ion{S}{II} column densities. The observed strong correlations of the depletion sequences of the metals all the way from low-metallicity quasi-stellar object DLAs to the Milky Way suggest that cosmic dust has a common origin, regardless of the star formation history, which, in contrast, varies significantly between these different galaxies. This supports the importance of grain growth in the ISM as a significant process of dust production.}

\authorrunning{C. Konstantopoulou et al.}
\titlerunning{Dust depletion of metals from local to distant galaxies I}
\maketitle

\section{Introduction}
Interstellar dust plays an important role in various physical and chemical processes that contribute to galactic evolution \citep{Dwek1998, Draine2011}. However, the composition and origin of dust grains in distant galaxies are not yet well understood.
The properties of dust have been studied more extensively in the Milky Way and in the Small Magellanic Cloud (SMC) and the Large Magellanic Cloud (LMC) \citep{Fitzpatrick&Massa1986ApJ, Pei1992, Gordon2003}. Characterizing dust at high redshift is more challenging, but crucial for improving our understanding of the early Universe \citep{Watson2015}. Recent observational constrains on the dust properties at early stages of chemical evolution are fundamental for complementing existing models of dust production and evolution \citep{Calura2008, Zhukovska2008, Arrigoni2010, Triani2020}.

Absorption-line spectroscopy has proven to be the most powerful tool of the various observational techniques for determining the composition of the interstellar medium (ISM) of a galaxy \citep{Prochaska1999, Savage1996}. It provides measurements of the column densities of different ions in the gas phase, from which the chemical abundances in the ISM of a galaxy can be determined. However, part of the metals are incorporated into dust grains. This effect is called dust depletion. It alters the relative abundances of the gas phase and challenges a comprehensive study of the ISM \citep{Field1974, Savage1996, Savaglio2003, Jenkins2009, DeCia2016, Roman-Duval2021}. It is therefore crucial to have a thorough understanding of the depletion of metals from the Local Group to high redshifts to be able to enhance our knowledge about the origin of cosmic dust and its composition \citep{Mattsson2019}.

In the Local Group, UV and optical absorption-line spectroscopy toward background stars can provide a detailed insight into the metal abundances of the elements in the ISM \citep{Field1974, York1983,Savage1996,Jenkins2017,Roman-Duval2021}. For distant galaxies, damped Lyman-$\alpha$ absorbers (DLAs, defined by log\,N$(\ion{H}{i})$ $\geq$ 20.3~cm$^{-2}$, e.g., \citealt{Wolfe2005}) toward bright background sources such as quasars (quasi-stellar objects, QSOs) and gamma-ray bursts (GRBs) are studied in absorption. DLAs predominantly trace galaxies at the low-mass end of the distribution \citep{Ledoux2006, Christensen2014}, and studying them contributes to our understanding of dust depletion, nucleosynthesis, metallicity, and molecular content across cosmic time \citep{Prochaska1999, Ledoux2002, Vladilo2002, Prochaska2007, Petitjean2008, Kulkarni2015, DeCia2018, Bolmer2019, Heintz2019a, Heintz2019b}. DLAs remain the best means for estimating elemental abundances in distant galaxies, allowing access to the study of elements in a wide range of environments and physical conditions, regardless of the luminosity of the absorbing galaxy.

Damped Lyman-$\alpha$ absorbers are the largest reservoirs of \ion{H}{i} in the Universe and carry the majority of metals at high-z \citep{Peroux2020}. Among the elements that can be observed in absorption, singly ionized titanium (\ion{Ti}{ii}) is unique. While Ti is an $\alpha$-element, it is also highly refractory and thus heavily depleted into dust grains, much more even than iron \citep[e.g.,][]{Jenkins2009, Ledoux2002}. These particular properties mean that Ti can be a tracer of dust, but also a tracer of $\alpha$-element enhancement. Titanium is dominant in $\ion{H}{i}$-dominated gas, with negligible contribution from ionized $\ion{H}{ii}$ gas because its ionization potential is almost equal to that of neutral hydrogen \citep{Viegas1995}. Most of the Ti in the Galactic ISM appears to be depleted into dust grains \citep{Jenkins2009}, but much less is known about Ti depletion in lower-metallicity systems, such as those found among QSO-DLAs. By constraining Ti at high redshifts, we can better understand the chemistry of the gas and the properties of dust in these galaxies.

\citet{Jenkins2009} used abundances of several elements in Milky Way stars to estimate the overall dust depletion in each line of sight. However, the main assumption of this method is that the metallicity of the gas is solar. This assumption cannot be made for DLAs because the gas metallicity in these systems can vary and can be significantly lower \citep{Pettini1994}. In addition, \citet{DeCia2021} showed that the ISM metallicity in galaxies is not necessarily uniform.
Instead, \citet{DeCia2016} characterized the dust depletion of several metals in QSO-DLAs and the Milky Way without any assumption on the gas metallicity. We expand this study and add more elements to the current dust depletion scheme. We study the depletion properties of C, P, O, Cl, Kr, S, Ge, Mg, Si, Cu, Co, Mn, Cr, Ni, Al, Ti, Zn, and Fe using the \citet{DeCia2016} formalism, which is based on the observations of relative abundances, in different environments, from the Local Group to distant galaxies. In Section 2 we present the samples that we used for our analysis. In Section 3 we describe the method for estimating the dust depletions. We discuss our results in Section 4, and we finally summarize and conclude in Section 5.

Throughout the paper we use a linear unit for the column densities N in terms of ions cm$^{-2}$. We refer to relative abundances of elements X and Y as $[X/Y] \equiv$ log$\frac{N(X)}{N(Y)}  -$ log$\frac{N(X)_{\odot}}{N(Y)_{\odot}}$, where reference solar abundances are reported in Table \ref{solabund}. We report 1\,$\sigma$ and 3\,$\sigma$ significance levels for the errors and limits, respectively.

\begin{table}[!h]
\caption{Solar abundances.}
\centering
\begin{tabular}{c c c c}
\hline\hline
Element & 12 + log(X/H)$_{\odot}$ $^{\rm a}$ & Source $^{\rm b}$\\
\hline
C  & 8.43 & s \\
P & 5.42 & a \\
O  & 8.69 & s \\
Cl & 5.23 & m \\
Kr & 3.12 & t \\
S  & 7.135 & a \\
Zn & 4.61 & m \\
Ge & 3.60 & a \\
Mg & 7.54 & a \\
Si & 7.51 & m \\
Cu & 4.25 & m \\
Co & 4.91 & a \\
Mn & 5.47 & m \\
Cr & 5.63 & a \\
Ni & 6.20 & a \\
Fe & 7.46 & a \\
Al & 6.43 & a \\
Ti & 4.90 & m \\
\hline
\label{solabund}
\end{tabular}
\vspace{-3mm}
\flushleft
{\footnotesize \textbf{Notes:}$^{\rm a}$ Reference solar abundances are adopted from \citet{Asplund2021}. $^{\rm b}$Photospheric (s), meteoritic (m), or average (a) abundances are used based on the recommendations from \citet{Lodders2009}.}
\end{table}

\section{Samples}
\label{sec:samples}
The full sample in this work consists of new column density measurements of QSO-DLAs as well as literature measurements of column densities in the neutral ISM in different galaxies: the Milky Way, the Magellanic Clouds, and QSO- and GRB-DLAs.
We used the QSO-DLA sample of \citet{DeCia2016} to make new measurements of Ti and Ni column densities, whenever possible. We only considered systems observed with high spectral resolution (R = 35\,000-58\,000) when measuring Ti and Ni column densities. For the Milky Way, we used ISM abundances from \citet{Jenkins2009}, \citet{DeCia2021}, \citet{Welty2010}, and \citet{Phillips1982}. For the LMC, we used \citet{Roman-Duval2021}, and for the SMC, we used \citet{Welty2010}, \citet{Tchernyshyov2015}, \citet{Jenkins2017}, the abundances of QSO-DLAs were taken from \citet{Berg2015}, \citet{DeCia2018}, and those for DLAs toward GRBs from \citet{Bolmer2019}.

Our sample probes gas in different regions of galaxies spanning a diversity of galaxy types and properties. GRB-DLAs select gas-rich galaxies, such as QSO-DLAs, but with active star formation and preferentially piercing through inner parts of these galaxies, while QSO-DLAs can be more peripheral \citep[e.g.,][]{Prochaska2007,Fynbo2008}. The Milky Way and Magellanic Cloud samples are also observations through the inner parts of galaxies. The Magellanic Clouds are in a different phase of their star formation history, with recent bursts of star formation \citep{HarrisZari2009, Indu2011, Crowther2016,Joshi2019, Bestenlehner2020, Hasselquist2021}, probably due to the proximity to the Milky Way and close encounters between the two Magellanic Clouds.

We collected measurements of 18 elements (C, P, O, Cl, Kr, S, Ge, Mg, Si, Cu, Co, Mn, Cr, Ni, Al, Ti, Zn, and Fe) from all the samples in order to characterize the dust depletion by estimating the relative abundances \citep{DeCia2016, Jenkins2017} of these elements. Different literature samples use different values for the oscillator strengths of the different ions. We homogenized them by correcting the column densities to the newest possible oscillator strengths for all the samples. We adopted the oscillator strengths from \citet{Cashman2017} for \ion{Kr}{i}, \ion{Mg}{ii}, \ion{Si}{ii}, \ion{Cu}{ii}, \ion{Co}{ii}, \ion{Mn}{ii}, \ion{Cr}{ii}, \ion{Al}{ii,} and \ion{Ti}{ii}, those from \citet{Boisse2019} for \ion{Ni}{ii}, those from \citet{Kisielius2015Zn} for \ion{Zn}{ii}, those from \citet{Kisielius2014S} for \ion{S}{ii,} and those from \citet{Kurucz2017} for P. When there was no recent change to the oscillator strength, the value from \citet{Morton2003} was used for the other metals.

\subsection{QSO- and GRB-DLAs}

The QSO-DLA sample is a compilation of new column density measurements and literature measurements \citep{DeCia2018, Noterdaeme2010, Ma2015, Fynbo2017, Noterdaeme2017}. The new QSO-DLA column density measurements were made using a large sample of 70 QSO-DLAs observed at high spectral resolution with the Very Large Telescope (VLT) Ultraviolet and Visual Echelle Spectrograph (UVES) from \citet{DeCia2016}. The procedure that we followed is described in Section \ref{sec:uves_sample}. Additionally, we used the constrained preexisting measurements of these QSO-DLAs to study the depletion of O, S, Si, Mg, Mn, and Cr. The literature sample includes one large QSO-DLA and one GRB-DLA sample. The main QSO-DLA sample was published in \citet{DeCia2018}. This contains a QSO-DLA sample observed with UVES from \citet{DeCia2016}, a large QSO-DLA sample from \citet{Berg2015}, and a few metal-rich QSO-DLAs from \citet{Ma2015}, \citet{Fynbo2017}, \citet{Noterdaeme2010} and \citet{Noterdaeme2017}. The \citet{Berg2015} catalog includes 395 QSO-DLAs, 44 of which are selected from the \citet{Herbert-Fort2006} catalog as metal-rich systems, and the rest are a literature compilation of column densities of all QSO-DLAs published between 1994 and 2014, with medium- or high- resolution requirement (R~\textgreater~10\,000). In the full QSO-DLA sample, 237 systems have both Zn and Fe constrained measurements.

Finally, we use a sample of 22 GRB-DLAs at redshifts z > 2 observed with VLT/X-Shooter with medium resolution (R~$\sim$~10\,000) from \citet{Bolmer2019}, 9 of which have both Zn and Fe constrained measurements. Our final QSO-DLA sample consists of 139 systems, of which 15 literature  and 13 new measurements were used to constrain the Ti depletion, 73 literature and 23 new measurements were used to constrain the Ni depletion, and the rest of the systems were used for the other elements.

\subsection{Milky Way and the Magellanic Clouds}

We included a sample of ISM metal column densities from the Milky Way, the LMC, and the SMC.
The Milky Way sample includes a compilation of elemental abundances from 243 lines of sight in the Milky Way \citep{Jenkins2009}, 25 bright O and B stars observed with the Hubble Space Telescope (HST) Space Telescope Imaging Spectrograph (STIS) with a resolving power of R~$\sim$~30\,000 \citep{DeCia2021} and Al measurements from \citet{Phillips1982}. The SMC sample is composed of elemental abundances toward 19 lines of sight from \citet{Welty2010}, \citet{Tchernyshyov2015}, \citet{Jenkins2017}. The LMC sample counts 32 abundances of massive stars observed with HST/STIS from \citet{Roman-Duval2021}. Of these, 39 galactic, 16 SMC, and 31 LMC lines of sight have Zn and Fe constrained measurements.

\begin{table}[!h]
\caption{Number of Ti and Ni measurements for the Milky Way, QSO-DLAs, the LMC, SMC, and GRB-DLAs. References: ($\star$) This work; (1) \citet{Berg2015} ;(2) \citet{Jenkins2009}; (3) \citet{DeCia2021}; (4) \citet{Welty2010}; (5) \citet{Tchernyshyov2015}; (6) \citet{Jenkins2017}; (7) \citet{Roman-Duval2021}; (8) \citet{Bolmer2019}.}
\centering
\begin{tabular}{c c c c}
\hline\hline
Environments & Ti & Ni & References\\
\hline
Milky Way & 31  & 23 & (2), (3)\\
DLAs & 28 & 98 & ($\star$), (1) \\
LMC & 10 & 29 & (7) \\
SMC & 13 & 12 & (4), (5), (6) \\
GRBs & 1 & 8 & (8) \\
\hline
\label{numberofmeasurements}
\end{tabular}
\end{table}

\section{Methods and results}
\label{sec:results}

\subsection{New column density measurements. Voigt-profile fitting}
\label{sec:uves_sample}

We determined the column densities of Ti and Ni in QSO-DLA spectra from UVES/VLT using the Voigt-profile fit method, which decomposes the line profiles into their individual velocity components and fits all transitions simultaneously. Using the Python package VoigtFit\footnote{\url{https://github.com/jkrogager/VoigtFit}} \citep{Krogager2018}, we modeled \ion{Ti}{ii} ($\lambda$1910 $\AA$, $\lambda$3067 $\AA$, $\lambda$3073 $\AA$, $\lambda$3230 $\AA$, $\lambda$3242 $\AA$, and $\lambda$3384 $\AA$) and \ion{Ni}{ii} lines ($\lambda$1317 $\AA$, $\lambda$1370 $\AA$, and $\lambda$1741 $\AA$) when they were covered by the UVES spectra. To better constrain the fit, we also modeled transitions of \ion{Fe}{ii}, \ion{Si}{ii,} and \ion{Cr}{ii}. In some cases, we included \ion{Zn}{ii}, \ion{S}{ii}, \ion{Mn}{ii}, \ion{Mg}{ii,} and \ion{P}{ii}. All the velocity profiles of the modeled transitions are shown in Appendix \ref{appsec: fits}.

We adopted the individual velocity components, namely the redshift z and the broadening parameter b, that are listed in Table F.1 of \citet{DeCia2016}, using \ion{Fe}{ii} as reference. We visually inspected the \ion{Fe}{ii} lines, and when they appeared to be saturated, we used \ion{Si}{ii} instead. For Q0347-383, Q0528-250a, Q0528-250b, Q0551-366, Q0841+129c, Q2138-444b, and Q2231-002, we added additional components that we identify in the line profile.  In addition to \ion{Ti}{ii} and \ion{Ni}{ii},
we measured \ion{Zn}{ii} using the transitions $\lambda$2026 $\AA$ and $\lambda$2062 $\AA$ for Q0102-190a and Q0841+129a, respectively.
The contaminating features were masked out, and the continuum level was automatically placed using Chebyshev polynomials. Contaminated lines can bias the  continuum placement and the resulting fit. When needed, the interactive masking was repeated until the fit was reasonable. Due to the high column density of neutral hydrogen in QSO-DLAs, which shields metals against higher ionizations, no ionization corrections are needed to derive the metal abundances. We adopted the newest oscillator strengths from \citet{Cashman2017} for \ion{Ti}{ii}, those from \citet{Boisse2019} for \ion{Ni}{ii,} and those from \citet{Kisielius2015Zn} for \ion{Zn}{ii}. For \ion{Fe}{ii,} there is no recent change to the oscillator strength, and therefore the value from \citet{Morton2003} was used.

When \ion{Ti}{ii} or \ion{Ni}{ii}  were not detected, we estimated the 3$\sigma$ upper limits because the lines are not strongly contaminated. We used the line profile of a well-constrained transition (\ion{Fe}{ii} or \ion{Si}{ii}) as reference. The reference was used to determine the limits of the line profile containing 99$\%$ of the optical depth, within which the upper limit was calculated.

For some ions (in systems Q0405-443c, Q1117-134, and Q1223+178), the fit did not converge properly although there was a detection. To constrain the fit in these cases, we fixed the line profile using redshift z- and b-values from a fit of another previously well-constrained ion (e.g., \ion{Fe}{ii} or \ion{Si}{ii}).

The total number of Ti and Ni measurements from all the samples that we used are summarized in Table \ref{numberofmeasurements}.
The total column densities as well as the upper limits are reported in Table \ref{uvestab}. Out of the 70 QSO-DLAs that were studied, we were able to obtain a constrained measurement of \ion{Ti}{ii} from only 13 systems, and 18 limits were estimated. \ion{Ni}{ii} was measured in 23 systems, and 7 limits were estimated. Combining these with the QSO-DLA literature sample, we have 28 Ti and 98 Ni constrained measurements in total.

The velocity profiles of all the modeled \ion{Ti}{ii} and \ion{Ni}{ii} transitions are shown in Appendix \ref{appsec: fits}. The total column densities were used to characterize the depletion of Ti and Ni with the relative method. They are reported in Table \ref{dlastable}.

\begin{table}[!h]
\caption{New UVES QSO-DLA Ti and Ni measurements}
\centering
\small
\setlength\tabcolsep{1pt}
\begin{tabular}{l c c c c c}
\hline\hline
QSO & $z_{\rm abs}$ & log N(\ion{Fe}{II})&log N(\ion{Zn}{II})&log N(\ion{Ni}{II})&log N(\ion{Ti}{II})\\\hline
Q0010-002  & 2.02484 & 15.18$\pm$0.04 & 12.09$\pm$0.04 & 13.69$\pm$0.03    & \textless{12.38}    \\
Q0058-292  & 2.67142 & 14.75$\pm$0.02 & 12.14$\pm$0.02 & 13.55$\pm$0.08    & 12.30$\pm$0.11      \\
Q0100+130  & 2.30903 & 15.05$\pm$0.02 & 12.34$\pm$0.01 & ... & 12.72$\pm$0.11      \\
Q0102-190a & 2.36966 & 14.44$\pm$0.02 & 11.61$\pm$0.08 & 13.16$\pm$0.07    & 12.20$\pm$0.12      \\
Q0216+080a & 1.76873 & 14.47$\pm$0.02 & 11.87$\pm$0.06 & \textless{13.08}  & \textless{12.22}    \\
Q0347-383  & 3.02485 & 14.35$\pm$0.02 & 12.13$\pm$0.05 & \textless{13.70} & \textless{11.92}    \\
Q0405-443a & 1.91267 & 15.16$\pm$0.02 & 12.34$\pm$0.03 & 13.87$\pm$0.01    & 12.48$\pm$0.02      \\
Q0405-443b & 2.55000 & 15.05$\pm$0.01 & 12.36$\pm$0.04 & 13.76$\pm$0.01    & \textless{12.22}    \\
Q0405-443c & 2.59466 & 15.08$\pm$0.02 & 12.50$\pm$0.01 & 13.80$\pm$0.01    & 12.55$\pm$0.05      \\
Q0458-020  & 2.03956 & 15.33$\pm$0.05 & 13.05$\pm$0.02 & 14.06$\pm$0.03    & \textless{12.72}   \\
Q0528-250a & 2.14105 & 14.78$\pm$0.02 & 12.19$\pm$0.03 & 13.65$\pm$0.01    & 12.62$\pm$0.07      \\
Q0528-250b & 2.81111 & 15.48$\pm$0.01 & 13.01$\pm$0.01 & 14.29$\pm$0.01    & 12.76$\pm$0.06      \\
Q0551-366  & 1.96221 & 15.05$\pm$0.03 & 12.92$\pm$0.03 & 14.09$\pm$0.01    & \textless{12.56}    \\
Q0841+129a & 1.86384 & 14.89$\pm$0.04 & 11.94$\pm$0.11 & 13.60$\pm$0.02    & 12.38$\pm$0.05      \\
Q0841+129b & 2.37452 & 14.72$\pm$0.02 & 12.03$\pm$0.02 & 13.55$\pm$0.01    & \textless{13.06}    \\
Q0841+129c & 2.47622 & 14.48$\pm$0.02 & 11.77$\pm$0.03 & 13.30$\pm$0.02    & \textless{11.94}    \\
Q1111-152  & 3.26552 & 14.82$\pm$0.02 & 12.22$\pm$0.09 & 13.56$\pm$0.01    & ...\\
Q1117-134  & 3.35046 & 14.82$\pm$0.03 & 12.11$\pm$0.05 & \textless{13.53}  & 12.49$\pm$0.05      \\
Q1157+014  & 1.94349 & 15.47$\pm$0.01 & 12.93$\pm$0.01 & 14.19$\pm$0.01    & 12.99$\pm$0.01      \\
Q1209+093  & 2.58437 & 15.36$\pm$0.02 & 12.96$\pm$0.02 & 14.16$\pm$0.04    & \textless{12.60}    \\
Q1223+178  & 2.46607 & 15.21$\pm$0.01 & 12.34$\pm$0.02 & 13.86$\pm$0.05    & 12.73$\pm$0.06      \\
Q1331+170  & 1.77635 & 14.62$\pm$0.02 & 12.44$\pm$0.03 & 13.41$\pm$0.02    & \textless{11.95}    \\
Q1409+095a & 2.01881 & 14.34$\pm$0.02 & 11.60$\pm$0.13 & 13.15$\pm$0.08    & \textless{12.88}    \\
Q1444+014  & 2.08679 & 13.99$\pm$0.03 & 12.02$\pm$0.06 & \textless{12.81}  & \textless{12.43}    \\
Q1451+123a & 2.25466 & 14.42$\pm$0.05 & 11.85$\pm$0.11 & \textless{13.60}  & \textless{12.97}    \\
Q2116-358  & 1.99615 & 14.72$\pm$0.05 & 12.33$\pm$0.08 & 13.61$\pm$0.02    & \textless{12.43}    \\
Q2138-444a & 2.38279 & 14.54$\pm$0.01 & 12.02$\pm$0.08 & \textless{12.90}   & \textless{12.64}    \\
Q2138-444b & 2.85234 & 14.63$\pm$0.02 & 11.81$\pm$0.02 & 13.33$\pm$0.02    & \textless{11.83}    \\
Q2206-199a & 1.92061 & 15.34$\pm$0.01 & 12.70$\pm$0.01 & 14.16$\pm$0.02    & 12.86$\pm$0.02      \\
Q2243-605  & 2.33062 & 14.95$\pm$0.01 & 12.38$\pm$0.02 & 13.83$\pm$0.02    & 12.64$\pm$0.03      \\
Q2332-094a & 2.28749 & 14.23$\pm$0.03 & 12.31$\pm$0.02 & 13.25$\pm$0.04    & ...         \\
Q2343+125  & 2.43123 & 14.51$\pm$0.02 & 12.05$\pm$0.02 & 13.36$\pm$0.03    & \textless{11.97} \\
Q2359-022a & 2.09510  & 14.47$\pm$0.01 & 12.38$\pm$0.08 & \textless{13.59}  & \textless{12.51}   \\
\hline
\label{uvestab}
\end{tabular}
\vspace{-3mm}
\flushleft
{\footnotesize \textbf{Notes:} Upper limits are given at the 3$\sigma$ confidence level. The redshifts and the column densities for Fe and Zn are taken from \citet{DeCia2016}, except for Q0102-190a and Q0841+129a, for which we made new Zn measurements.}
\end{table}

\subsection{Estimating the dust depletions}

Metals are missing from the observable gas-phase and are instead depleted into dust grains. The depletion of an element X, $\delta_{x}$, is defined as
\begin{equation}
    \delta_{X} = [X/H]_{\rm{gas}} - [X/H]_{\mathrm{tot}},
    \label{deltax1}
\end{equation}
where [X/H]$_{\rm{gas}}$ are the observed gas-phase abundances, and [X/H]$_{\mathrm{tot}}$ are the total (gas + dust) abundances. \citet{DeCia2016} showed that the depletion of an element X can be purely estimated from some relative abundances. In general, the relative abundance of an element X with respect to another element Y ([X/Y]) is more sensitive to the depletion of X if Y is a volatile element. Thus, the depletion of X can be derived from the observed [X/Zn] after correcting for the mild depletion of Zn, which we discuss below. The depletion of various metals X correlates with the overall strength of the depletion, or dust tracer, for instance, [Zn/Fe]. These correlations are referred to as depletion sequences \citep{DeCia2016}.

To derive the depletion of X from the observed [X/Zn], we assumed a depletion of Zn, $\delta_{\rm{Zn}}$ = -0.27 $\times$ [Zn/Fe], as derived from the Milky Way and QSO-DLAs by \citet{DeCia2016}. We also corrected for $\alpha$-element enhancement and Mn underabundance, which is further discussed in the next section.
Taking these assumptions into account, $\delta_{X}$ was then derived as
\begin{equation}
    \delta_X = \mathrm{[X/Zn]} + \delta_{Zn} - \alpha_{X},
    \label{deltax}
\end{equation}
where $\alpha_{\rm{X}}$ is the intrinsic [X/Fe]$_{\rm{nucl}}$ due to stellar nucleosynthesis, for example, $\alpha$-element enhancement or Mn underabundance.

A good dust tracer can be any relative abundance of two elements ([X/Y]) that have very different refractory properties (i.e., they deplete very differently), but follow each other in nucleosynthesis. In this paper, we use [Zn/Fe] as a dust tracer (but other relative abundances, e.g., [Si/Ti] or [O/Si], can be used). [Zn/Fe] is the ratio of two elements with very different degrees of incorporation into dust grains, with Zn being a volatile element and Fe a refractory element. [Zn/Fe] is also called the depletion factor and is often used as a tracer of the overall dust depletion in QSO-DLAs \citep{Noterdaeme2008}. Zn and Fe trace each other \citep{Timmes1995} in the metallicity range considered here (-2 $\leq$ [M/H] $\leq$ 0). Stellar measurements show that [Zn/Fe]~$\sim$~0 in the metallicity range  -2 $\leq$ [M/H] $\leq$ 0 \citep[e.g.,][]{Saito2009,Barbuy2015}, which is appropriate for most QSO-DLAs \citep{DeCia2018}. Outside this metallicity range, [Zn/Fe] appears to be supersolar in metal-poor halo stars \citep{Primas2000, Nissen2004, Nissen2007} and subsolar in the bulge \citep{Barbuy2015}. Moreover, [Zn/Fe] correlates well with [Si/Ti] \citep[see Fig. A.1. of][]{DeCia2016}. Si and Ti follow each other nucleosynthetically, and if Zn did not trace Fe, then we would see a deviation from the linear relation, but this is not observed. More details about the reliability of [Zn/Fe] as a dust tracer can be found in \citet{DeCia2018}.

\subsection{Corrections for $\alpha$-element enhancement and Mn underabundance}
\label{nucl_corr}

\begin{figure*}[!t]
    \centering
    \includegraphics[width=\textwidth]{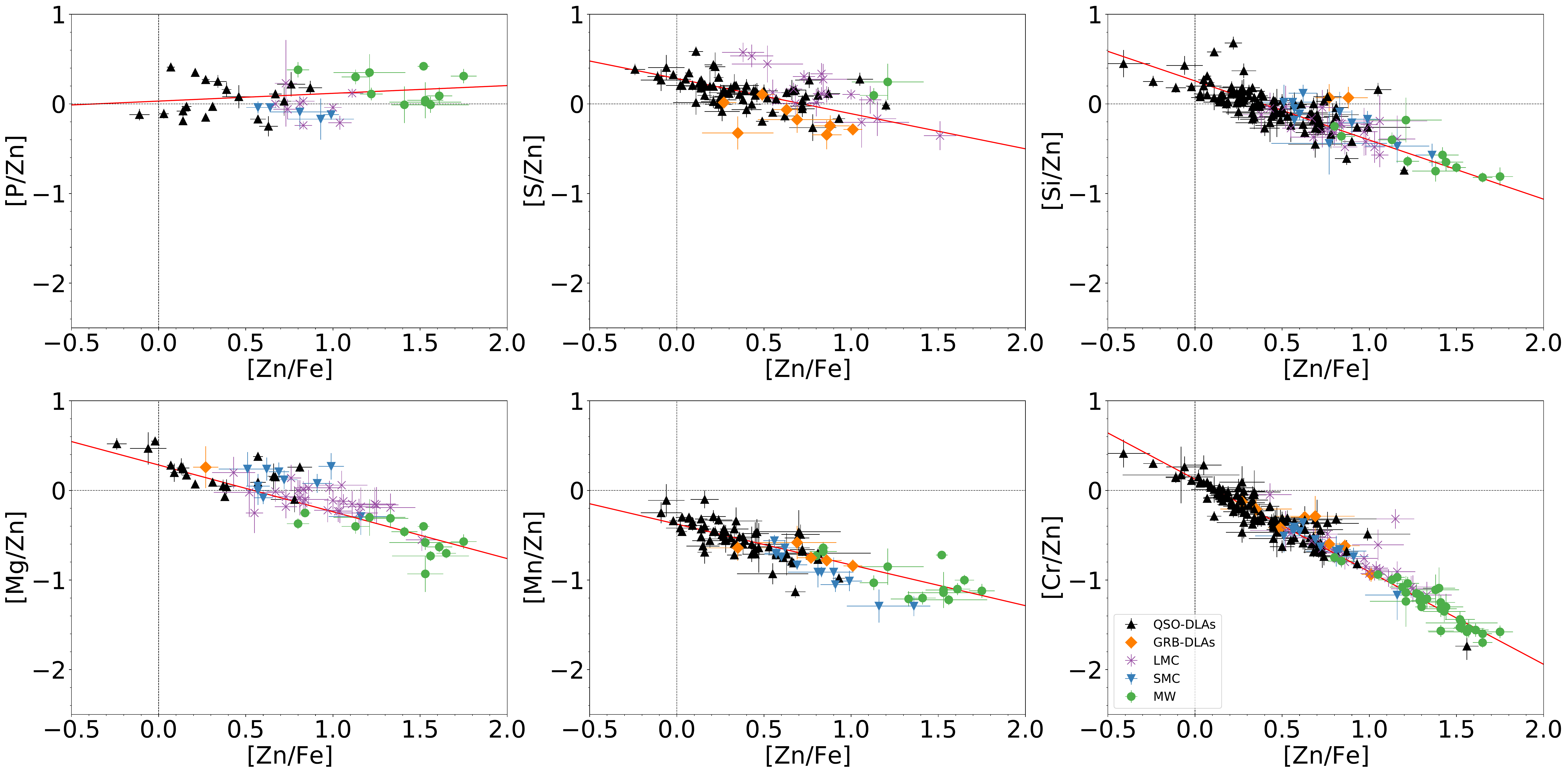}
    \caption{Relative abundances of element X with respect to Zn against [Zn/Fe]. The black triangles show QSO-DLAs, the green dots show the Milky Way, the purple stars show the LMC, the blue triangles show the SMC, and the orange diamonds show the GRB-DLAs. The red line shows the linear fit to the data.The slope becomes steeper for elements that are more strongly depleted into dust grains.}
    \label{fig:RelAbund}
\end{figure*}

The depletion $\delta_{X}$ is by definition a negative value (eq. \ref{deltax1}). Figure \ref{fig:RelAbund} shows the observed gas-phase relative abundances [X/Zn] against [Zn/Fe] before any dust or $\alpha$-element enhancement corrections. Fig. \ref{fig:RelAbund} shows that at the offsets of [X/Zn] with [Zn/Fe] = 0 for some elements, [X/Zn] is positive. This is due to an enhancement of the $\alpha$-elements observed in the gas phase as a result of the chemical enrichment of the ISM by supernovae. Core-collapse supernovae produce large amounts of $\alpha$-elements, such as O, Si, S, Mg, and Ti, with only a smaller production of Fe-peak elements \citep{Nomoto2006}. This results in an observed overabundance of the $\alpha$-elements with respect to Fe ([$\alpha$/Fe]). On the other hand, type~Ia supernovae mostly produce the Fe-peak elements and a smaller amount of $\alpha$-elements, leading to a linear decrease of [$\alpha$/Fe] until [$\alpha$/Fe] $\sim$ 0. Mn is mostly produced by type~Ia supernovae, leading to a Mn underabundance compared to Fe before the contribution of SNe~Ia becomes strong \citep{Nomoto1997}. We refer to these effects as nucleosynthesis effects because they are the product of explosive nucleosynthesis in stars that enrich the ISM with metals.

We applied corrections to account for effects such as $\alpha$-element enhancement and Mn underabundance. To do this, we adopted the shapes of the nucleosynthetic curves from \citet{DeCia2016} (see their Fig. 7). In particular, we assumed that at lower metallicities and low [Zn/Fe], there is an $\alpha$-element enhancement and Mn underabundance ($\alpha_{\rm{X},0}$), while at higher metallicities and at high [Zn/Fe], there is a plateau at $\alpha_{\rm{X},0}$ = 0.05. In between, $\alpha_{\rm{X},0}$ decreases linearly with [Zn/Fe]. The shapes of the curves are consistent with the nucleosynthetic patterns observed in the Galaxy (see \citealt{Lambert1987}, \citealt{McWilliam1997}, for O, Si, Mg, and S in galactic stars and in \citealt{Wheeler1989}, \citealt{Mishenina2015} and \citealt{Battistini2015} for Mn). The values of $\alpha_{\rm{X},0}$ are obtained by the intercepts of the relative abundances [X/Zn] at [Zn/Fe] = 0. They are $\alpha_{\mathrm{O},0}$ = 0.38 $\pm$ 0.10, $\alpha_{\mathrm{S},0}$ = 0.25 $\pm$ 0.03, $\alpha_{\mathrm{Si},0}$ = 0.26 $\pm$ 0.03, $\alpha_{\mathrm{Mg},0}$ = 0.30 $\pm$ 0.04, and $\alpha_{\mathrm{Mn},0}$ = -0.39 $\pm$ 0.03. For Ti, we used $\alpha_{\mathrm{Ti},0}$ = 0.29 $\pm$ 0.08, estimated by \citet{McWilliam1997} for the Galaxy, which is identical to the intercept of [Ti/Zn] with [Zn/Fe] = 0 that we measure, $\alpha_{\mathrm{Ti},0}$ = 0.29 $\pm$ 0.03.

While the plateau and the overall shape of the $\alpha$-element enhancement and Mn underabundance distribution with metallicity is rather well constrained, the position of the knee (the metallicity at which $\alpha_{\rm{X},0}$ starts to decrease) is unknown for QSO-DLAs, and we may expect it at lower metallicities \citep{deBoer2014} because we expect QSO-DLA galaxies to have lower stellar masses overall \citep{Christensen2014, Krogager2017}. For the LMC, the $\alpha$-element knee is poorly constrained and might be at similar metallicities as in the Milky Way or at [M/H] $\sim$ -1.5~dex \citep{deBoer2014}.
For the Magellanic Clouds and for QSO-DLAs, we assumed the middle point (at [Fe/H = -1.3]) of the knee position range estimated by \citet{deBoer2014} (see their Fig. 3). The approximate assumption of the position of the $\alpha$-element knee for these environments can be improved with future observations. Nevertheless, the exact positioning of the $\alpha$-element knee has a very small effect on the $\alpha$-element correction and therefore does virtually not affect our results. In addition, we explored the possibility of a different assumption for the Magellanic Clouds, a constant distribution of $\alpha$-element enhancement (and Mn underabundance), as we discuss in Section \ref{subsec: nucl_assump}.

Using these assumptions on the effects of $\alpha$-element enhancement and Mn underabundance, we derived the depletions using eq. \ref{deltax}. Corrections for these effects were taken into account for the derivations of the depletions, but contribute very little ($\leq$ 0.4~dex) compared to the strong dust depletion effects on the gas-phase metal abundances, which can reach up to 3~dex for highly depleted metals such as Ti \citep{Jenkins2009}.

\subsection{Dust-depletion sequences}
We followed the method of \citet{DeCia2016} to characterize the depletion of several elements in different environments (QSO-DLAs, GRB-DLAs, Milky Way, LMC, and SMC) using a literature compilation that includes systems with Zn and Fe constrained measurements. In their method, \citet{DeCia2016} use the relative abundance of an element X with respect to a lightly depleted element, such as Zn, P, or S, to estimate the dust depletion of element X ($\delta_{\rm{X}}$). The depletion of X correlates linearly with the dust tracer (e.g., [Zn/Fe]), and thus the dust depletion can then be expressed by a simple linear relation. We characterize the dust depletion of P, S, Si, Mg, Mn, and Cr in the above environments and further focus on the dust depletion of Ti and Ni, which have not been characterized before with this method and for which we provide new column density measurements. We extend our study to additionally characterize the dust depletion of  O, C, Cl, Kr, Ge, Cu, Co, and Al for the Milky Way and QSO-DLAs, but only a few measurements are available at low levels of depletion (QSO-DLAs), and only two LMC measurements are added to the O depletion sequence.

We fit the dust depletion sequences as derived from the observed relative abundances and after correcting for the depletion of Zn and for $\alpha$-element enhancement or Mn underabundance with a linear least-squares approximation that took the errors on both x-axis ([Zn/Fe]) and y-axis ($\delta_{X}$) ($\sigma_{[Zn/Fe]}$ and $\sigma_{\delta_{X}}$) into account. We used the Python package {\tt scipy.odr}, which performs orthogonal distance regression to the data and determines the parameters that minimize the sum of the squared error for each data point. It uses the ODRPACK library \citep{Boggs1992}, which implements a modified trust-region Levenberg-Marquardt-type algorithm to estimate the function parameters. The dust depletion can be represented by linear fits to the data as
\begin{equation}
    \delta_X = A2_X +B2_X \times \mathrm{[Zn/Fe]},
    \label{deltaxcoeff}
\end{equation}
where the A2$_{X}$ and B2$_{X}$ coefficients for all the elements are reported in Table \ref{coefficients}. These are given for two different assumptions on the nucleosynthetic curves of the Magellanic Clouds, one with the knee at [Fe/H] = -1.3, and one assuming a constant over- or underabundance plateau. Although the differences are small, we recommend that the second assumption coefficients are used.

\begin{figure*}[!t]
    \centering
    \includegraphics[width=\textwidth]{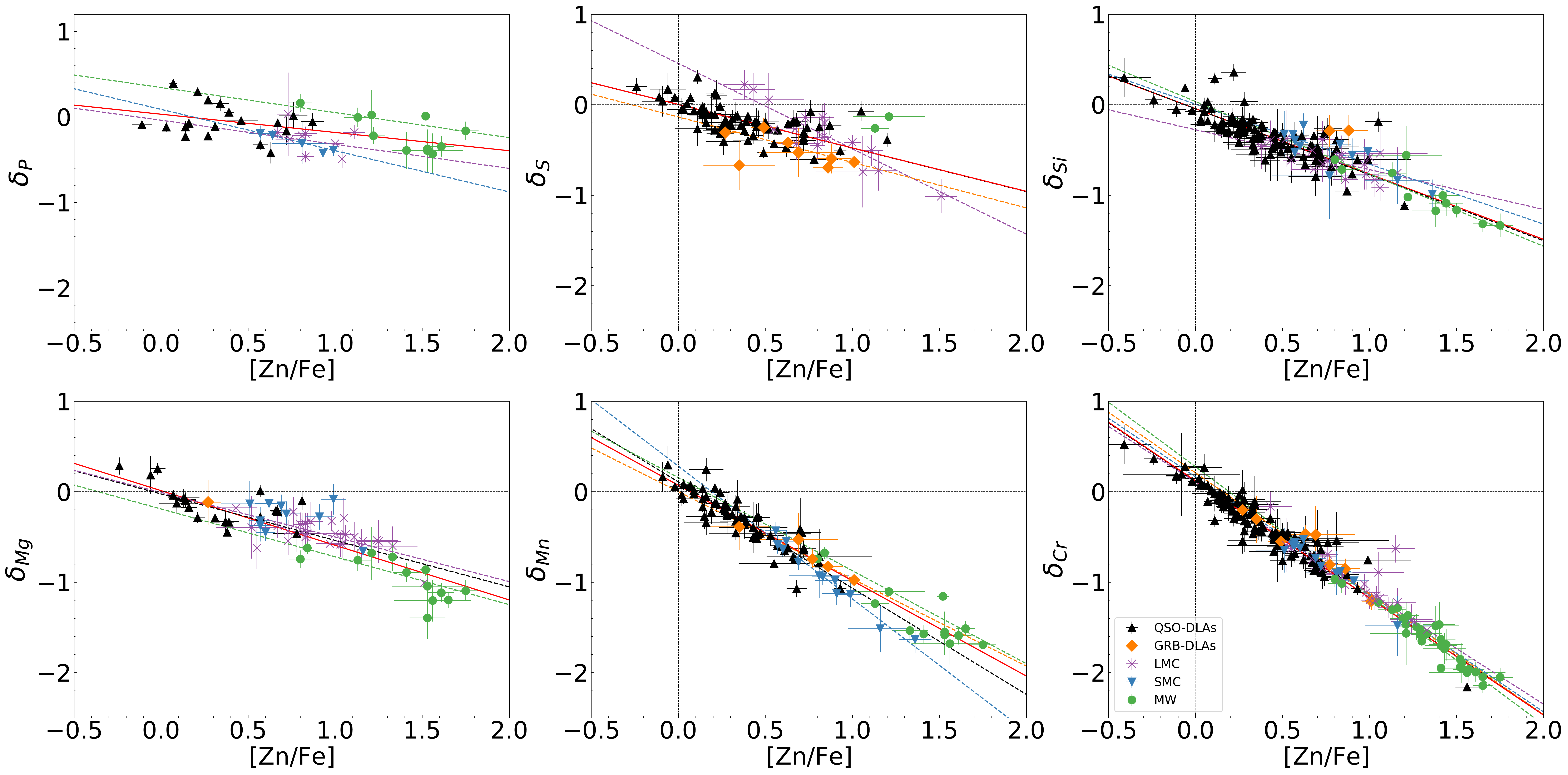}
    \caption{Dust depletion of element X ($\delta_{X}$) against [Zn/Fe]. The black triangles show QSO-DLAs, the green dots show the Milky Way, the purple stars show the LMC, the blue triangles show the SMC, and the orange diamonds show the GRB-DLAs. The red line shows the linear fit to the data. The slope becomes steeper for elements that are more strongly depleted into dust grains.}
    \label{fig:DeplSeq}
\end{figure*}

Figure \ref{fig:DeplSeq} shows the depletion sequences of the larger coverage elements (P, S, Si, Mg, Mn, and Cr). We discuss the elements with limited coverage in Appendix \ref{limited_cov}. For Ti, Ni, P, S, Si, Mg, Mn, and Cr, we performed fits to the individual environments, but also to all the data merged together. We also fit the relative abundances of element X with respect to Zn against [Zn/Fe]. The coefficients A1$_{X}$ and B1$_{X}$ were derived from the linear fit of the observed sequences of relative abundances,
\begin{equation}
    \mathrm{[X/Y]} = A1_X +B1_X \times \mathrm{[Zn/Fe]}.\end{equation}

The best merged fit is shown with solid red lines in Fig. \ref{fig:TiNi}, from which we derived the coefficients A1$_{X}$, B1$_{X}$, A2$_{X}$ , and B2$_{X}$. We report the coefficients A2$_{X}$ and B2$_{X}$ for the merged fit and the individual environments, the intrinsic scatter $\sigma_{\delta_{\rm{X}}}$ of the data, estimated as the standard deviation of the x and y-component residuals for each data point, and the statistical properties of the fits to the individual sample and the merged data  in Tables \ref{titable} and \ref{nitable}.

\begin{table*}[!t]
\caption{Coefficients A2$_{X}$ and B2$_{X}$ resulting from the linear merged fit $\delta_{\rm{X}}$ = A2$_{X}$ +B2$_{X}$ $\times  \mathrm{[Zn/Fe]}$ of the depletion sequences shown in Figures~\ref{fig:TiNi}, \ref{fig:DeplSeq}, \ref{fig:DeplSeq_volatile}, the degrees of freedom $\nu,$ and the Pearson correlation coefficients r. Exception are Zn and Fe, which are taken from \citet{DeCia2016}. $^a$ r = -1 by construction. The coefficients are given for two assumptions on the knee position for the Magellanic Clouds. We recommend that the one with constant $\alpha_{\rm{X}}$ is used.}
\label{coefficients}
\centering
\begin{tabular}{c|ccc|ccc|c}
 \hline
 \multirow{3}{*}{X} &
 \multicolumn{3}{c|}{SMC/LMC knee at [$\rm{Fe/H}$] = -1.3} & \multicolumn{3}{c|}{Constant $\alpha_{\rm{X}}$} &  \multirow{3}{*}{$\nu$}\\
 & &&&&&\\
 & A2$_{x}$ & B2$_{x}$ & r & A2$_{x}$ & B2$_{x}$ & r &  \\ \hline\hline
C  & 0.00   & -0.10$\pm$0.10 & -0.10             &         0.00   & -0.10$\pm$0.10 & -0.10  & 3   \\
P & 0.03$\pm$0.05   & -0.21$\pm$0.08 & -0.46 & 0.03$\pm$0.05   & -0.21$\pm$0.08 & -0.46   & 43 \\
O  & 0.00$\pm$0.00   & -0.20$\pm$0.05 & -0.31                        & 0.00$\pm$0.00  & -0.20$\pm$0.05 & -0.50 & 22  \\
Cl & 0.00   & -0.12$\pm$0.09 & -0.80                        & 0.00   & -0.12$\pm$0.09 & -0.80   & 6   \\
Kr & 0.00  & -0.04$\pm$0.09 & -0.53                        & 0.00  & -0.04$\pm$0.09 & -0.53  & 3   \\
S  & 0.01$\pm$0.02   & -0.48$\pm$0.04 & -0.68  & 0.01$\pm$0.02  & -0.48$\pm$0.04 & -0.68 & 99  \\
Zn & 0.00$\pm$0.01   & -0.27$\pm$0.03 & -1.00$^{\rm a}$ & 0.00$\pm$0.01   & -0.27$\pm$0.03 & -1.00$^{\rm a}$ & ... \\
Ge & 0.00   & -0.40$\pm$0.04 & -0.91                        & 0.00   & -0.40$\pm$0.04 & -0.91   & 1   \\
Mg & 0.01$\pm$0.03 & -0.60$\pm$0.04 & -0.86    & 0.01$\pm$0.03  & -0.66$\pm$0.04 & -0.91 & 69  \\
Si & -0.04$\pm$0.02  & -0.72$\pm$0.03 & -0.87   & -0.04$\pm$0.02 & -0.75$\pm$0.03 & -0.88 & 151 \\
Cu & 0.00   & -0.73$\pm$0.04 & -0.87                        & 0.00   & -0.73$\pm$0.04 & -0.87   & 4   \\
Co & 0.00  & -0.89$\pm$0.19 & -0.87                        & 0.00  & -0.89$\pm$0.19 & -0.87   & 6   \\
Mn & 0.07$\pm$0.02   & -1.06$\pm$0.03 & -0.96 & 0.07$\pm$0.02  & -1.03$\pm$0.03 & -0.96 & 93  \\
Cr & 0.12$\pm$0.01   & -1.30$\pm$0.01 & -0.98 & 0.12$\pm$0.01   & -1.30$\pm$0.01 & -0.98   & 192 \\
Ni & 0.07$\pm$0.02   & -1.31$\pm$0.03 & -0.96 & 0.07$\pm$0.02   & -1.31$\pm$0.03 & -0.96   & 168 \\
Fe & -0.01$\pm$0.03  & -1.26$\pm$0.04 & -1.00$^{\rm a}$ &-0.01$\pm$0.03  & -1.26$\pm$0.04 & -1.00$^{\rm a}$ & ...\\
Al & 0.00   & -1.66$\pm$0.35 & -0.69                        & 0.00   & -1.66$\pm$0.35 & -0.69  & 7 \\
Ti & -0.06$\pm$0.03  & -1.64$\pm$0.04 & -0.97                        & -0.07$\pm$0.03 & -1.67$\pm$0.04 & -0.97 & 81
\\ \hline
\end{tabular}
\end{table*}


\begin{figure*}
    \centering
        \includegraphics[width=\textwidth]{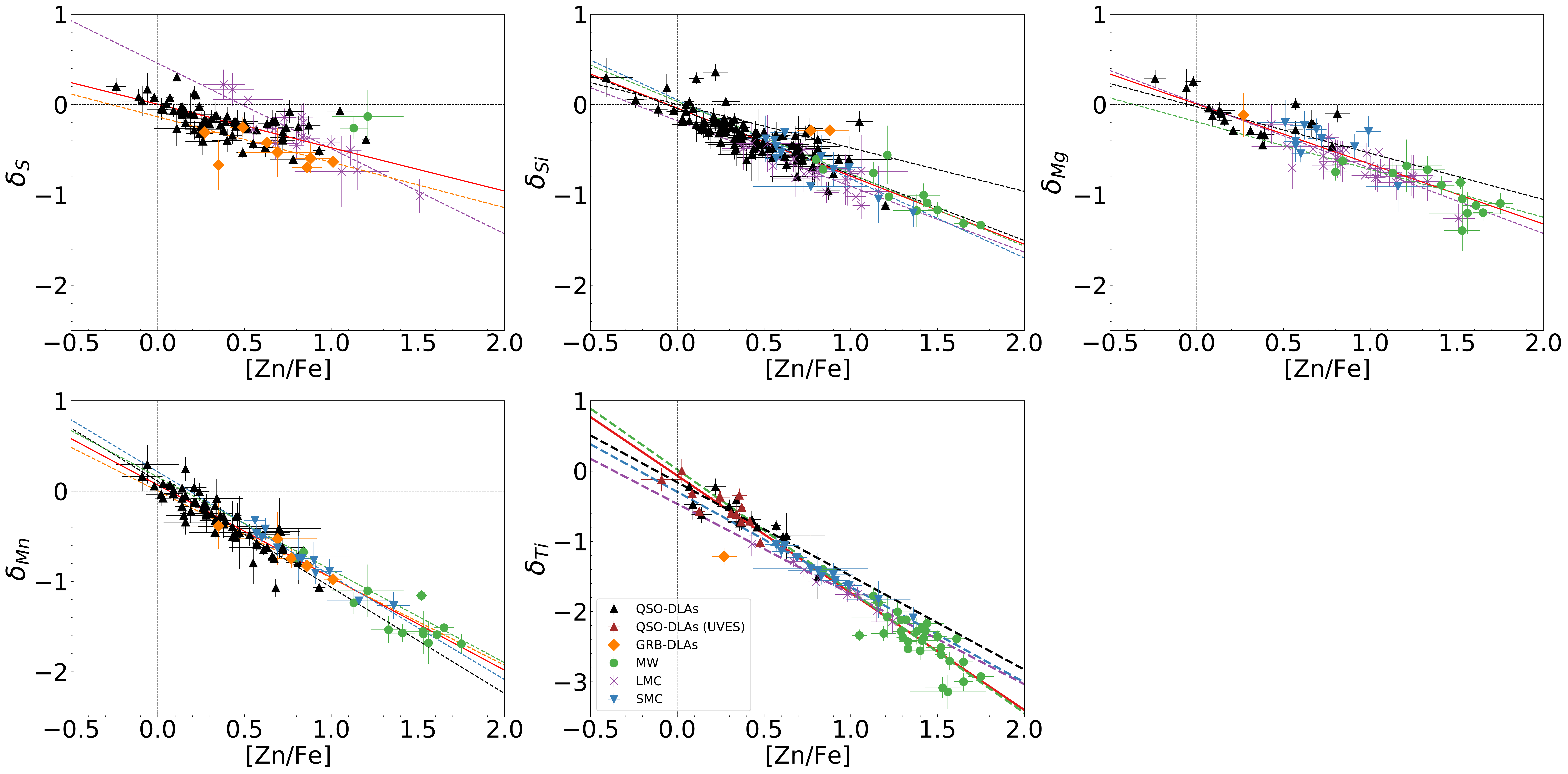}
    \caption{Same as Fig. \ref{fig:DeplSeq} for the $\alpha$-elements and Mn after assuming a constant overabundance for the $\alpha$-elements and an underabundance plateau for Mn in the Magellanic Clouds.}
    \label{fig:constMCs}
\end{figure*}

\begin{figure*}[!ht]
    \centering
    \includegraphics[width=\textwidth]{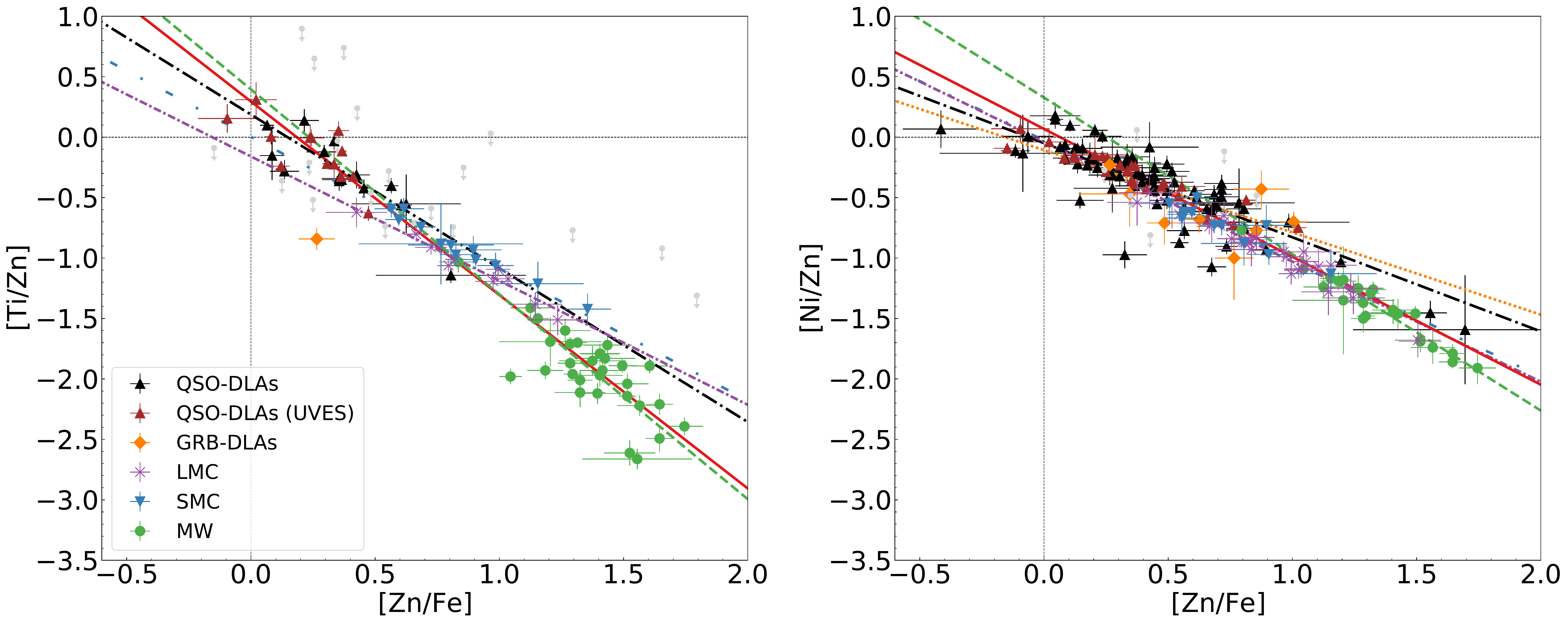}
\end{figure*}
\begin{figure*}[!ht]
    \centering
    \includegraphics[width=\textwidth]{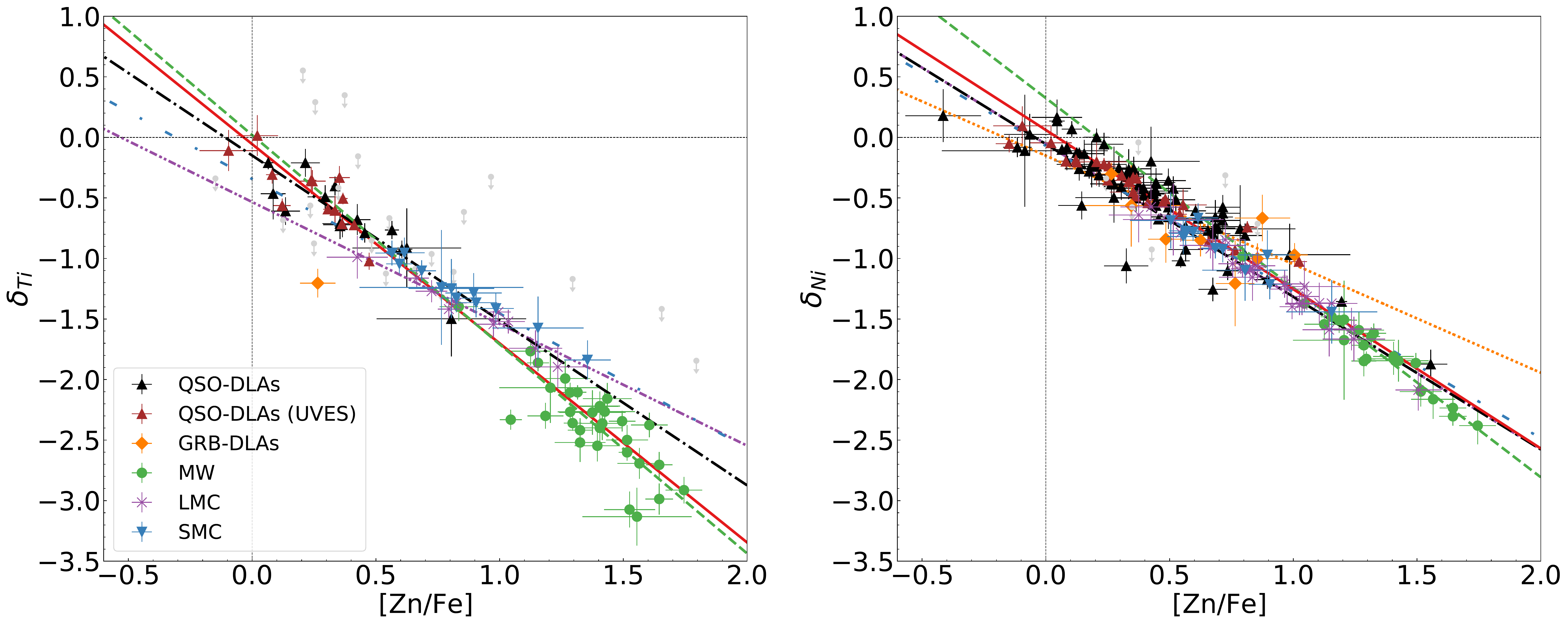}
    \caption{Relative abundances of Ti and Ni with respect to Zn against [Zn/Fe] (top). Dust depletion of Ti and Ni against [Zn/Fe] and the linear fits for the QSO-DLAs (black triangles and black dashdotted line) for the Milky Way (green points and dashed green line), for the LMC (purple stars and dash-dot-dotted purple densely line), for the SMC (blue triangles and dash-dotted blue line), and for the GRB-DLAs (orange diamonds and dotted orange line; bottom). The 3$\sigma$ upper limits are shown with gray arrows. The solid red line shows the linear fit to the merged data.}
    \label{fig:TiNi}
\end{figure*}

\begin{table*}[!ht]
\caption{Orthogonal distance regression results of the fits for the depletion of Ti in the different environments. The columns list the $\chi^{2}$, the reduced $\chi_{\nu}^{2}$, the degrees of freedom $\nu$, the $\chi{_{\nu}}^{2}$ difference between the fit to the individual samples, and the merged fit to the individual sample ($\Delta\chi{_{\nu}}^{2}$), the Pearson correlation coefficients r, the p - value, and intrinsic scatter $\sigma_{\delta_{\rm{X}}}$.}
\centering
\begin{tabular}{c c c c c c c c c c}
\hline\hline
Ti & A2$_{\rm{X}}$ & B2$_{\rm{X}}$ &$\chi^{2}$ & $\chi^{2}_{\nu}$ &$\nu$ & $\Delta \chi^{2}_{\nu}$ & r & p~-~value & $\sigma_{\delta_{\rm{Ti}}}$ \\
\hline
Milky Way& 0.02$\pm$0.30 & -1.73$\pm$0.22 & 57.84 & 2.00 & 29 & 0.07 & -0.81 & 4.13e-08 & 0.12\\
QSO-DLAs & -0.17$\pm$0.07& -1.36$\pm$0.19 & 61.24 & 2.36  & 26 & 0.16 & -0.87 & 2.32e-09 & 0.13\\
LMC & -0.53$\pm$0.12  & -1.01$\pm$0.13 &  2.83 & 0.35 & 8 & 0.96 & -0.97&3.00e-06 & 0.05\\
SMC &-0.35$\pm$0.04&-1.11$\pm$0.06 & 0.79 & 0.07 & 11 & 0.86 & -0.99 & 2.90e-10 & 0.02\\
GRB-DLAs & - & -  & - & - & - & - &-&- & -\\
Merged fit  & -0.06$\pm$0.03 & -1.64$\pm$0.04 & 164.49 & 2.03 & 81 & - & -0.97 & $\ll$ & 0.13\\
\hline
\label{titable}
\end{tabular}
\end{table*}

\begin{table*}[!ht]
\caption{Same as Table \ref{titable}, but for Ni.}
\centering
\begin{tabular}{c c c  c  c  c cccc}
\hline\hline
Ni & A2$_{\rm{X}}$ & B2$_{\rm{X}}$ &$\chi^{2}$ & $\chi^{2}_{\nu}$ & $\nu$& $\Delta \chi^{2}_{\nu}$ & r & p~-~value & $\sigma_{\delta_{\rm{Ni}}}$ \\
\hline\hline
Milky Way& 0.33$\pm$0.12  & -1.57$\pm$0.09 & 10.16 & 0.48  & 21 & 0.59 &-0.97& 2.09e-14 & 0.03\\
QSO-DLAs & -0.01$\pm$0.02 &  -1.12$\pm$0.04& 279.04 & 2.91 & 96 & 0.60 &-0.90 & $\ll$ & 0.11 \\
LMC & -0.05$\pm$0.05 & -1.26$\pm$0.05& 4.31 & 0.16 & 27 & 0.25 &-0.98 & 1.07e-19 & 0.03\\
SMC & -0.07$\pm$0.10 &-1.21$\pm$0.16& 1.93 & 0.19 & 10 & 0.18 & -0.94 & 4.53e-06 & 0.04\\
GRB-DLAs & -0.15$\pm$0.15 &-0.90$\pm$0.20& 6.50 & 1.08 & 6 & 0.80 & -0.71 &5.00e-02 & 0.10\\
Merged fit  & 0.07$\pm$0.02 & -1.31$\pm$0.03 & 328.53 & 1.96 & 168 & - & -0.96 &$\ll$ & 0.10\\
\hline
\label{nitable}
\end{tabular}
\end{table*}

\subsection{Titanium depletion}
\label{sec: Tidepletion}

In this and the next section, we focus on titanium and nickel, for which we characterize the depletion based on the technique of relative abundances for the first time. We used a relatively large dataset that included our new measurements.

The depletion of Ti was derived from the observed relative abundances of [Ti/Zn] and [Zn/Fe] (Fig. \ref{fig:TiNi}) in 28 QSO-DLAs, one GRB-DLA, ten LMC, 13 SMC, and 31 Milky Way systems. Only one constrained measurement of Ti is available for GRB-DLAs, which was included in the merged fit, but the individual fit was not performed for this group. The linear regression fits to the Ti depletions in the different environments and the merged fit are shown in Fig. \ref{fig:TiNi}.

To evaluate the goodness of the fits, we estimated the $\chi^{2}$, the reduced $\chi_{\nu}^{2}$, the Pearson correlation coefficients r, the p-value for each fit, and the difference between the individual datasets and the merged fit ($\Delta \chi^{2}_{\nu}$). This is reported in Table \ref{titable} and further discussed in Appendix \ref{appsec: stats}.

The strongest titanium depletion can be observed for the Milky Way, where 90 to $\sim$ 99.9$\%$ of titanium is depleted into dust.  $\delta_{\rm{Ti}}$ lies between $\sim$ -1.4 and -3.1~dex, which is similar to the Ti depletion that \citet{Jenkins2009} estimated for the Milky Way ISM using fewer data. They estimated the depletion with an independent method ($\sim$ -1.1 to -3.2~dex). This is an important comparison because the depletion of Ti is so strong that it dominates other potential effects that may affect different methods in different ways, such as $\alpha$-element enhancement, ionization, or the metallicity of the ISM. In particular, the consistency of our estimates of the strongest depletion of Ti in the Milky Way with those of \citet{Jenkins2009} confirms that our assumption on the depletion of Zn (to estimate $\delta_{\rm{Ti}}$ from [Ti/Zn]) is in the correct range.
The titanium depletion in QSO-DLAs is considerably weaker and ranges from about 0.003 to -1.5~dex. QSO-DLAs can probe systems with low metallicity and dust content, and in turn, very low dust depletions, all the way down to zero for dust-free QSO-DLAs. Even for the most refractory elements such as Ti, dust depletion can be zero. In the Magellanic Clouds, the range is in between that of the Milky Way and the QSO-DLAs, ranging from about -1 to -1.9~dex for the LMC and -1 to -1.8~dex for the much smaller sample of the SMC.

The best merged-fit coefficients for Ti are A2$_{\rm{Ti}}$ = -0.07 $\pm$ 0.03 and B2$_{\rm{Ti}}$ = -1.67 $\pm$ 0.04. \citet{Wiseman2017} made an approximate estimate on the coefficients of Ti and Ni based only on Milky Way data and are also consistent with our results (private communication). Other studies also find that the strong Ti depletion is most relevant in higher-metallicity environments \citep[e.g.,][]{Jenkins2009, Welty2010}. This is consistent with our results because Ti depletion appears to be stronger in the Milky Way and decreases with decreasing metallicity (Magellanic Clouds and DLAs). Increasing depletion with increasing metallicity is also typically the case for the other metals. We observe, however, a potential departure of the Ti depletion from linearity in the medium range of metallicities, in the Magellanic Clouds, which is discussed in Sect. \ref{subsec: nucl_assump}.

GRB120327A slightly deviates from the linear fit of the depletion sequence with a low Ti content (log\,N($\ion{Ti}{ii})$ = 12.69 $\pm$ 0.08) (see the orange diamond in Fig. \ref{fig:TiNi}). This might be due to dust depletion, ionization, nucleosynthesis, or a measurement problem, such as line contamination. A low Ti abundance might be caused by dust depletion, but the effect would be less prominent than what we observe and would be observed for the other metals as well. However, all other metals (Zn, S, Cr, Ni, and Fe) for this system fit the expected depletion pattern well.
\ion{Ti}{ii} has a low-ionization potential, making it more sensitive to ionization than the other ions (e.g., \ion{Fe}{ii} \ion{or Si}{ii}). There is no evidence for ionization for the DLA toward GRB120327A,
which has a very high \ion{H}{i} column density (log\,$N(\ion{H}{i})$ = 22.07 from \citealt{Bolmer2019}), so that \ion{H}{i} is expected to be quite efficient at shielding metals. Nucleosynthesis effects are also ruled out because we observe an underabundance, in contrast to an overabundance that we would expect from $\alpha$-element enhancement. Finally, this does not appear to be a measurement problem (because of the way in which the Ti column density was measured) because the Ti line profile was well modeled by \citet{Bolmer2019}, but was only based on one transition, which is the weak $\lambda$1910$\AA$ \ion{Ti}{ii} line.

\subsection{Nickel depletion}

We used measurements of Ni from 98 QSO-DLAs, eight GRB-DLAs, 23 Milky Way, 29 LMC and 12 SMC targets to characterize the depletion of Ni.
The depletions of Ni for the different environments are shown in the right panels of Fig. \ref{fig:TiNi}, and the goodness-of-fit parameters are reported in Table \ref{nitable}.

Within the Milky Way, about 90$\%$ to 99$\%$ of Ni is depleted into solid dust grains, similarly to Fe. The highest depletion is observed for the Milky Way, where it ranges from about -1 to -2.4~dex. The depletion in the LMC ranges from -0.6 to -2~dex. Finally, for the QSO-DLAs, it is 0.2 to -1.9~dex, for the SMC, it is -0.7 to -1.4~dex, and for GRB-DLAs, it is -0.3 to -1.2.

The correlation between $\delta_{\rm{Ni}}$ and [Zn/Fe] is tight in all the environments (QSO-DLAs, GRB-DLAs, Milky Way, and LMC) when they are merged, but also individually. The strong linear relation is reflected in the r value of the merged fit (r = -0.96), as well as the high statistical significance (p $\ll$ 10$^{-1}$) and the relatively small intrinsic scatter of $\sigma_{\delta_{\rm{Ni}}}$ = 0.10.

The best merged-fit coefficients for Ni are A2$_{\rm{Ni}}$ = 0.07 $\pm$ 0.02 and B2$_{\rm{Ni}}$ = -1.31 $\pm$ 0.03. \citet{Jones2018} also based their work on the observations of relative abundances to estimate the depletion of Ni, but in sub-DLA absorption systems, with measurements tabulated by \citet{Quiret2016}. They reported a depletion sequence that is characterized by A2$_{\rm{Ni}}$ = -0.03 $\pm$ 0.06 and B2$_{\rm{Ni}}$ = -1.19 $\pm$ 0.11, but with a relatively large intrinsic scatter $\sigma_{\delta_{\rm{Ni}}}$ = 0.18.

\begin{figure*}
    \centering
        \includegraphics[width=\textwidth]{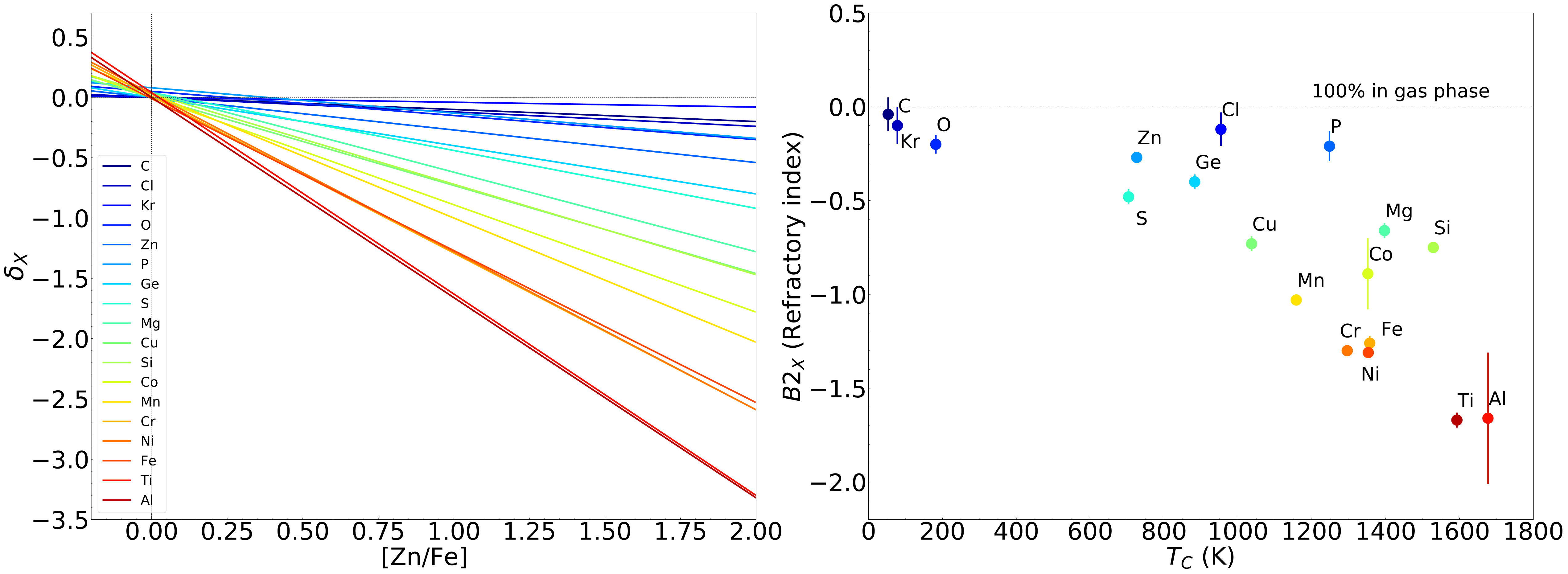}
    \caption{Merged fits of the dust depletion ($\delta_{X}$) as a function of [Zn/Fe] for all the elements (left). The elements are sorted by depletion strength. The slopes of the fits of the depletion sequences B2$_{X}$ (or refractory index) as a function of the respective element condensation temperature T$_{c}$ (right). The dust depletion relation with B2$_{X}$ is described by $\delta_X = A2_X +B2_X \times \mathrm{[Zn/Fe]}$. The condensation temperatures T$_{c}$ are compiled by \citet{Lodders2003}. Elements with higher condensation temperatures are more strongly depleted into dust.}
    \label{fig:Tcwfits}
\end{figure*}

\section{Discussion}
\label{sec:disc}

The sequences of the dust depletion for the majority of the elements from low-metallicity QSO-DLAs to the the Milky Way are linear. We note that while there is a dependence on [X/Zn] and [Zn/Fe] because both ratios have Zn in common, the same linear correlations are also found when  [X/Si] and [X/P] is used instead of [X/Zn] \citep[see][]{DeCia2016}. The slopes of these linear relations depend on the condensation temperature of the elements X. Thus, these relations are a result of dust depletion and not a dependence on the x-y axes.

The linear fits of the depletion sequences to all the available data of each element from all the environments (Milky Way, Magellanic Clouds, and QSO- and GRB-DLAs) we considered are presented in the left panel of Fig. \ref{fig:Tcwfits}. The fits become steeper for elements that are more strongly depleted into dust grains (refractory elements). In addition, the slope of the depletion sequences, B2$_{X}$, or refractory index, which is a measure of how strongly depleted an element is, decreases with the condensation temperature \citep{Field1974, Savage1996, Jenkins2009, Mattsson2012, Tchernyshyov2015, DeCia2016}. This relation can be observed in the right panel of  Fig. \ref{fig:Tcwfits}, where the condensation temperatures T$_{c}$ are compiled by \citet{Lodders2003}. The slope dependence on the refractory properties of metals confirms that the cause of the observed linear correlation is indeed dust depletion.

\subsection{Chromium}

Figure \ref{fig:RelAbund} shows that the zero intercept (at [Zn/Fe] = 0) of the Cr depletion sequence is not exactly zero, but instead it is higher by $\sim$ 0.13~dex $\pm$ 0.01. This may suggest that chromium might partly be affected by an $\alpha$-element enhancement.

Chromium belongs to the iron-group elements, and like half of the iron-group elements, it is mainly produced by type~Ia supernovae (SNe~Ia) \citep{Nomoto1997, Travaglio2004, KobayashiNomoto2009,Kobayashi2020}. However, Cr is also a product of the $\alpha$ process, synthesized in the incomplete Si-burning regions of SN~II during explosive nucleosynthesis \citep{Kobayashi2006, Nomoto2006}.
\citet{Nomoto2006} showed that Cr can be copiously produced by core-collapse supernovae, with [X/Fe] $\sim$ 0.1 for progenitor masses up to 20~M$_{\odot}$.
Thus, core-collapse supernovae, which produce the $\alpha$-elements, produce a significant amount of Cr after radioactive decay, leading to a small Cr overproduction. This would explain a small $\alpha$-element enhancement, but it would not reach the enhancement of the $\alpha$-elements (O, S, Si, and Mg). At no depletion ([Zn/Fe = 0]) we do observe [Si/Zn] $\sim$ [S/Zn] $\sim$ [Mg/Zn] $\sim$ [Ti/Zn] $\sim$ 0.3~dex and [O/Zn] $\sim$ 0.50~dex (which excludes a large $\alpha$-element enhancement amplitude for Zn), but instead, we observe [Cr/Zn] $\sim$ 0.13~dex $\pm$ 0.01. \citet{Prochaska&Wolfe2002} reported a slight overabundance of Cr at low levels of [Zn/Fe], and \citet{Dessauges2006} also reported an overabundance of Cr by $\sim$ 0.13~dex, as well as an overabundance of the $\alpha$-elements, consistent with our results. We did not take this small overabundance into account when we corrected for nucleosynthesis effects because Cr is not typically considered an $\alpha$-element.

The highly depleted QSO-DLA point at [Zn/Fe] = 1.555 of the Cr depletion sequence shown in Fig. \ref{fig:DeplSeq} is J1211+0833. The host galaxy of this system is found to be a chemically enriched, evolved, massive, and star-forming disk-like galaxy \citep{Ma2018}. This system is so chemically enriched that it has left the low-metallicity QSO-DLA regime and lies in the high-metallicity range of the Milky Way. It still follows the Cr dust depletion sequence, however. This further supports our results that the depletion sequences are consistent for all the environments.

\subsection{Ionization of sulfur}
\label{subsec: ionS}

Fig. \ref{fig:RelAbund} shows that \ion{S}{ii} column densities are higher when they are observed toward stellar sources (MW, SMC, and LMC) than in measurements toward background QSOs or GRBs. This difference between measurements of \ion{S}{ii} column densities in local and distant galaxies is clearly explained by ionization effects.

The ionization potential of \ion{S}{ii} is the highest (23.4~eV) of the elements considered in this work and is not easily ionized. This means that \ion{S}{ii} can be found in both \ion{H}{i} and \ion{H}{ii} regions, while other elements are easily ionized to their double-ionized state in the \ion{H}{ii} regions, therefore they do not contribute to what we measure in the neutral ISM. This \ion{S}{ii} in the \ion{H}{ii} regions can add to what we consider to be present in the foreground \ion{H}{i} gas. At the same time, the \ion{S}{ii} transitions are quite strong, which virtually prevents measurements of \ion{S}{ii}  in lines of sight with large amounts of N(\ion{H}{i}). Therefore \ion{S}{ii} is determined only in
lines of sight with low N(\ion{H}{i}), which makes the contribution from the \ion{H}{ii} region of the star relatively important. Overall, DLA gas probes only the neutral phase, while the absorbers toward \ion{H}{ii} regions, for instance, in the Milky Way or the Magelanic Clouds, have an increased contribution from \ion{S}{ii} of the \ion{H}{ii} regions.
The case of S is further discussed in \citet{Jenkins2009}, \citet{Jenkins2017}.

\subsection{Nucloesynthesis as the cause of the deviations observed in the Magellanic Clouds}
\label{subsec: nucl_assump}

Figures \ref{fig:TiNi} and \ref{fig:DeplSeq} show a potential bending of the dust depletion sequences for Ti, Mg, S, and Mn in the Magellanic Clouds. In particular, Ti, S, and Mg have a shallower and Mn a steeper slope in the Magellanic Clouds than in the Milky Way. This potential departure from a linear relation is more pronounced when the linear dependence of $\delta_{\rm{Zn}}$ on [Zn/Fe] is removed and nucleosynthesis is corrected for, that is, $\alpha$-element enhancement and Mn underabundance. The deviations are not strictly significant (1.8$\sigma$, 2$\sigma$, 2.5$\sigma,$ and 5$\sigma$ for Mg, Ti, Mn, and S, respectively). A statistical analysis to quantify these deviations is presented in Appendix \ref{appsec: stats}.
The small deviations in relative abundances that we observe in the neutral ISM in the Magellanic Clouds are not randomly distributed among the elements. The overabundance of $\alpha$-elements (Ti, Mg, and S, and difficult to tell for Si) and underabundance of Mn is systematic. Therefore, these tentative deviations might be due to $\alpha$-element enhancement and Mn underabundance in the Magellanic Clouds that we did not take into account in our initial assumptions.

The production of $\alpha$-elements is dominated by core-collapse SNe \citep{Kobayashi2006, KobayashiNomoto2009}, while the production of Mn is dominated by SN~Ia \citep{Gratton1989, Feltzing2007, KobayashiNomoto2009}. This makes it quite likely that the deviations we observe are caused by nucleosynthesis effects. In our initial analysis, we corrected for $\alpha$-element enhancement and Mn underabundance assuming a distribution of the [$\alpha$/Fe] with metallicity with a classical plateau - knee distribution (e.g., \citet{McWilliam1997}), also for the Magallanic Clouds.

As described in Section \ref{nucl_corr}, the exact position of the $\alpha$-element knee is only poorly constrained in the Magellanic Clouds. However, small changes in the knee position do not affect our results because the points that deviate from the fit are far from the knee position. In addition, the metallicities of the neutral ISM of the systems that show deviations are in the range in which we did not expect $\alpha$-element enhancement \citep{DeCia2022}.

\begin{figure}
    \centering
    \includegraphics[width=\columnwidth]{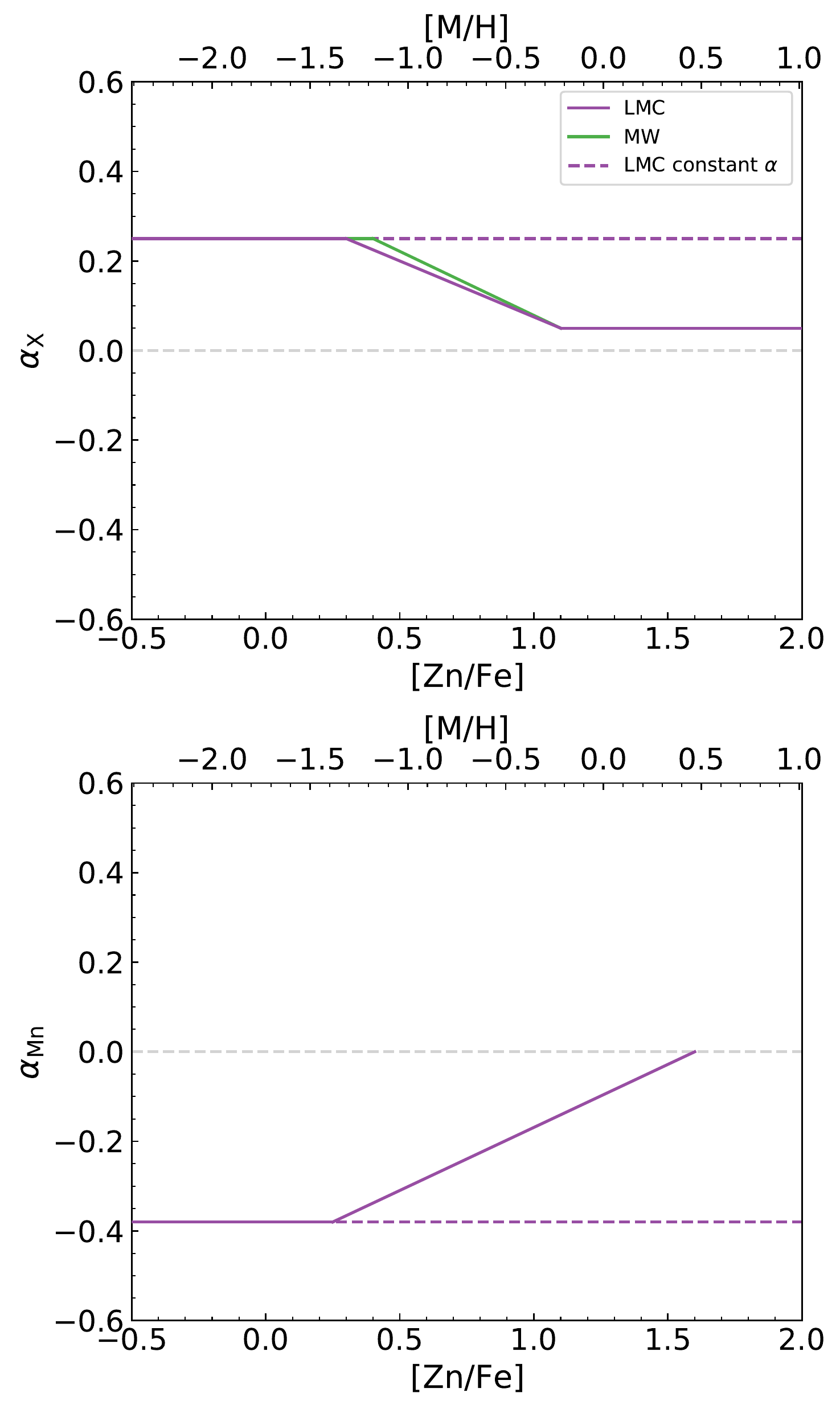}
    \caption{Assumed nucleosynthetic over- or underabundances $\alpha_{\rm{X}}$ for the $\alpha$-elements X (S, Si, Mg, and Ti) and Mn with respect to [Zn/Fe] and metallicity [M/H]. The solid green curve shows the Milky Way, the solid purple curve shows the LMC as adopted from \citet{deBoer2014}, and the dashed purple curve shows the LMC with a constant over- or underabundance plateau. For the plateau of the $\alpha$-elements, we used as reference $\alpha_{\rm{S}}$ = 0.25.}
    \label{fig:curves_constMCs}
\end{figure}

When we assume a constant over- or underabundance plateau instead, the previously deviating points follow the linear fit. The shapes of the different assumptions on the nucleosynthetic curves for the $\alpha$-elements and for Mn are shown in Fig. \ref{fig:curves_constMCs}. The dust depletion sequences for the $\alpha$-elements and for Mn after correcting for nucleosynthesis effects in the Magellanic Clouds assuming a constant over- or underabundance plateau are shown in Fig. \ref{fig:constMCs}. All elements after the new corrections follow linearity, except for S, which is known to be further enhanced toward \ion{H}{ii} regions by ionization. This might mean that the nucleosynthetic curves for the Magellanic Clouds might be much different from what was previously assumed, and there is only a constant plateau instead. This might be due to a recent burst of star formation in the Magellanic Clouds, which implies that their star formation history is likely to be different than in the Milky Way.

The overabundance of $\alpha$-elements and underabundance of Mn in the Magellanic Clouds was also observed and discussed in previous studies \citep{Vladilo2002, Guber2016, Jenkins2017, DeCia2018}. \citet{Hasselquist2021} observed stellar $\alpha$-element enhancement in the Magellanic Clouds and an upturn of the [$\alpha$/Fe] plateau, which they interpreted as due to a very recent burst of star formation. The $\alpha$-element enhancement may be local and follow the recent star formation activity. On the other hand, other works failed to find a clear [$\alpha$/Fe] plateau in the LMC \citep[e.g.,][]{VanDerSwaelmen2013}.
It is possible that we observe $\alpha$-element enhancement and Mn underabundance toward limited individual systems rather than systematically over the entire metallicity range. Our observations probe the neutral ISM toward hot young OB stars, which means that they may be showing $\alpha$-element enhancement and Mn underabundance. Additionally, it is possible that $\alpha$-element enhancement can be observed in the neutral ISM toward young stars following a recent star formation burst, while there was no time for a new generation of stars to form and fully show the $\alpha$-element enhancement.

\section{Summary and conclusions}
\label{sec:conc}
We characterized the dust depletion of several metals in different environments from the Milky Way, the Magellanic Clouds, and DLAs toward GRBs and QSOs based on the relative abundances of metals with different refractory properties (or how easily they incorporate into dust grains). We collected all the available literature measurements of metal column densities in these environments, making this a comprehensive study of relative chemical abundances of metals and for different environments. We measured new column densities of Ti and Ni in 70 QSO-DLAs from high-resolution UVES spectra from the sample of \citet{DeCia2016}.

We observed and fit the dust depletion sequences and the relative abundances for all the elements. We further focused on the dust depletion of Ti and Ni, for which we also provided new column density measurements.

We observe a slight deviation from the linear fits of the depletion $\delta_{\rm{X}}$ for Ti, S, Mg, and Mn in the Magellanic Clouds. A potential cause of the anomaly in these elements might be the $\alpha$-element enhancement and Mn underabundance in the Magellanic Clouds. When a constant over- (under-) abundance plateau of the $\alpha$-elements (Mn) is assumed, the deviations disappear and all environments for all the elements follow linearity. The observed $\alpha$-element enhancement and Mn underabundance in the neutral ISM at a wide range of metallcities might be due to recent bursts of star formation in the Magellanic Clouds.

We observe a Cr overabundance of $\sim$ 0.13~dex $\pm$ 0.01. This suggests that it might be affected by an $\alpha$-element enhancement, while Cr is not typically considered an $\alpha$-element.

The column densities of \ion{S}{ii} seem to be higher for local galaxies (Milky Way, SMC, and LMC). This can be explained by an additional contribution of their \ion{H}{ii} regions to the measured \ion{S}{II} column densities.

The tendency of each element to be depleted into dust varies depending on its condensation temperature. This dependence confirms that the the depletion sequences are indeed driven by dust depletion and not by other physical processes.

Overall, the dust depletion of the elements are well described with linear fits to the merged data, despite the small deviations that statistically are not significant for the merged fit. Nevertheless, the small deviations seem to be due to nucleosynthesis and not to dust depletion. The dust depletion sequences are virtually the same for different environments, which implies that dust depletion behaves in the same way, regardless of the environment. This might mean that the process that mostly causes the build-up of dust grains is independent of the star-formation history of the different galaxies, which likely changes dramatically from system to system. This suggests that the growth of dust grains in the ISM is an important process of dust formation.

\begin{acknowledgements}
We thank the referee Sandra Savaglio for the detailed and constructive report that improved this manuscript. CK, ADC, JKK and TRH acknowledge support by the Swiss National Science Foundation under grant 185692. KEH acknowledges support from the Carlsberg Foundation Reintegration Fellowship Grant CF21-0103. This paper makes use of the Python code VoigtFit available on \url{https://github.com/jkrogager/VoigtFit}. This research has made use of NASA’s Astrophysics Data System.
\end{acknowledgements}

\bibliographystyle{aa}
\bibliography{ref}

\begin{appendix}

\section{Statistical analysis for Mn and the $\alpha$-elements}
\label{appsec: stats}

We estimated the $\chi_{\nu}^{2}$ between the individual datasets (QSO-DLAs, GRB-DLAs, Milky Way, LMC, and SMC) and the merged fit (the fit assuming all datasets merged together) for Mn and for the $\alpha$-elements Ti, S, Si, and Mg, taking both x and y uncertainties into account. This was used to calculate the difference $\Delta \chi^{2}_{\nu}$ by subtracting the $\chi_{\nu}^{2}$ of the merged fit, which can quantify the observed deviations and their significance.

For Ti, the smallest difference from the merged fit is observed for the Milky Way with $\Delta \chi^{2}_{\nu}$ = 0.07, while the largest difference from the merged fit is observed for the LMC with $\Delta \chi^{2}_{\nu}$ = 0.97 with a 2$\sigma$ significance, but with only eight degrees of freedom used for the fit. For the Milky Way, QSO-DLAs, and the LMC, we find a $\chi_{\nu}^{2}$ > 1. This might be due to underestimated uncertainties, natural scatter of the data, or a poor fit. For the SMC, the uncertainties are probably overestimated, resulting in a $\chi_{\nu}^{2}$ < 1. The individual fits are well constrained, with r values ranging from -0.99 (SMC) to -0.81 (Milky Way) and high statistical significance, with low p-values (p $\ll$ 10$^{-1}$, see Table \ref{titable}). The lowest statistical significance is recorded for the LMC, but the fit was only performed on ten data points. The linear fit to the Ti depletions in all environments merged together has an r value of -0.97 and a statistical significance (p $\ll$ 10$^{-1}$).

For S, the smallest difference from the merged fit is estimated for QSO-DLAs with $\Delta \chi^{2}_{\nu}$ = 0.02, and the largest difference is measured for the LMC with $\Delta \chi^{2}_{\nu}$ = 2.02 with 5$\sigma$ significance. This deviation is due to ionization effects (see Section \ref{subsec: ionS}), however. Si does not significantly depart from the merged fit, and we do not report any statistically significant deviation for any of the environments. The merged fit remains a good description of the distribution. For Mg, we observe an offset between the merged fit and the Milky Way fit, with a difference of $\Delta \chi^{2}_{\nu}$ = 0.78 and 1.8$\sigma$ significance and a large scatter for all the environments. The merged fit remains the best description of the distribution.
For Mn, we observe an offset of the Milky Way fit from the merged fit with a difference of $\Delta \chi^{2}_{\nu}$ = 1.05 within 2.5$\sigma$, but only a few points were used for the fit with a large scatter. While the SMC points have a small intrinsic scatter, the difference from the merged fit is $\Delta \chi^{2}_{\nu}$ = 0.93 within 2.2$\sigma$. We conclude that while we see a slight tendency of some elements to depart from the linear fit, none of these deviations are statistically significant, and the overall distribution is best described by the merged linear fit.

\section{Elements with limited coverage}
\label{limited_cov}

\begin{figure*}[!h!t]
    \centering
    \includegraphics[width=\textwidth]{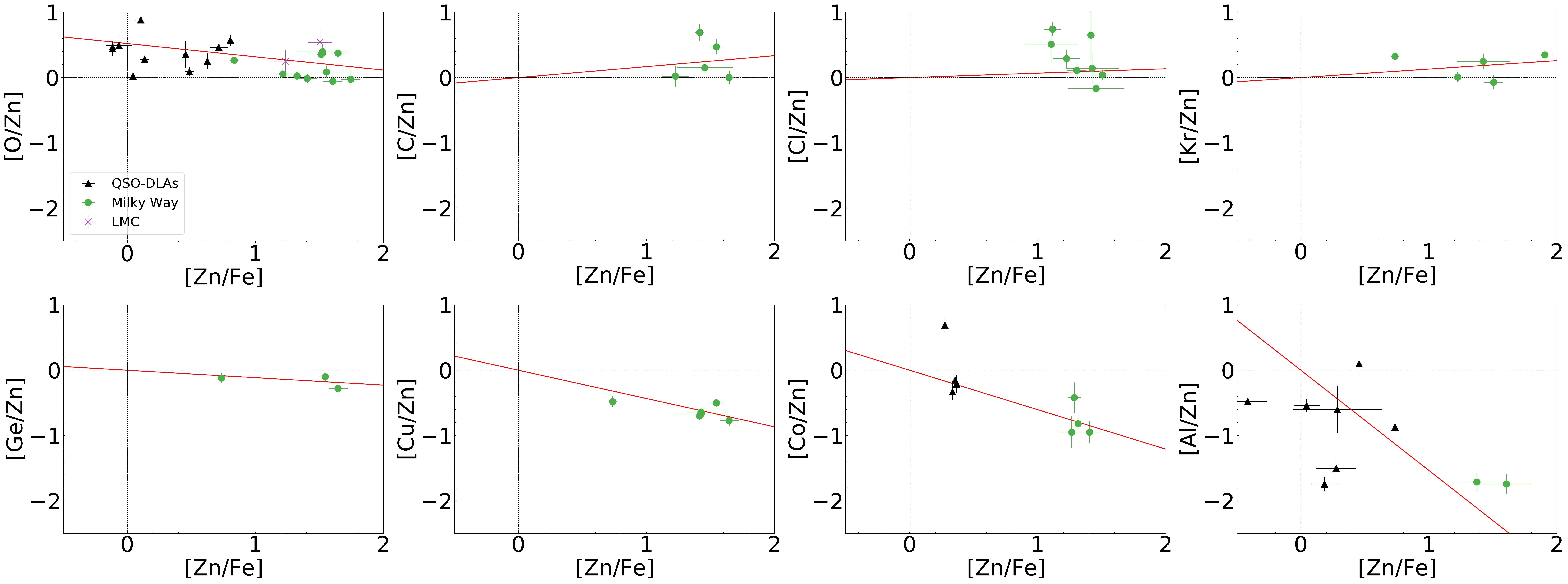}
    \caption{Relative abundances of element X with respect to Zn against [Zn/Fe]. The black triangles show QSO-DLAs, the green dots show the Milky Way, and purple stars show the LMC. The red line shows the linear fit to the data.}
    \label{fig:RelAbund_volatile}
\end{figure*}

\begin{figure*}[!h]
    \centering
    \includegraphics[width=\textwidth]{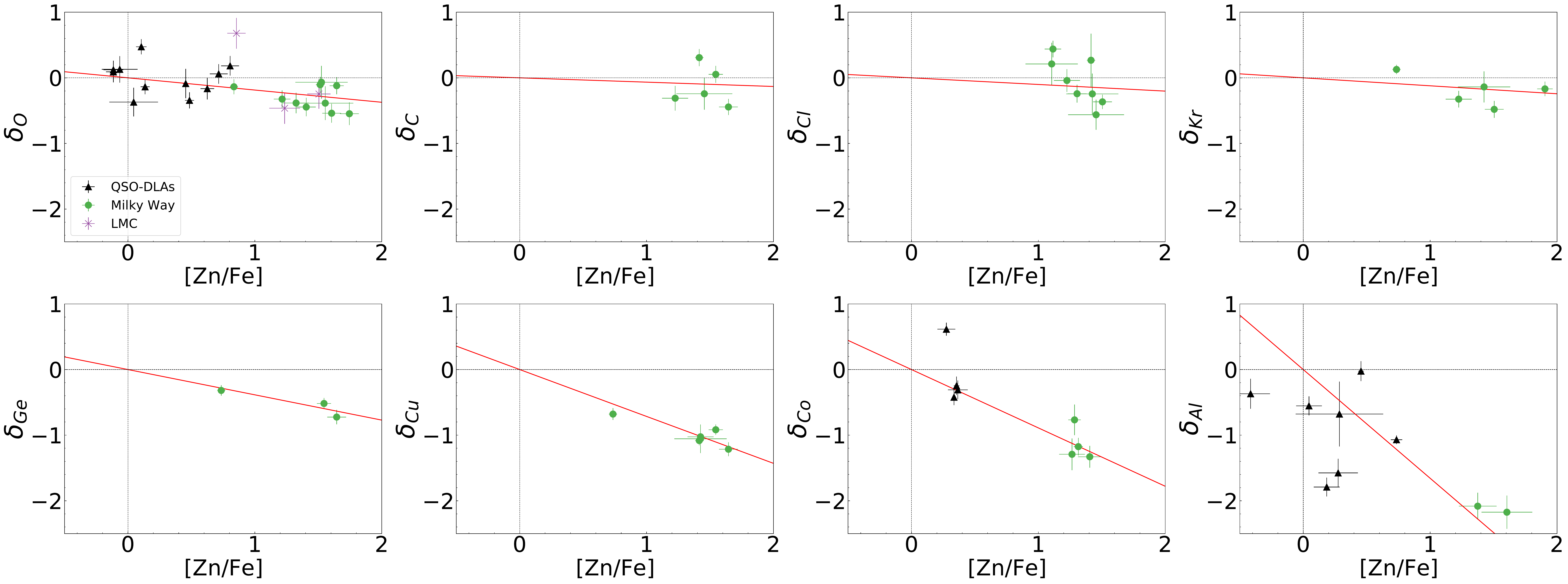}
    \caption{Dust depletion of element X ($\delta_{\rm{X}}$) against [Zn/Fe]. The black triangles show QSO-DLAs, green dots show the Milky Way, and purple stars show the LMC. The red line shows the linear fit to the data. These elements have limited coverage especially at low depletion levels (QSO-DLAs).}
    \label{fig:DeplSeq_volatile}
\end{figure*}

We added many more elements (C, Cl, Kr, Ge, Cu, Co, Al and O) to the current dust depletion scheme whose depletions have not been characterized before using this formalism. We used measurements for the Milky Way and QSO-DLAs, but only few data are available and only some LMC measurements for O. The relative abundances of the elements X as a function of [Zn/Fe] are shown in Fig. \ref{fig:RelAbund_volatile}, and their dust depletion sequences are presented in Fig. \ref{fig:DeplSeq_volatile}. Because of the small number of degrees of freedom, the fit was made assuming zero intercept (A2$_{x}$ = 0.00). This can be justified by the fact that there should be zero depletion at the lowest overall strengths of depletion, [Zn/Fe] = 0.

For O and Al, the largest number of QSO-DLA data points are available ofthe elements with limited coverage. However, their depletion coefficients should be used with caution because of potential issues with line saturation. The \ion{O}{I} measurements in QSO-DLAs were taken from \citet{DeCia2016}, where they were based on high-resolution data and mostly included the few systems for which \ion{O}{I} could be estimated based on multiple lines, including weak \ion{O}{I} lines in the Ly-$\alpha$ forest, and only in a very few cases using the \ion{O}{I} $\lambda$1302$\AA$ line alone when it was not saturated. For the Milky Way, the weak \ion{O}{I} lines are typically used \citep[e.g.,][and references therein]{Jenkins2009}. For \ion{Al}{ii}, the QSO-DLA measurements were taken from \citet{Berg2015}. This is a large collection from the literature of measurements made from the only available \ion{Al}{ii} $\lambda$1670$\AA$  line, which is often saturated. Therefore, these measurements are not very reliable. This is probably the cause of the large scatter in Fig. \ref{fig:DeplSeq_volatile}. Nevertheless, because Al is very highly depleted into dust grains, it is still useful to include the rough estimate on its depletion despite the very large uncertainties (e.g., B2$_{Al}$ = -1.66 $\pm$ 0.35).
The A2$_{X}$ (A2$_{X}$ = 0.00) and B2$_{X}$ coefficients along with the Pearson correlation coefficients of the dust depletion sequences of elements with limited coverage resulting from the fit are reported along with the rest of the elements in Table \ref{coefficients}.

\label{appsec: lim_cov}
\nopagebreak[4]

\newpage

\section{Velocity profiles of the new Ni and Ti DLA measurements}
\label{appsec: fits}

\begin{figure}[h!]
\includegraphics[keepaspectratio=true,width=\columnwidth]{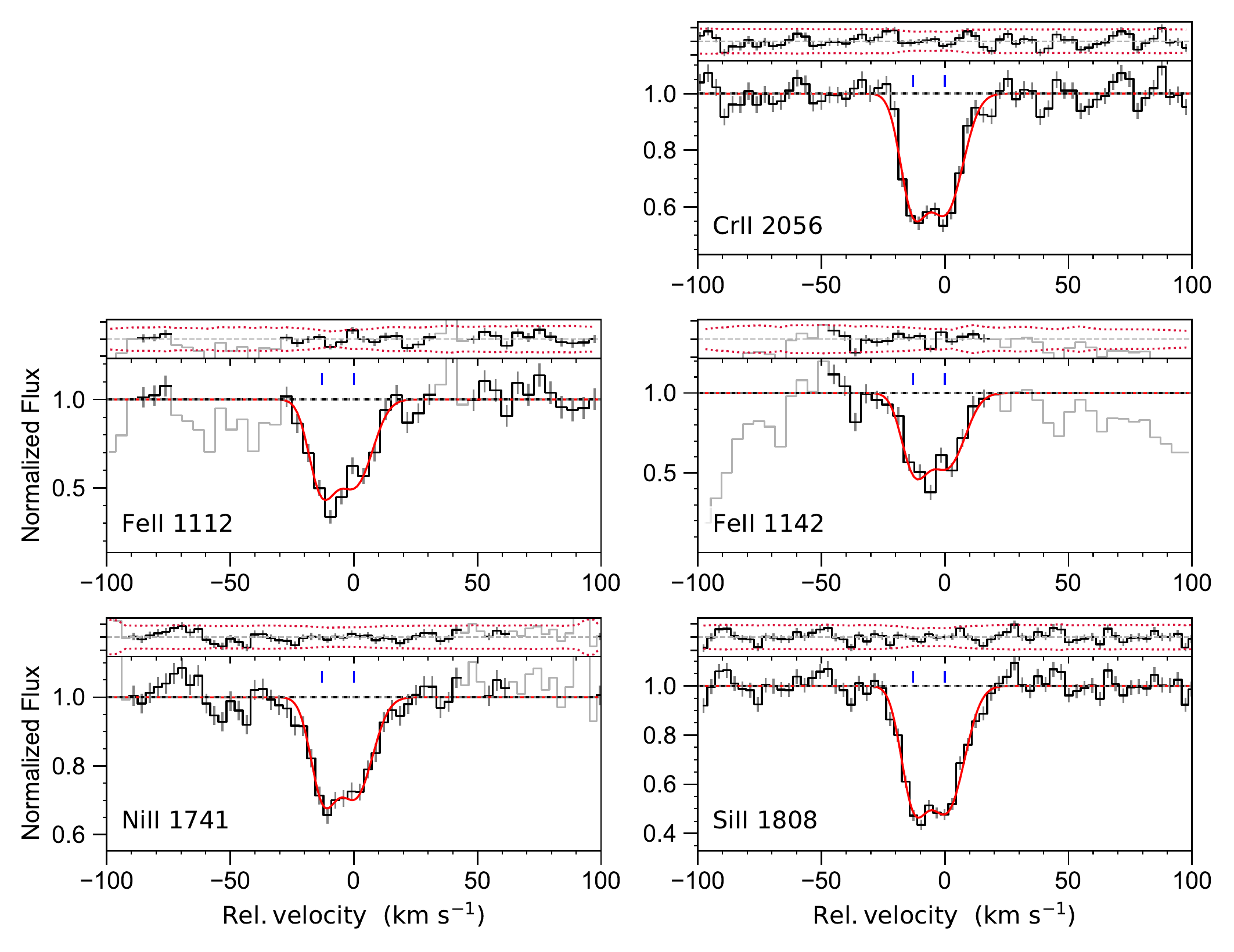}
\caption{Velocity profiles of selected low-ionization transition lines from the DLA system at $z_{\rm abs}$ = 2.02484 toward Q\,0010$-$002. Masked regions are shown in gray. The top subpanels show the residuals, and the blue tick marks show the individual velocity components.}
\end{figure}

\begin{figure}[h!]
\includegraphics[keepaspectratio=true,width=\columnwidth]{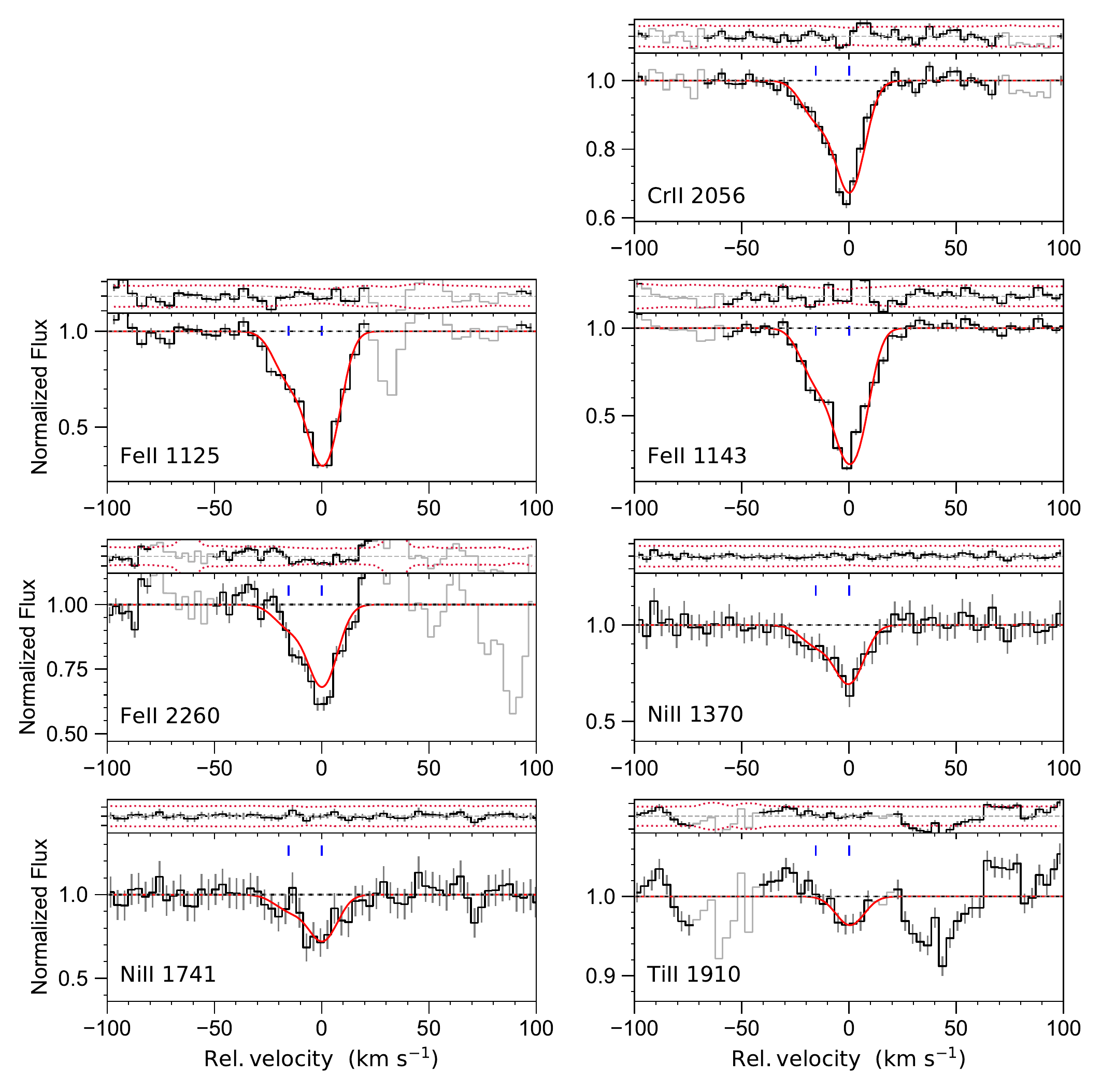}
\caption{Velocity profiles of selected low-ionization transition lines from the DLA system at $z_{\rm abs}=$ 2.67142 toward Q\,0058$-$292.}
\end{figure}

\begin{figure}[h!]
\includegraphics[keepaspectratio=true,width=\columnwidth]{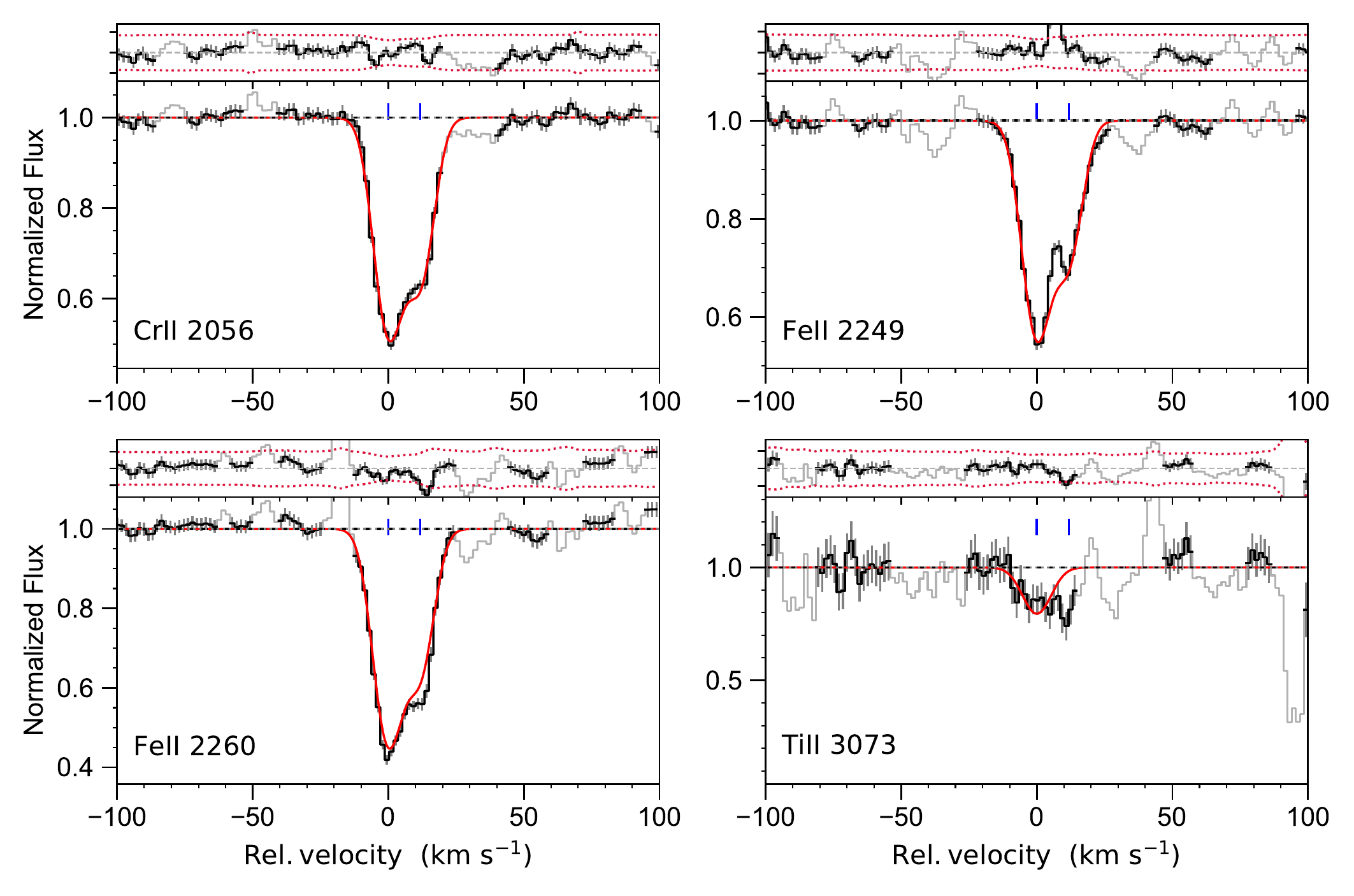}
\caption{Velocity profiles of selected low-ionization transition lines from the DLA system at $z_{\rm abs}=$ 2.30903 toward Q\,0100$+$030.}
\end{figure}

\begin{figure}[h!]
\includegraphics[keepaspectratio=true,width=\columnwidth]{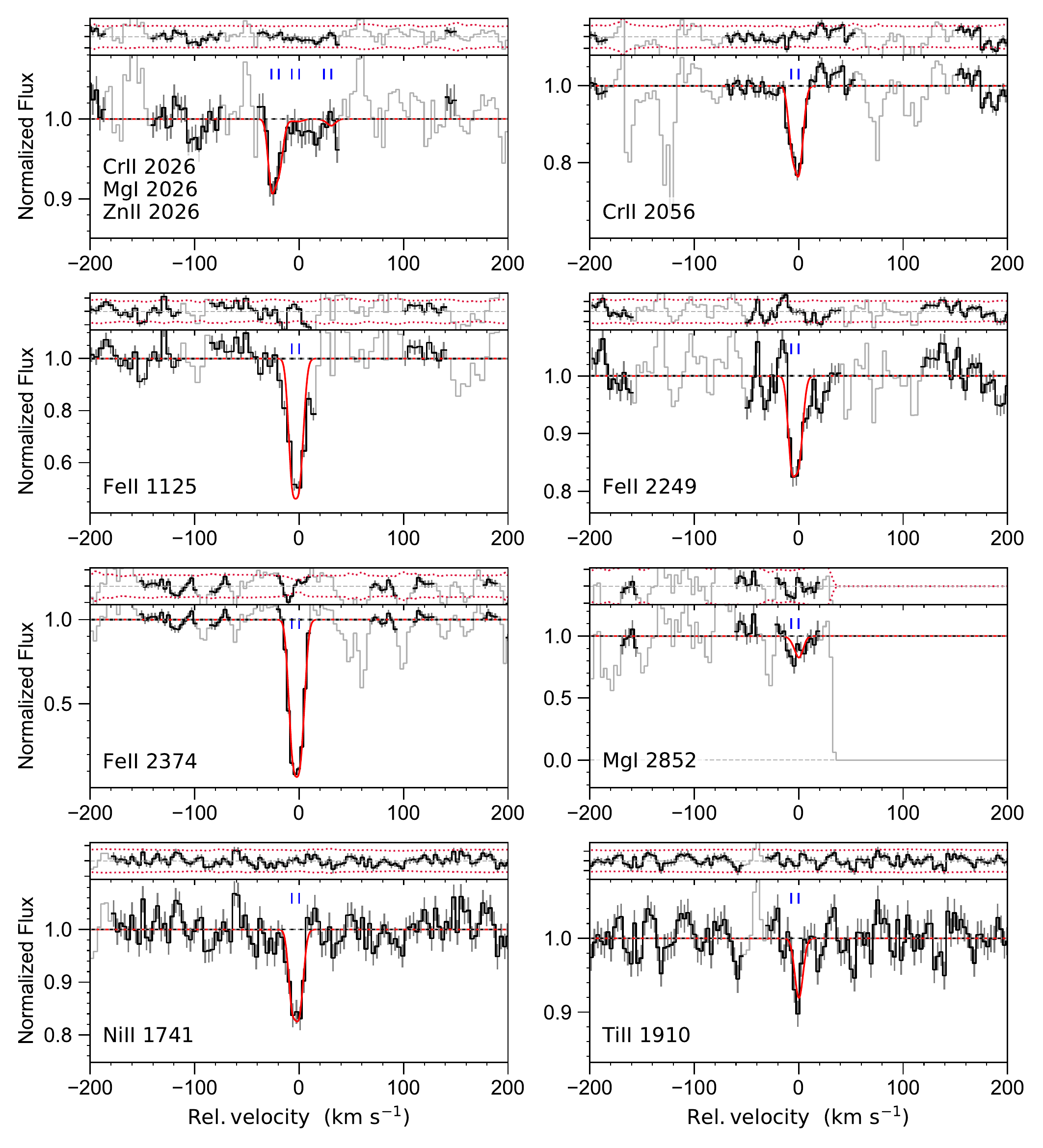}
\caption{Velocity profiles of selected low-ionization transition lines from the DLA system at $z_{\rm abs}=$ 2.36966 toward Q\,0102$-$190a.}
\end{figure}

\begin{figure}[h!]
\includegraphics[keepaspectratio=true,width=\columnwidth]{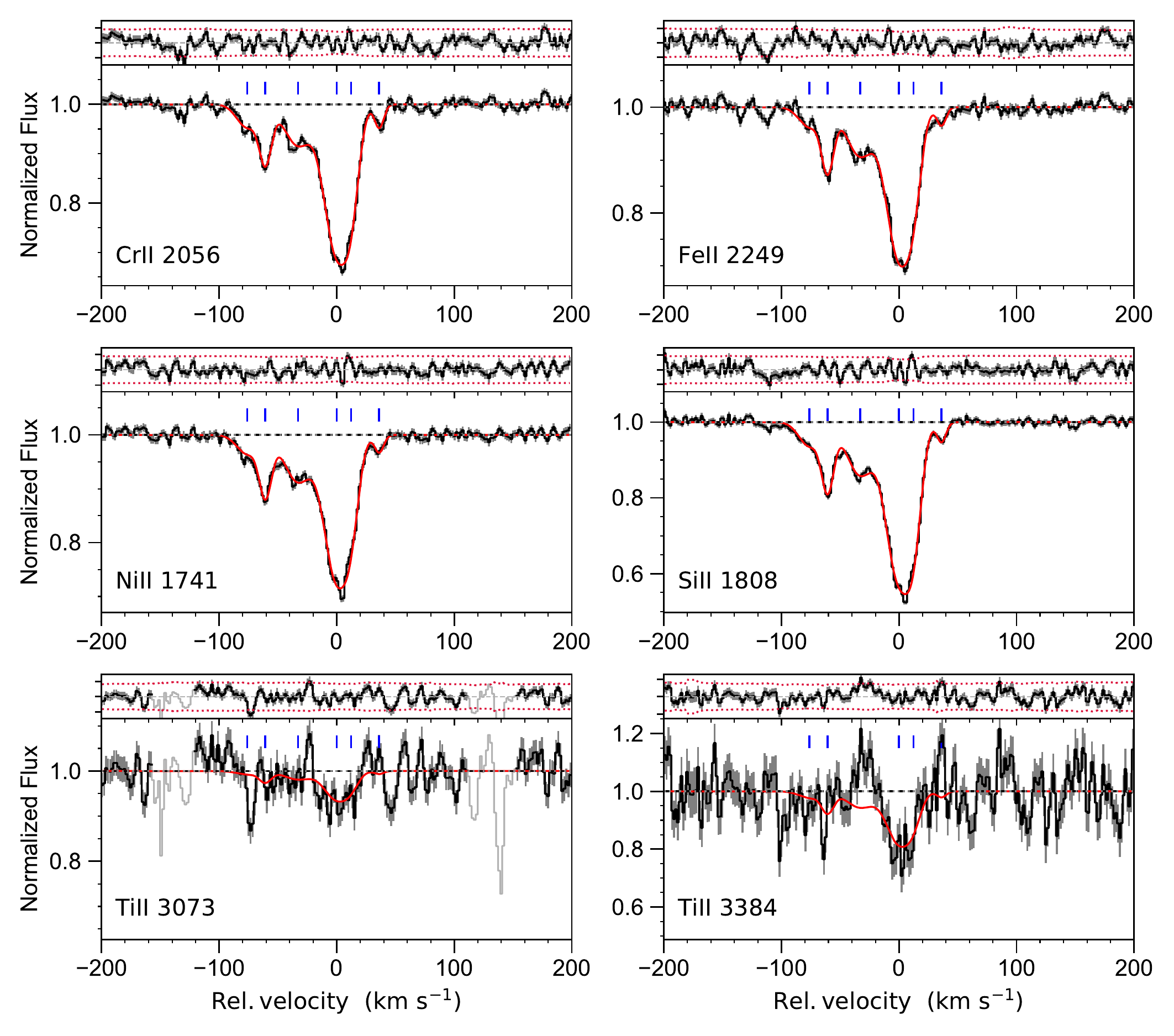}
\caption{Velocity profiles of selected low-ionization transition lines from the DLA system at $z_{\rm abs}=$ 1.91267 toward Q\,0405$-$443a.}
\end{figure}

\begin{figure}[h!]
\includegraphics[page=1,keepaspectratio=true,width=\columnwidth]{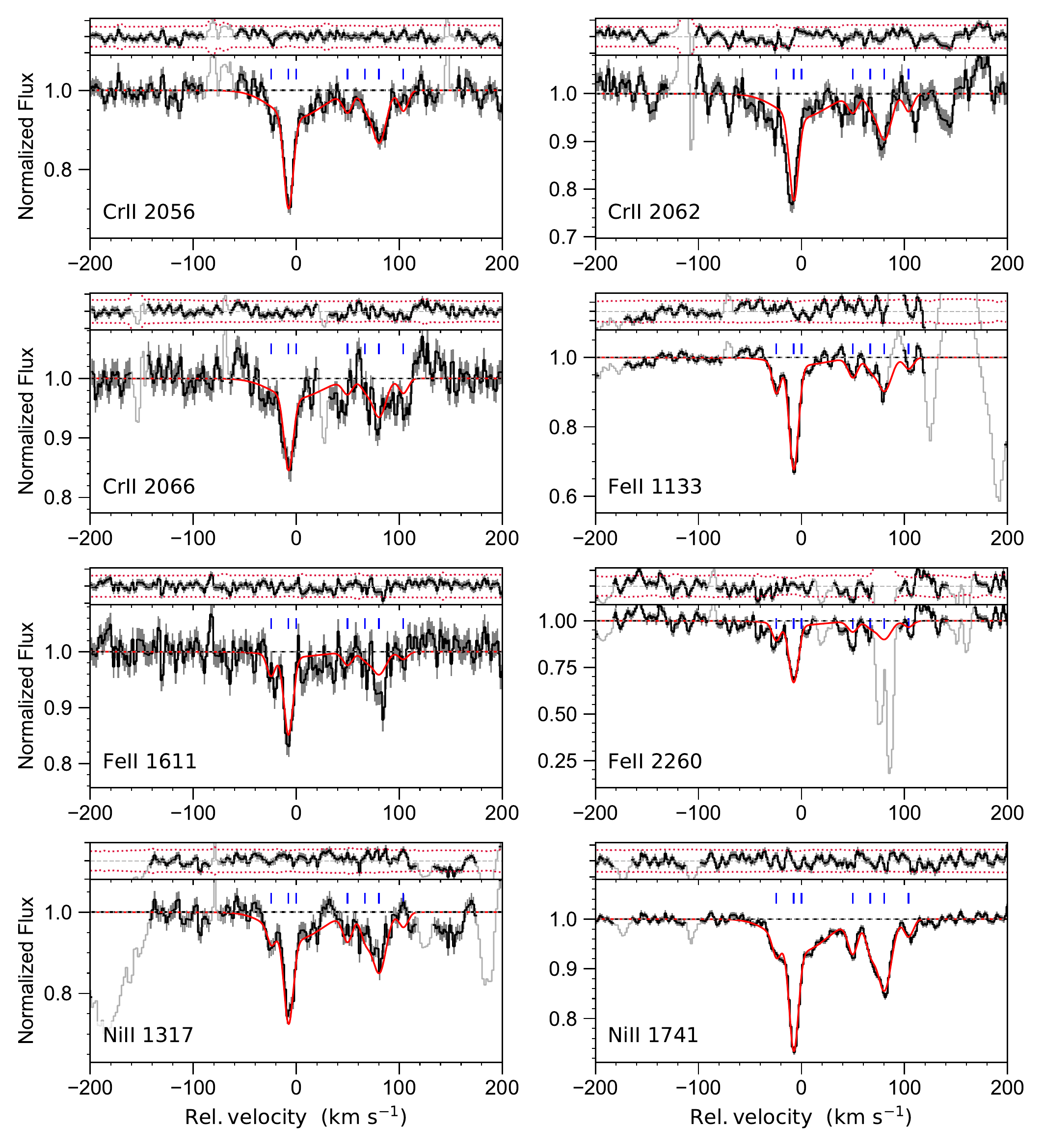}
\includegraphics[page=2,keepaspectratio=true,width=\columnwidth]{Figures/Fits/spec_addi_Q0405-443b.pdf}
\caption{Velocity profiles of selected low-ionization transition lines from the DLA system at $z_{\rm abs}=$ 2.55000 toward Q\,0405$-$443b.}
\end{figure}

\begin{figure}[h!]
\includegraphics[keepaspectratio=true,width=\columnwidth]{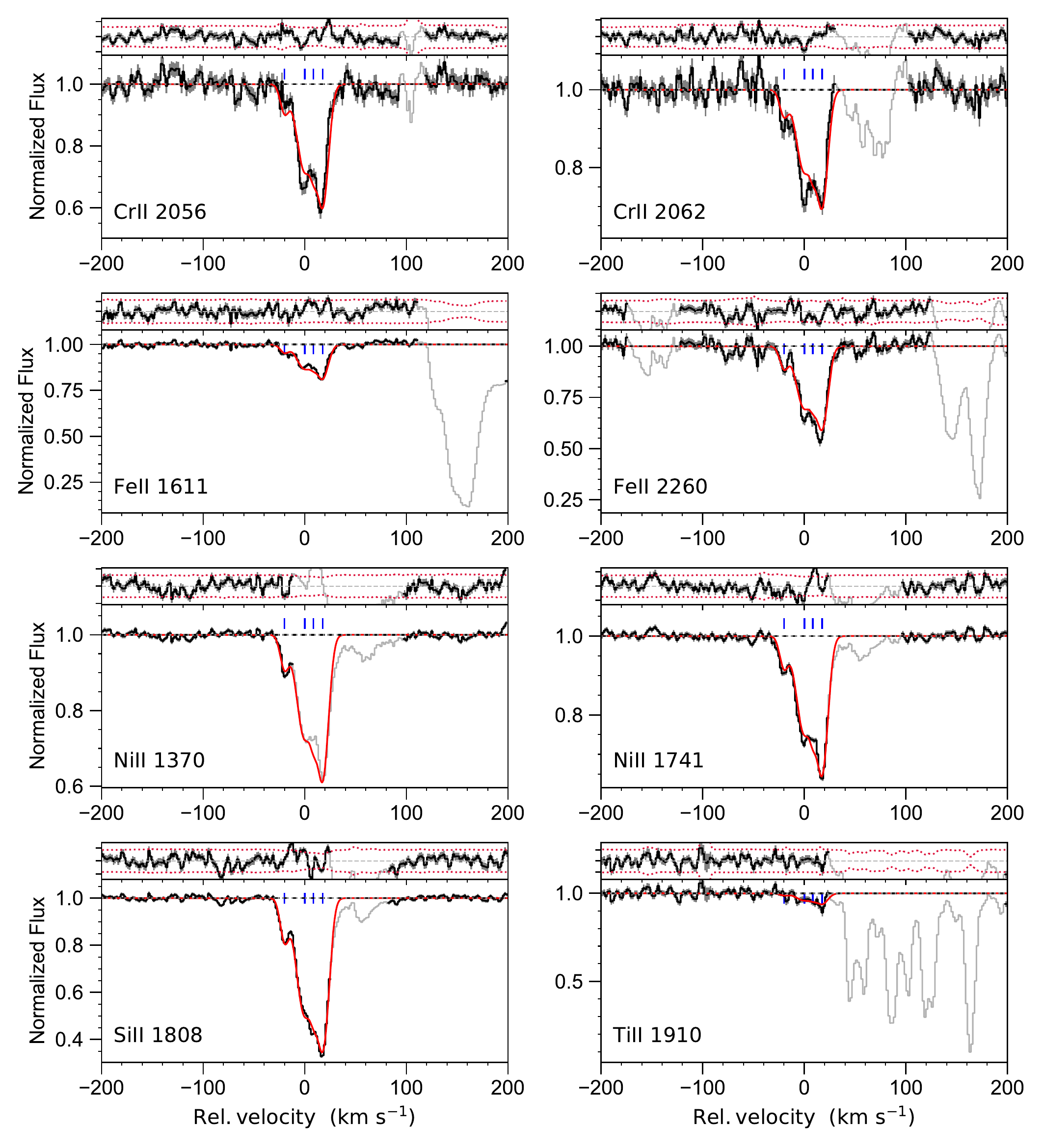}
\caption{Velocity profiles of selected low-ionization transition lines from the DLA system at $z_{\rm abs}=$ 2.59466 toward Q\,0405$-$443c.}
\end{figure}
\thispagestyle{empty}
\begin{figure}[h!]
\includegraphics[keepaspectratio=true,width=\columnwidth]{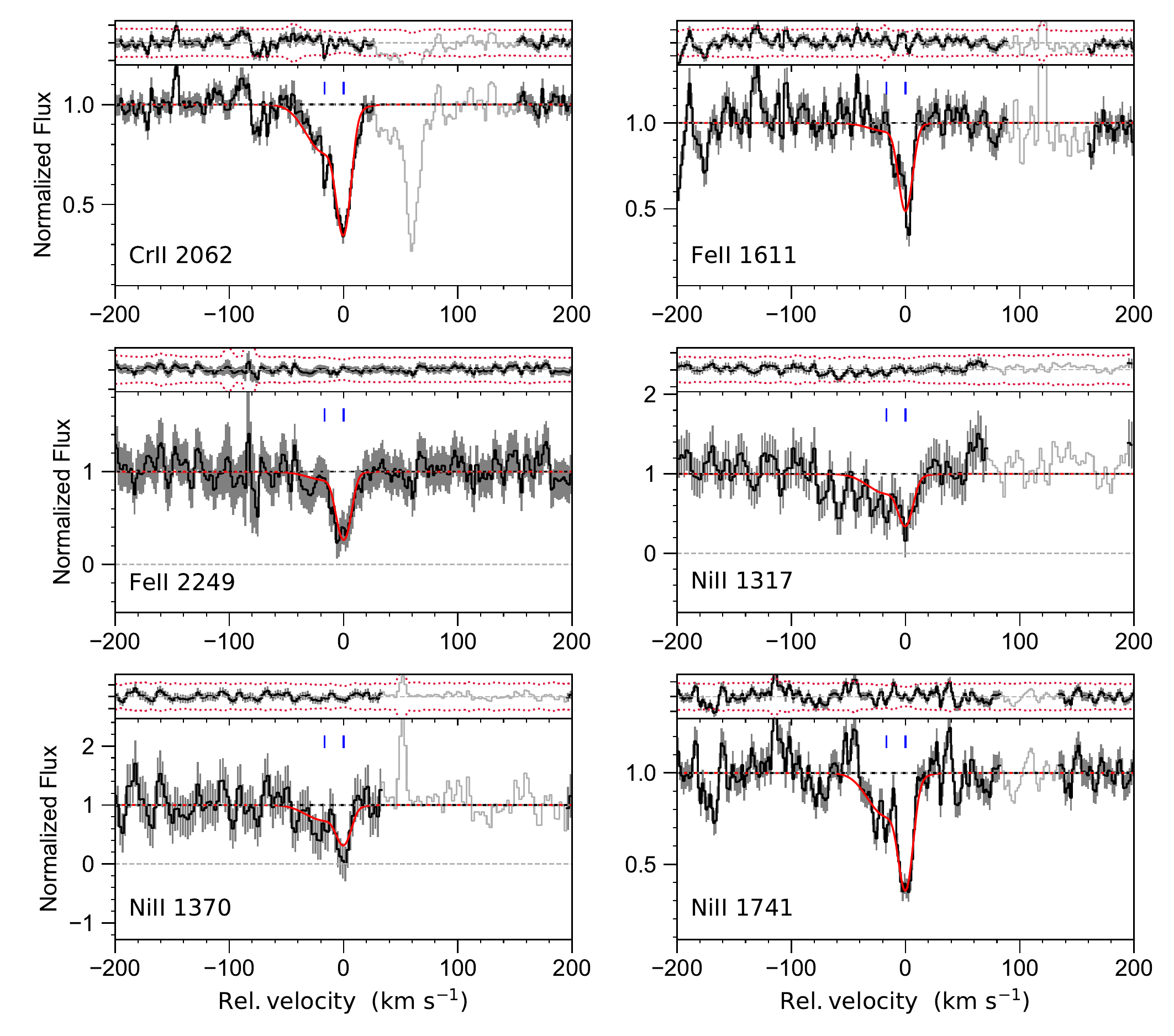}
\caption{Velocity profiles of selected low-ionization transition lines from the DLA system at $z_{\rm abs}=$ 2.03956 toward Q\,0458$-$020.}
\end{figure}

\begin{figure}[h!]
\includegraphics[page=1,keepaspectratio=true,width=\columnwidth]{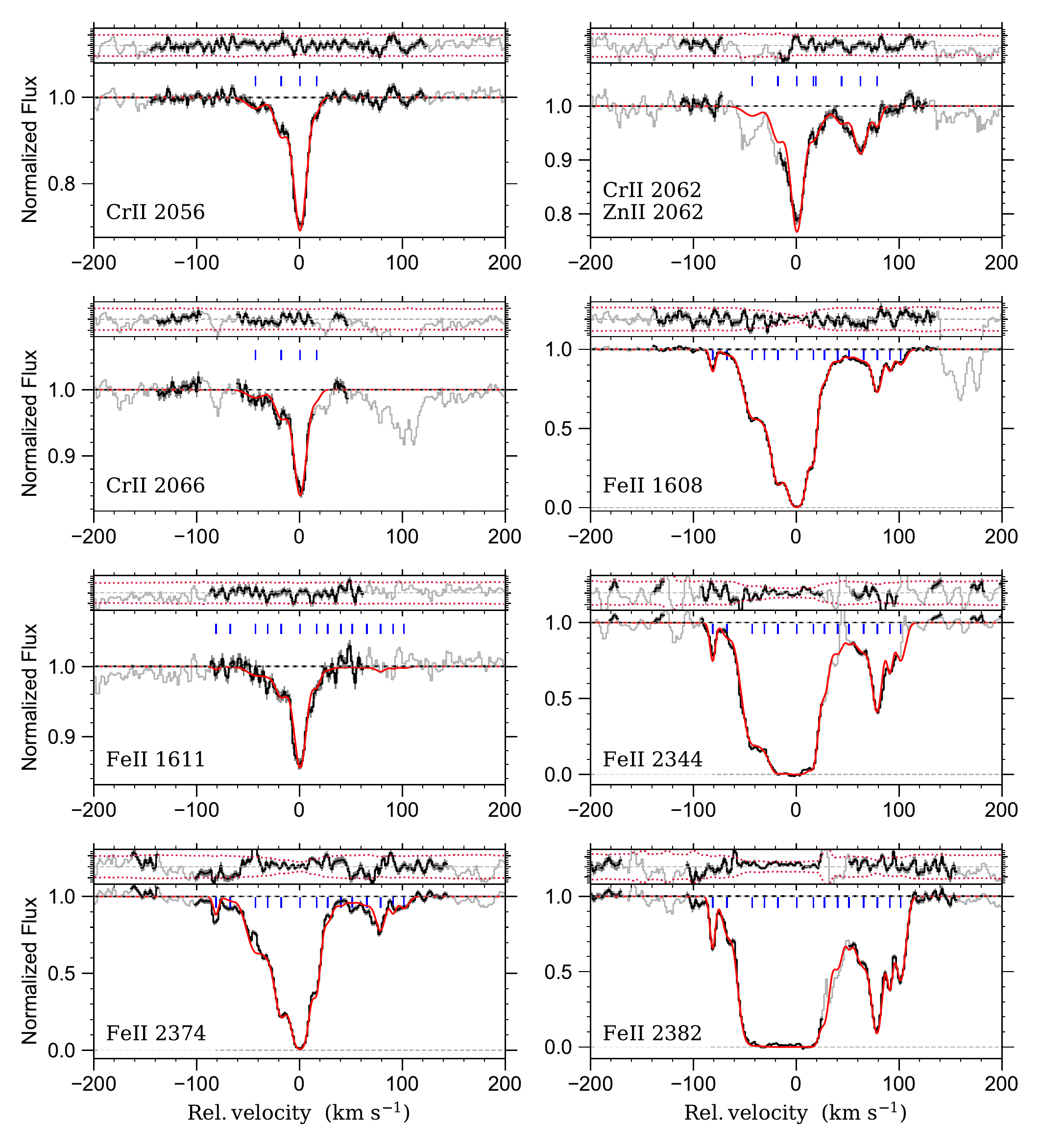}
\includegraphics[page=2,keepaspectratio=true,width=\columnwidth]{Figures/Fits/Q0528a.pdf}
\includegraphics[page=3,keepaspectratio=true,width=\columnwidth]{Figures/Fits/Q0528a.pdf}
\vspace{-7mm}
\caption{Velocity profiles of selected low-ionization transition lines from the DLA system at $z_{\rm abs}=$ 2.14105 toward Q\,0528$-$250a.}
\end{figure}

\begin{figure}[h!]
\includegraphics[page=1,keepaspectratio=true,width=\columnwidth]{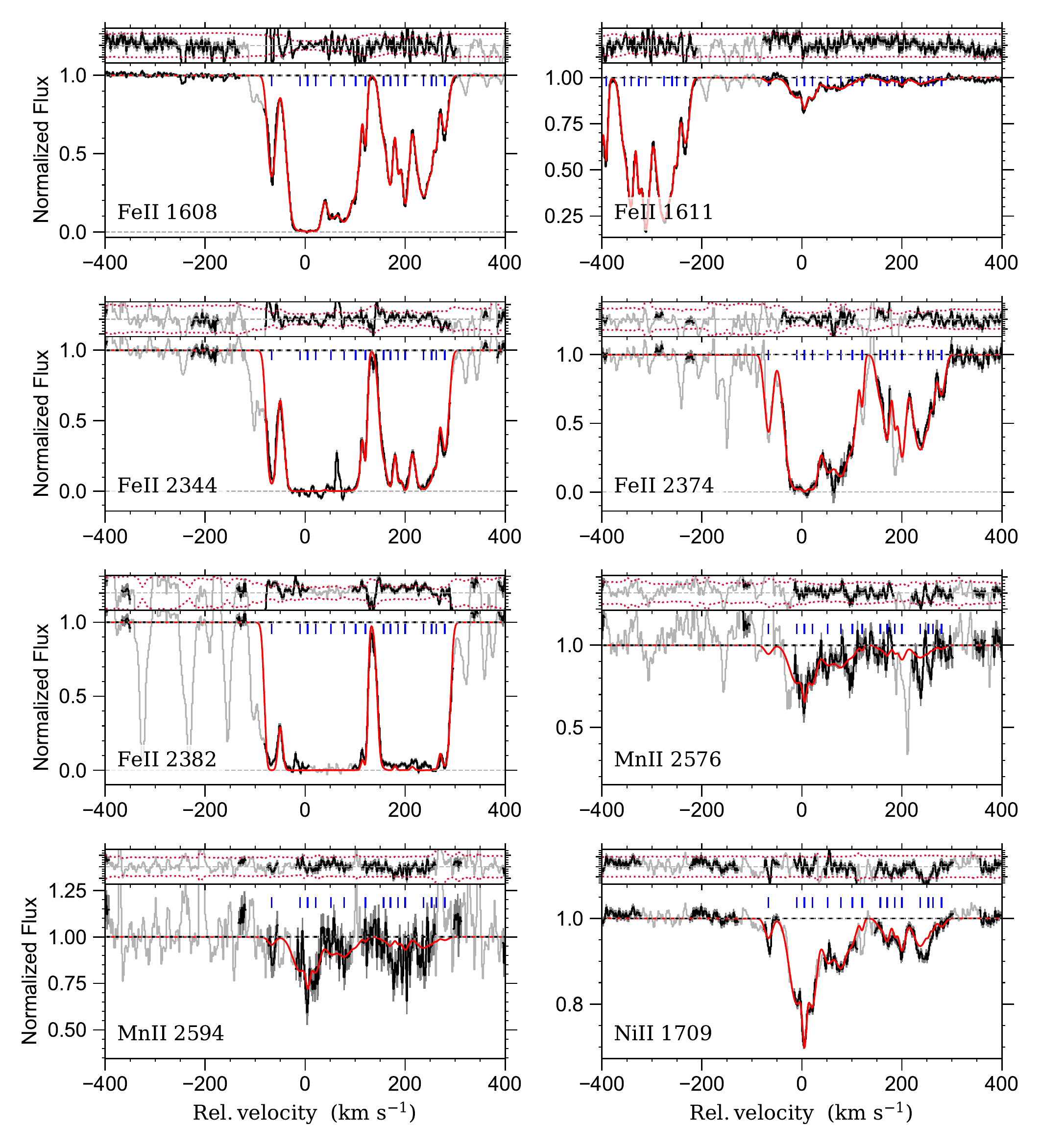}
\includegraphics[page=2,keepaspectratio=true,width=\columnwidth]{Figures/Fits/Q0528b_Ti.pdf}
\caption{Velocity profiles of selected low-ionization transition lines from the DLA system at $z_{\rm abs}=$ 2.81111 toward Q\,0528$-$250b.}
\end{figure}

\begin{figure}[h!]
\includegraphics[keepaspectratio=true,width=\columnwidth]{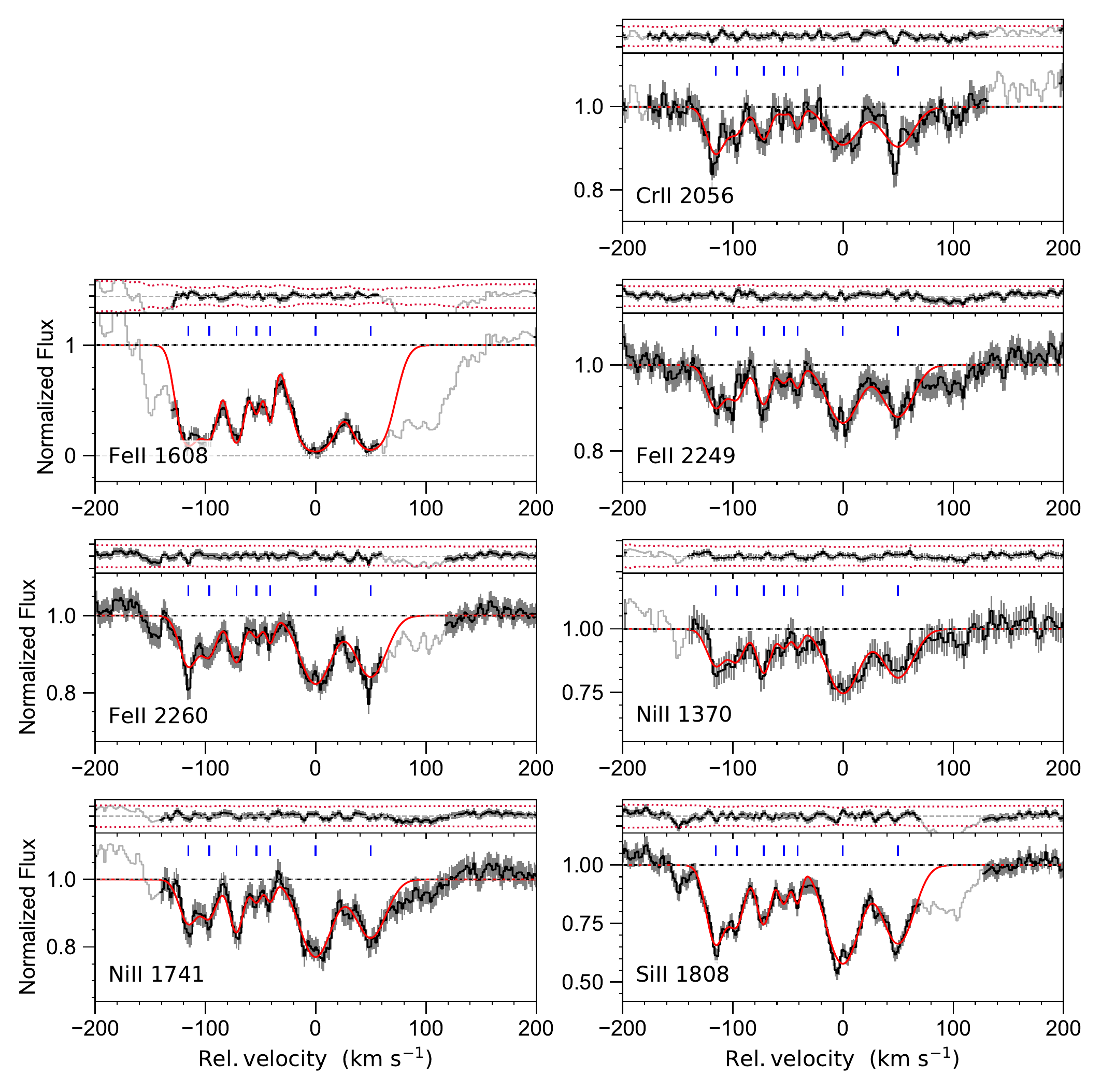}
\caption{Velocity profiles of selected low-ionization transition lines from the DLA system at $z_{\rm abs}=$ 1.96221 toward Q\,0551$-$366.}
\end{figure}

\begin{figure}[h!]
\includegraphics[page=1,keepaspectratio=true,width=\columnwidth]{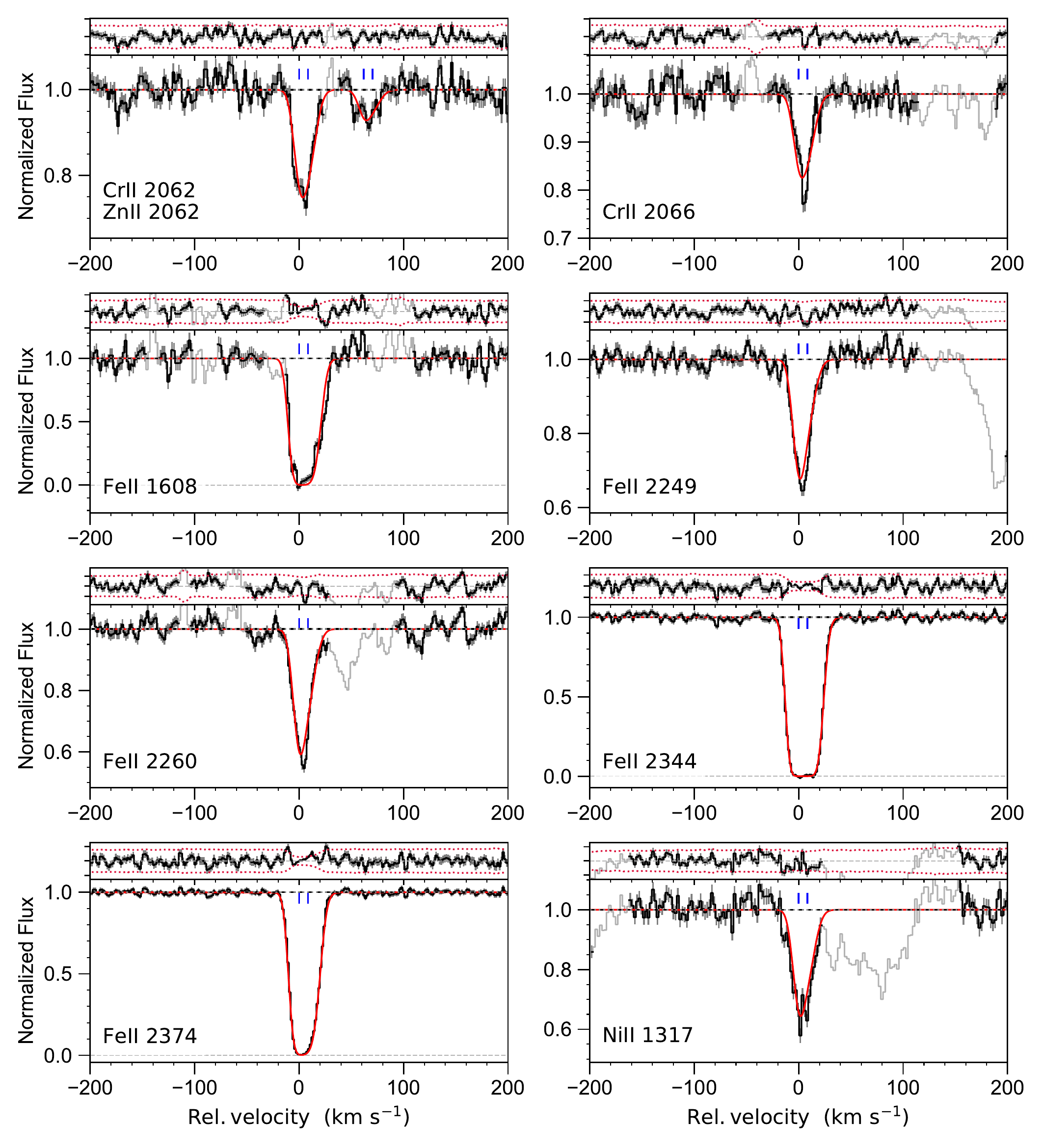}
\includegraphics[page=2,keepaspectratio=true,width=\columnwidth]{Figures/Fits/spec_addi_Q0841+129a.pdf}
\caption{Velocity profiles of selected low-ionization transition lines from the DLA system at $z_{\rm abs}=$ 1.86384 toward Q\,0841$+$129a.}
\end{figure}

\begin{figure}[h!]
\includegraphics[page=1,keepaspectratio=true,width=\columnwidth]{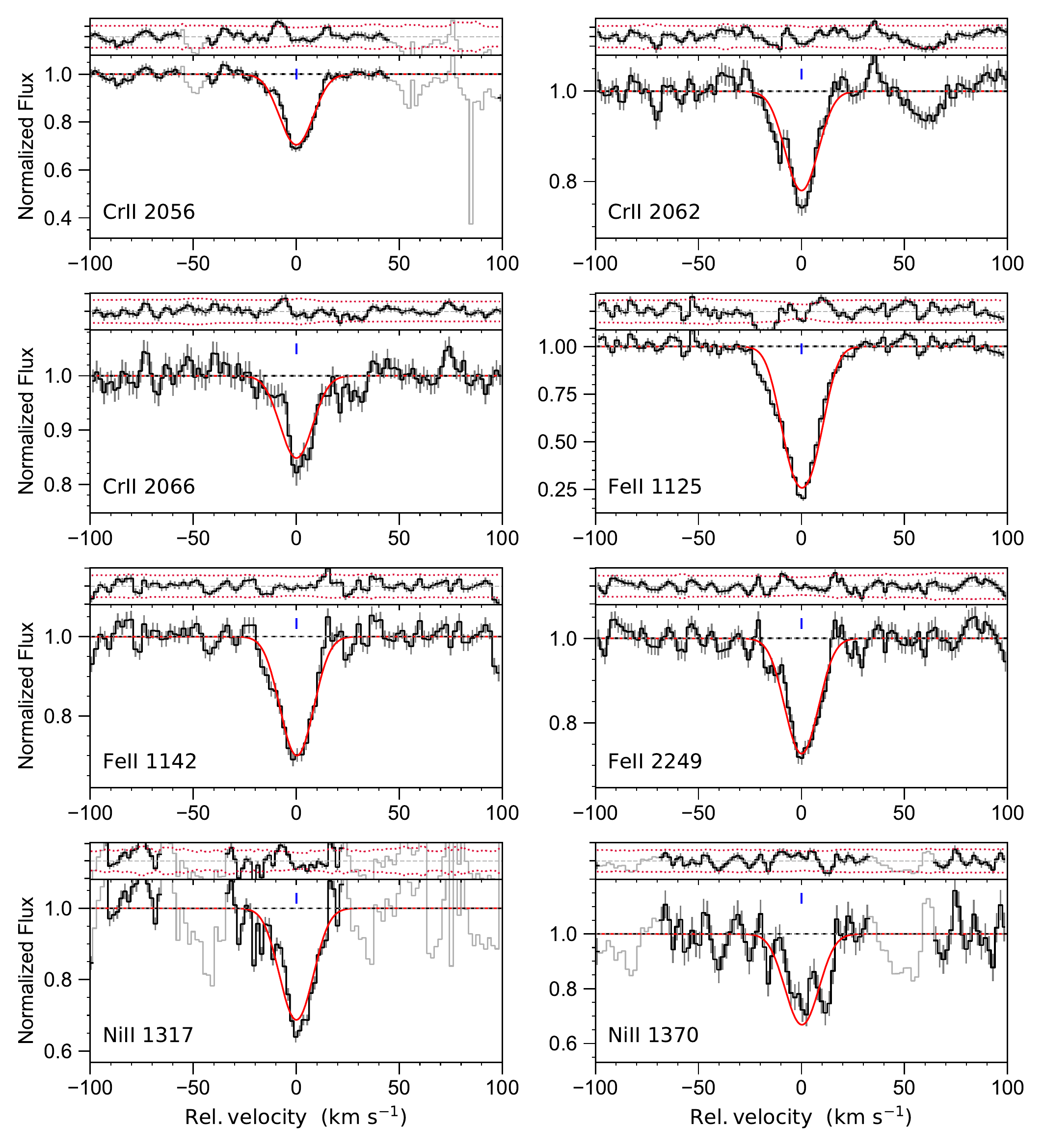}
\includegraphics[page=2,keepaspectratio=true,width=\columnwidth]{Figures/Fits/spec_addi_Q0841+129b.pdf}
\caption{Velocity profiles of selected low-ionization transition lines from the DLA system at $z_{\rm abs}=$ 2.37452 toward Q\,0841$+$129b.}
\end{figure}

\begin{figure}[h!]
\includegraphics[keepaspectratio=true,width=\columnwidth]{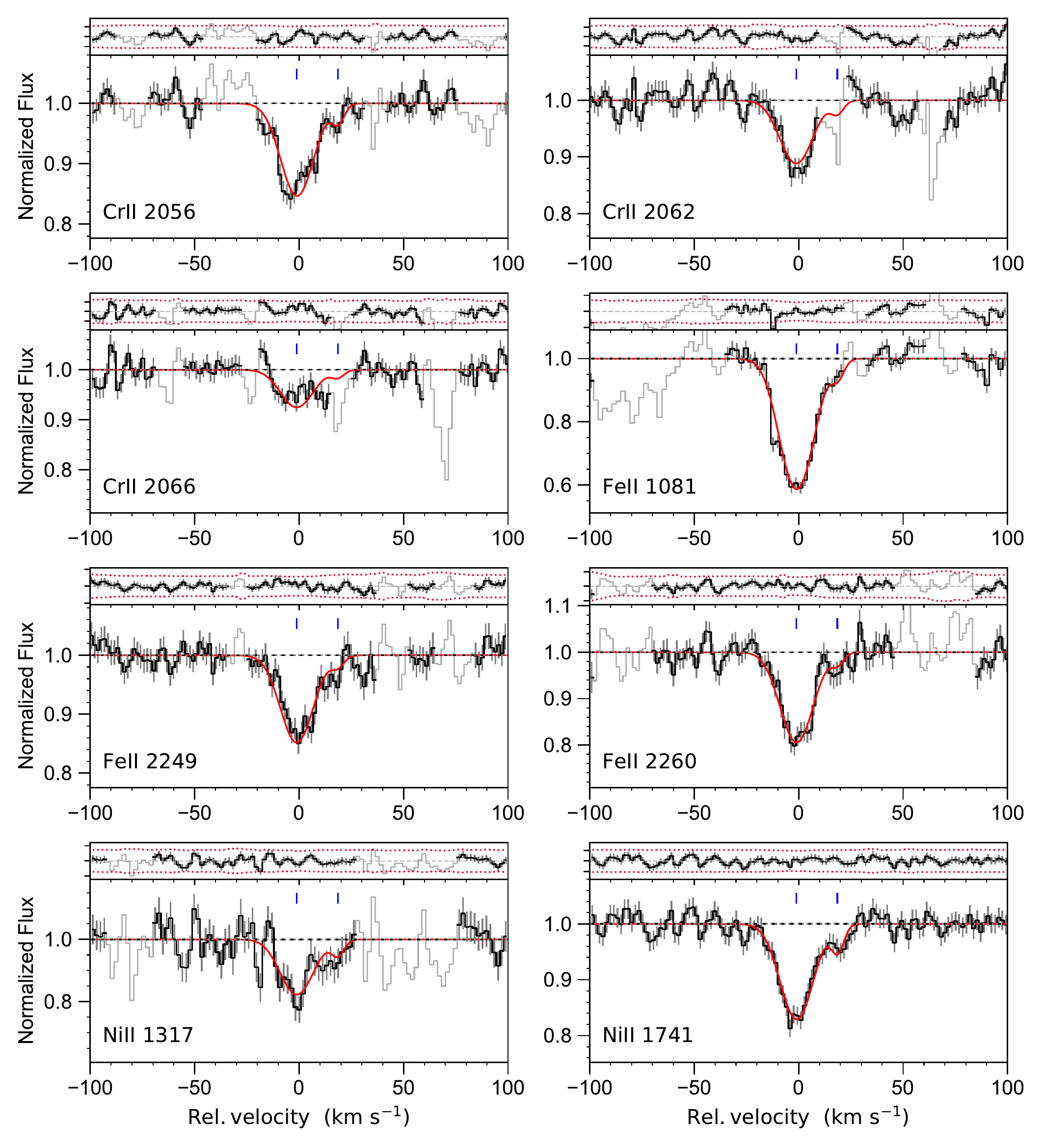}
\caption{Velocity profiles of selected low-ionization transition lines from the DLA system at $z_{\rm abs}=$ 2.47622 toward Q\,0841$+$129c.}
\end{figure}

\begin{figure}[h!]
\includegraphics[keepaspectratio=true,width=\columnwidth]{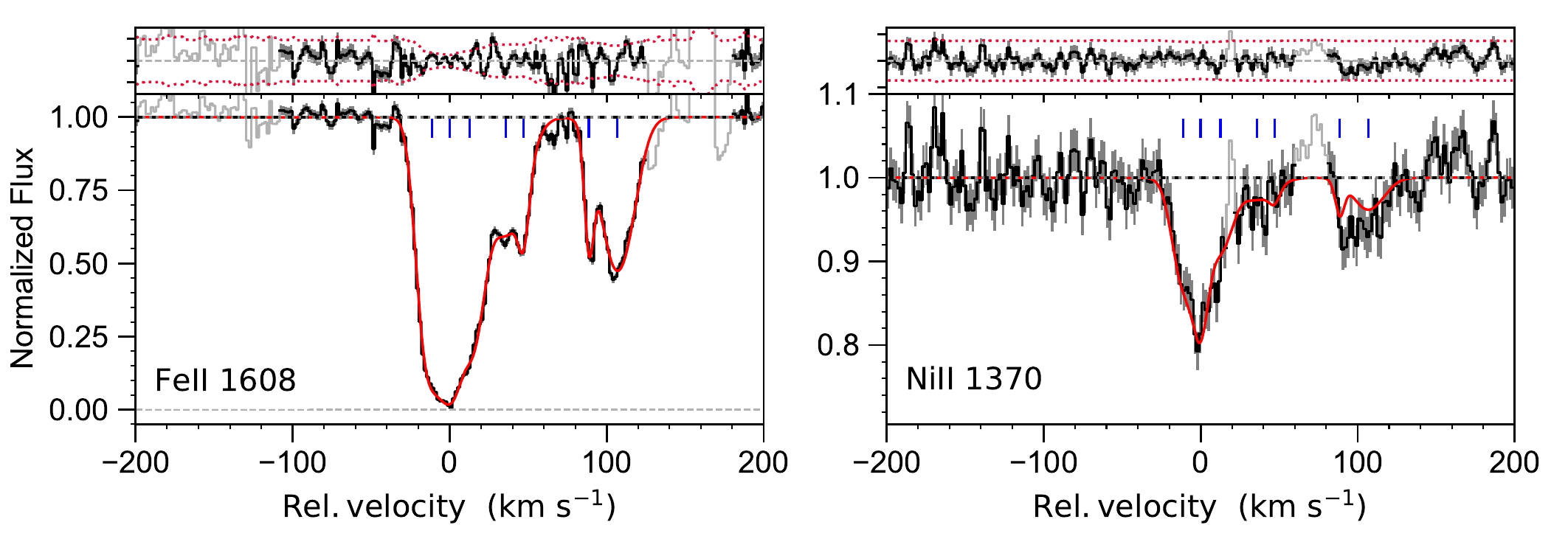}
\caption{Velocity profiles of selected low-ionization transition lines from the DLA system at $z_{\rm abs}=$ 3.26552 toward Q\,1111$-$152.}
\end{figure}

\begin{figure}[h!]
\includegraphics[keepaspectratio=true,width=\columnwidth]{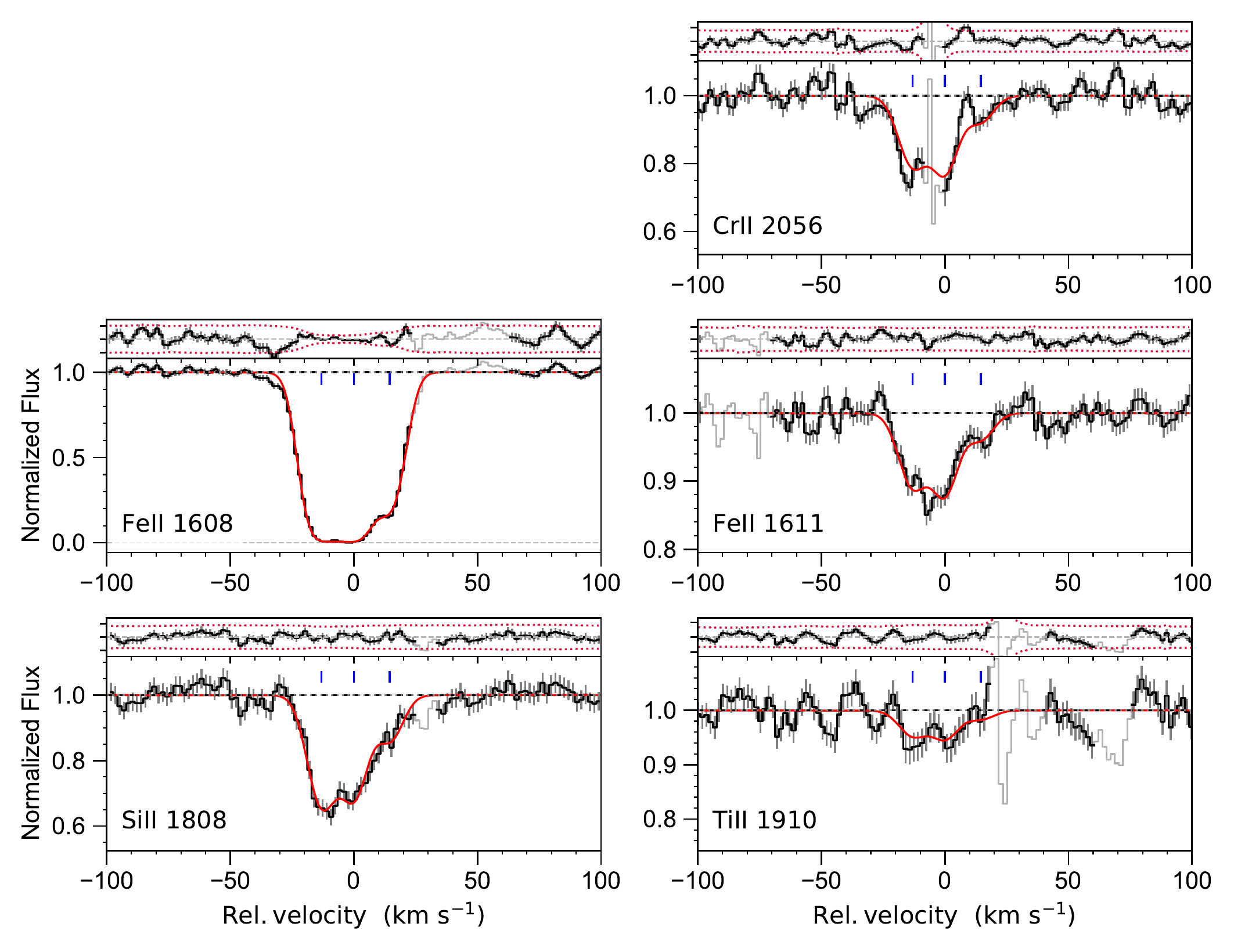}
\caption{Velocity profiles of selected low-ionization transition lines from the DLA system at $z_{\rm abs}=$ 3.35046 toward Q\,1117$-$134.}
\end{figure}

\begin{figure}[h!]
\includegraphics[page=1,keepaspectratio=true,width=\columnwidth]{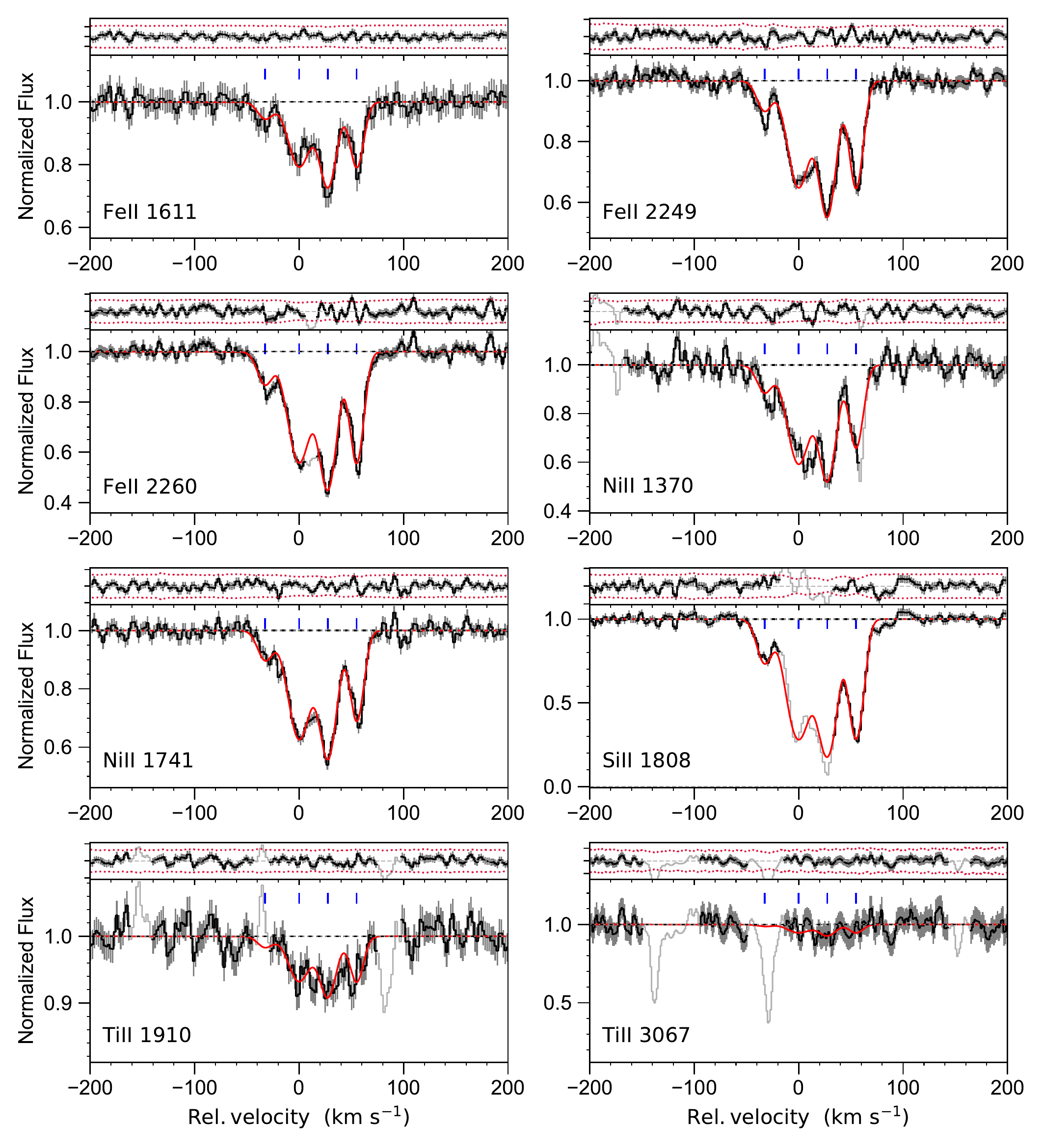}
\end{figure}
\begin{figure}
\includegraphics[page=2,keepaspectratio=true,width=\columnwidth]{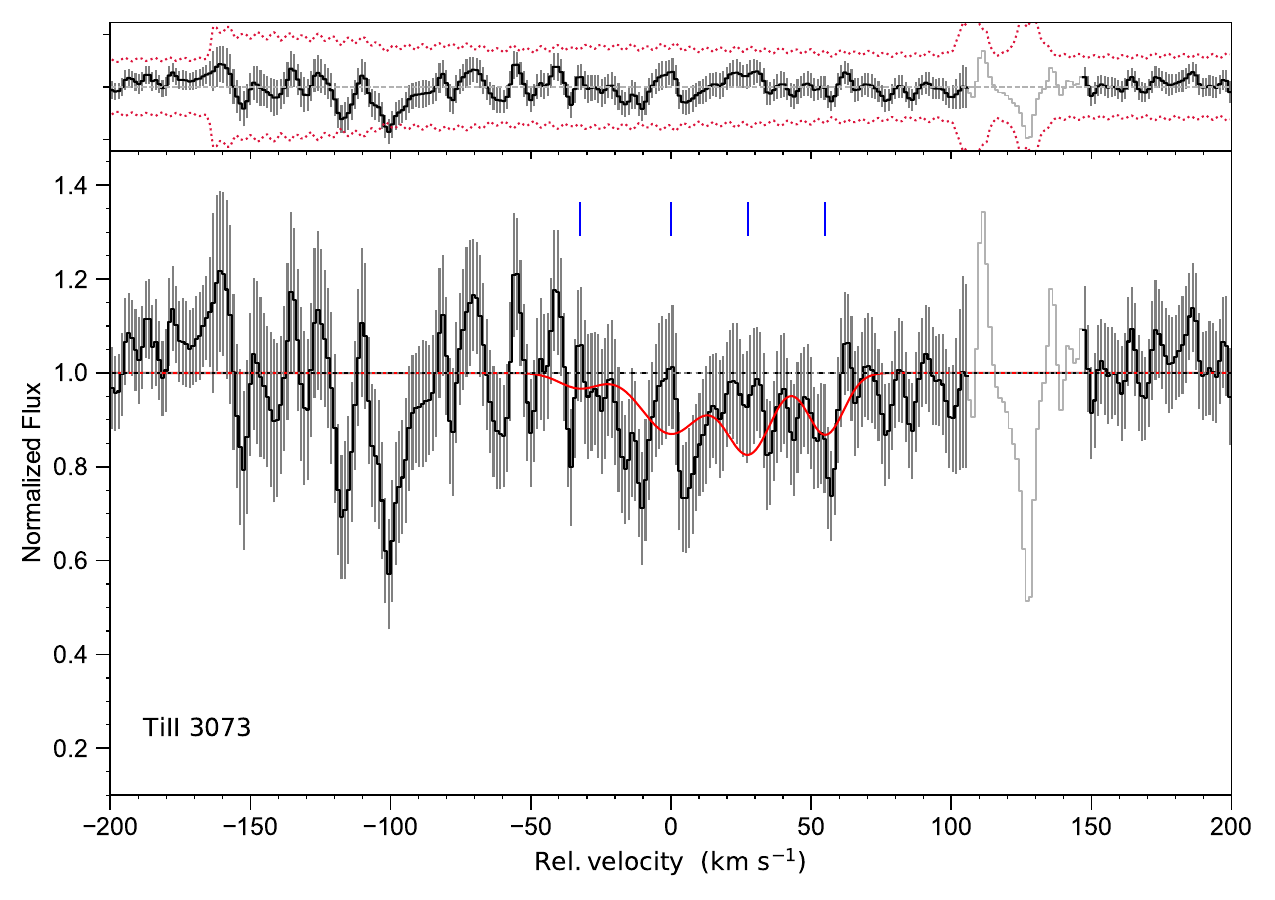}
\caption{Velocity profiles of selected low-ionization transition lines from the DLA system at $z_{\rm abs}=$ 1.94349 toward Q\,1157$+$014.}
\end{figure}

\begin{figure}[h!]
\includegraphics[keepaspectratio=true,width=\columnwidth]{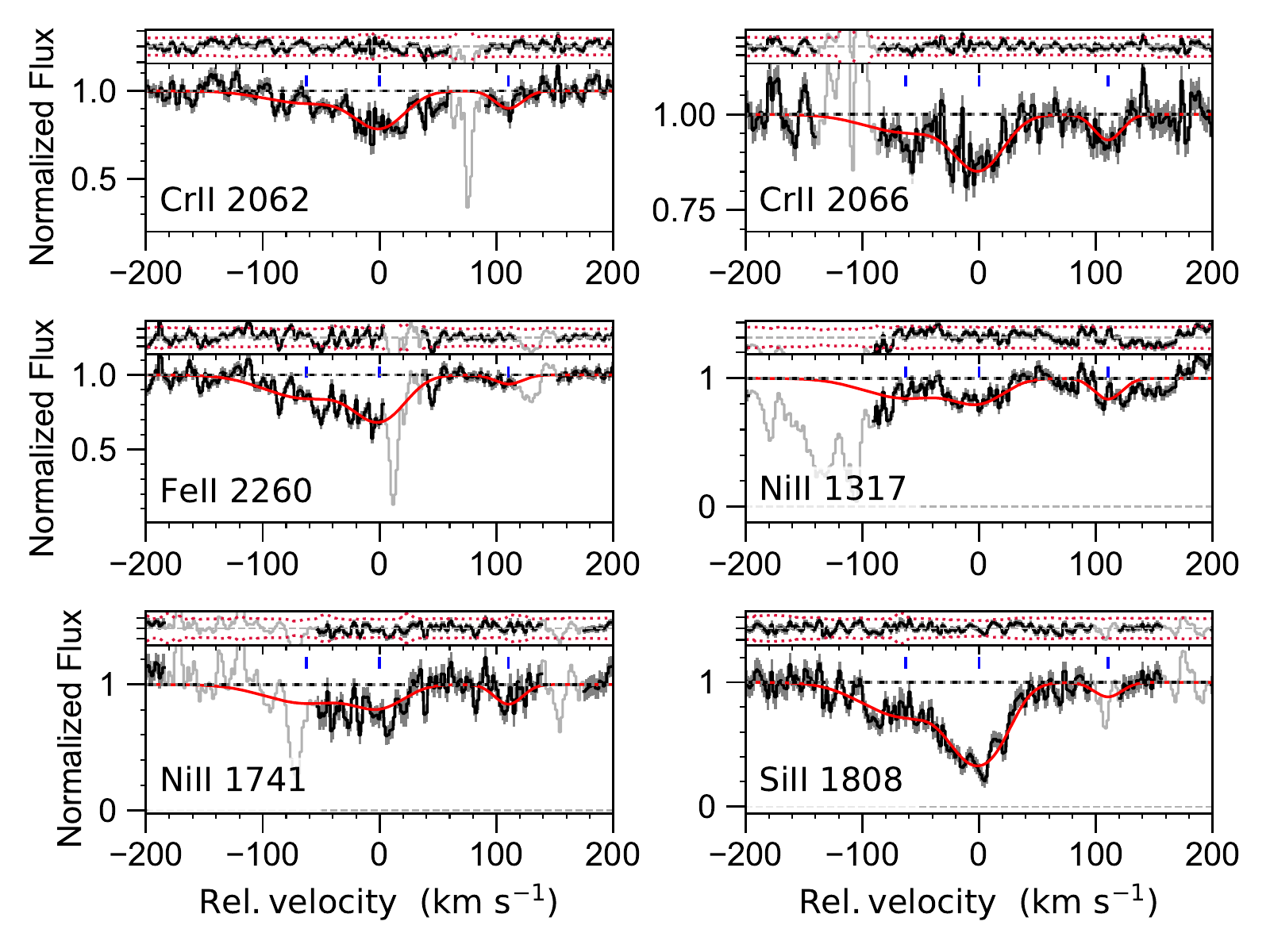}
\caption{Velocity profiles of selected low-ionization transition lines from the DLA system at $z_{\rm abs}=$ 2.58437 toward Q\,1209$+$093.}
\end{figure}

\begin{figure}[h!]
\includegraphics[keepaspectratio=true,width=\columnwidth]{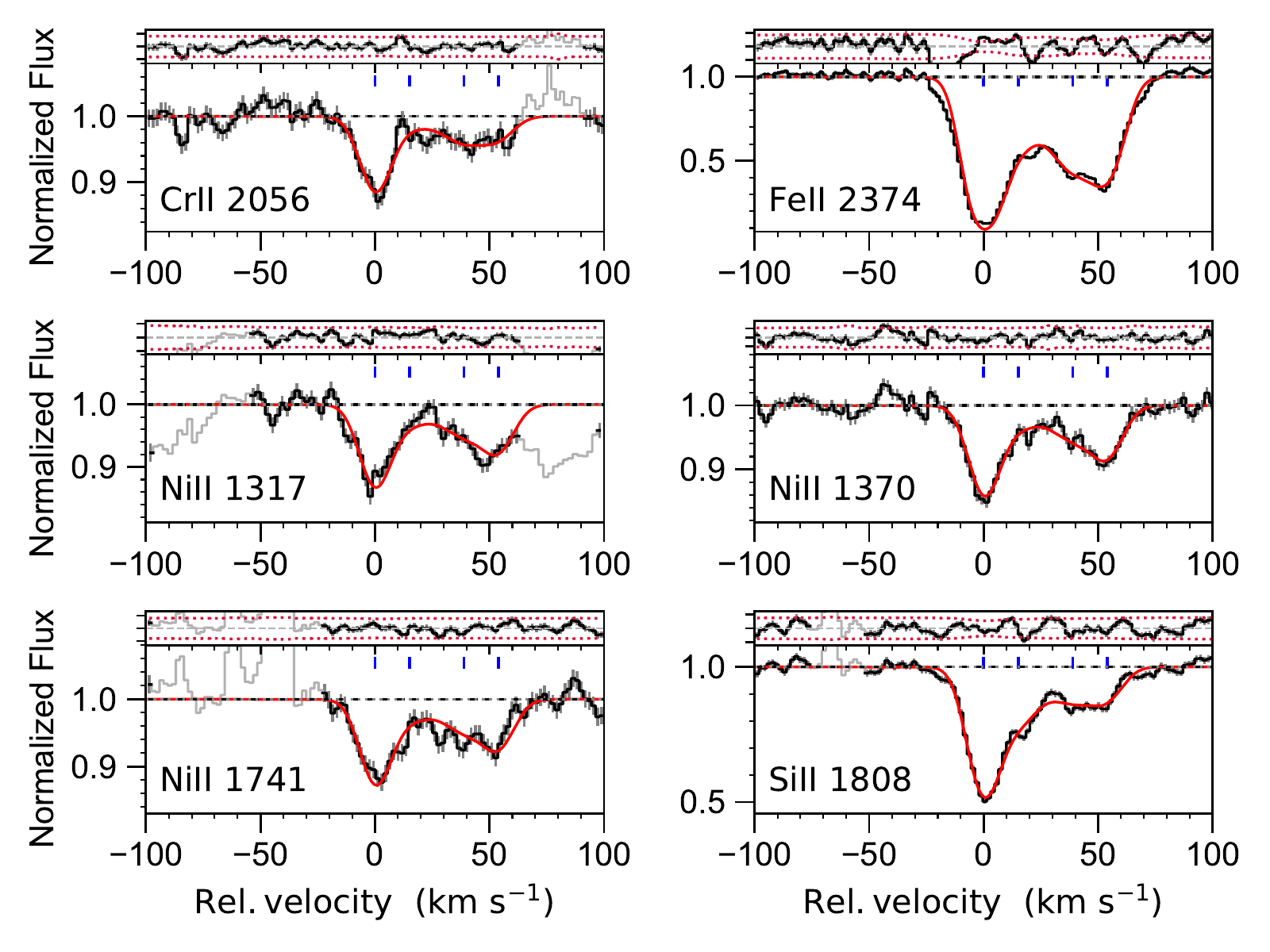}
\caption{Velocity profiles of selected low-ionization transition lines from the DLA system at $z_{\rm abs}=$ 1.77635 toward Q\,1331$+$170.}
\end{figure}

\begin{figure}[h!]
\includegraphics[keepaspectratio=true,width=\columnwidth]{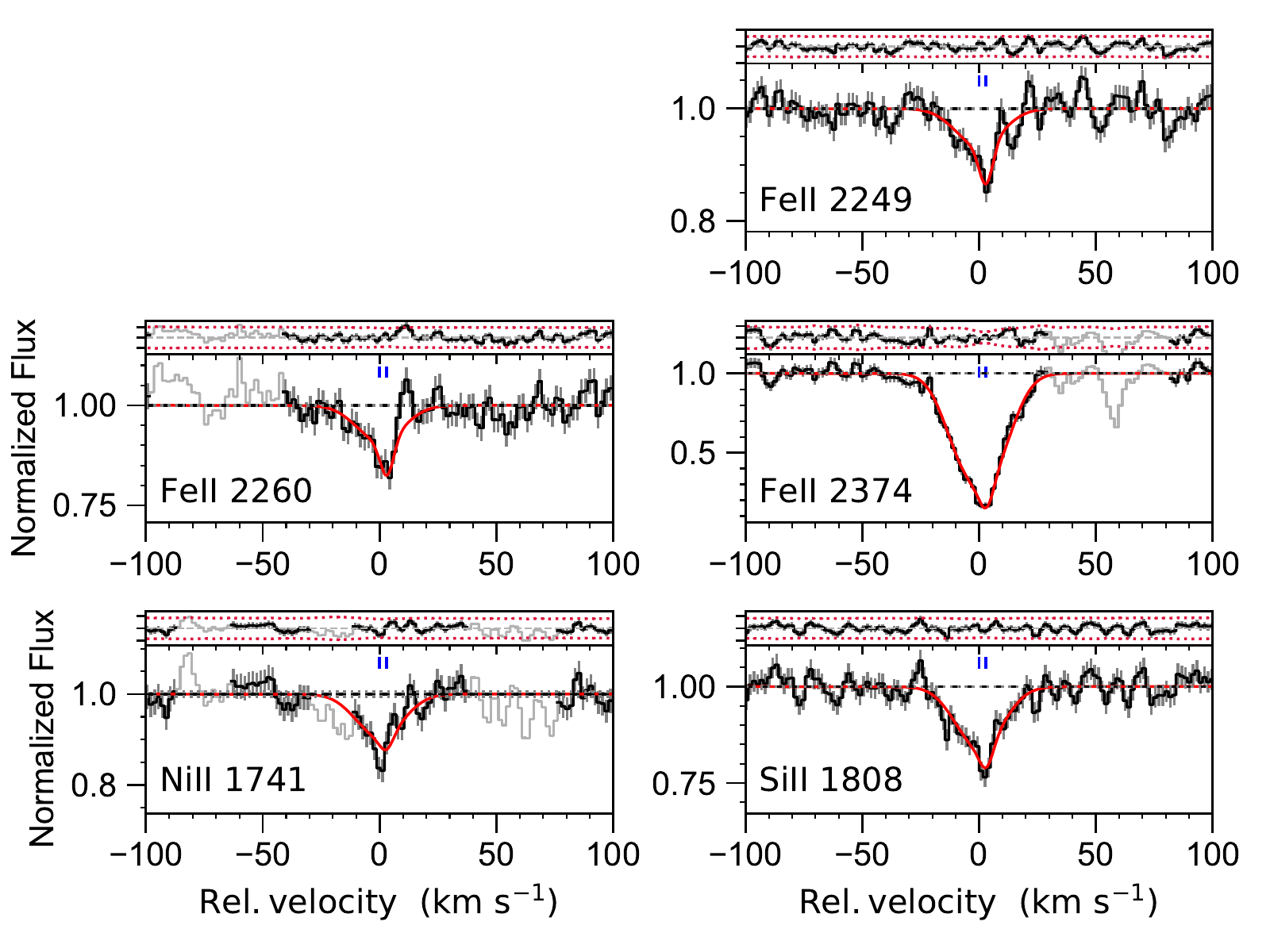}
\caption{Velocity profiles of selected low-ionization transition lines from the DLA system at $z_{\rm abs}=$ 2.01881 toward Q\,1409$+$095a.}
\end{figure}

\begin{figure}[h!]
\includegraphics[keepaspectratio=true,width=\columnwidth]{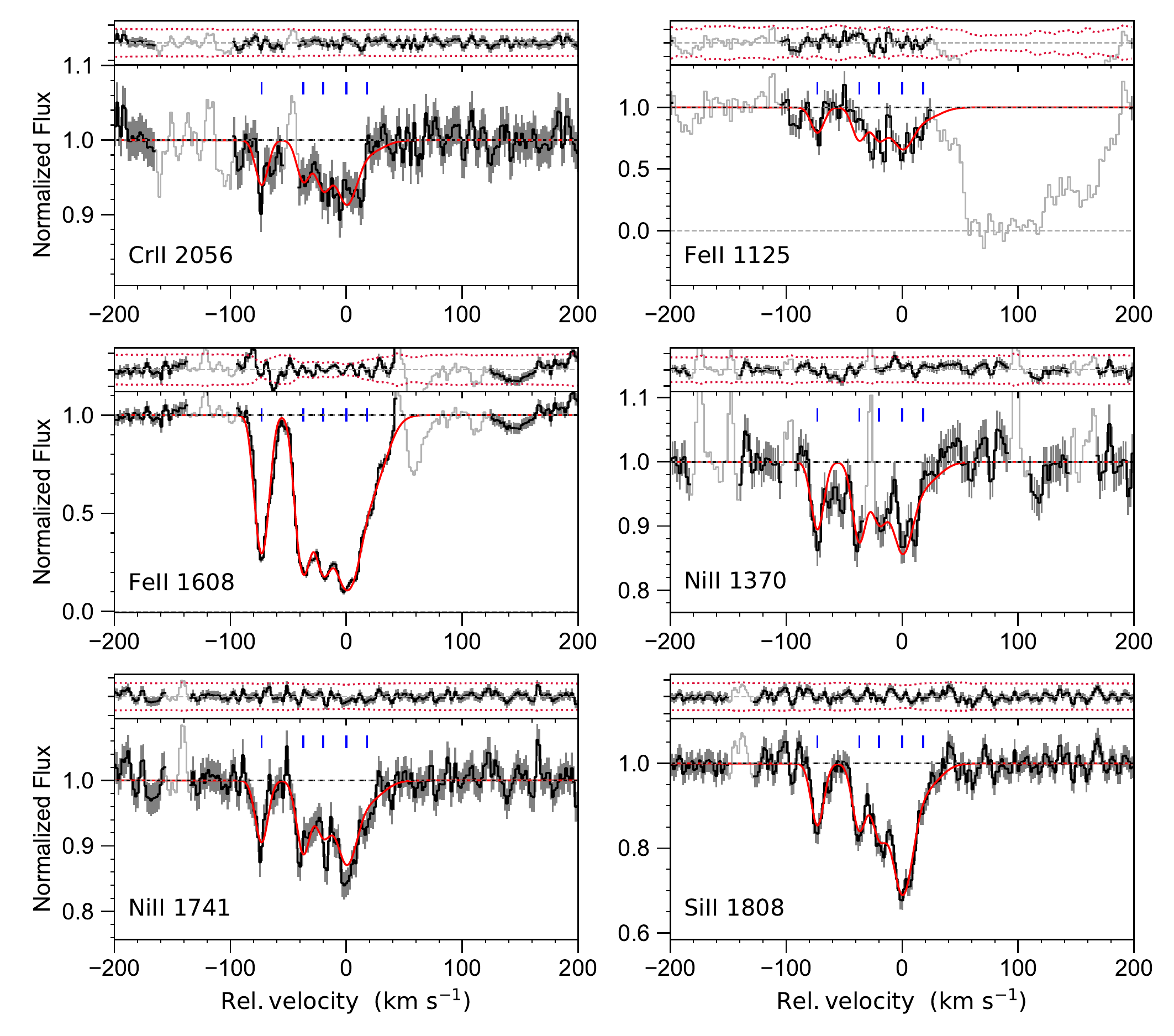}
\caption{Velocity profiles of selected low-ionization transition lines from the DLA system at $z_{\rm abs}=$ 1.99615 toward Q\,2116$-$358.}
\end{figure}

\begin{figure}[h!]
\includegraphics[keepaspectratio=true,width=\columnwidth]{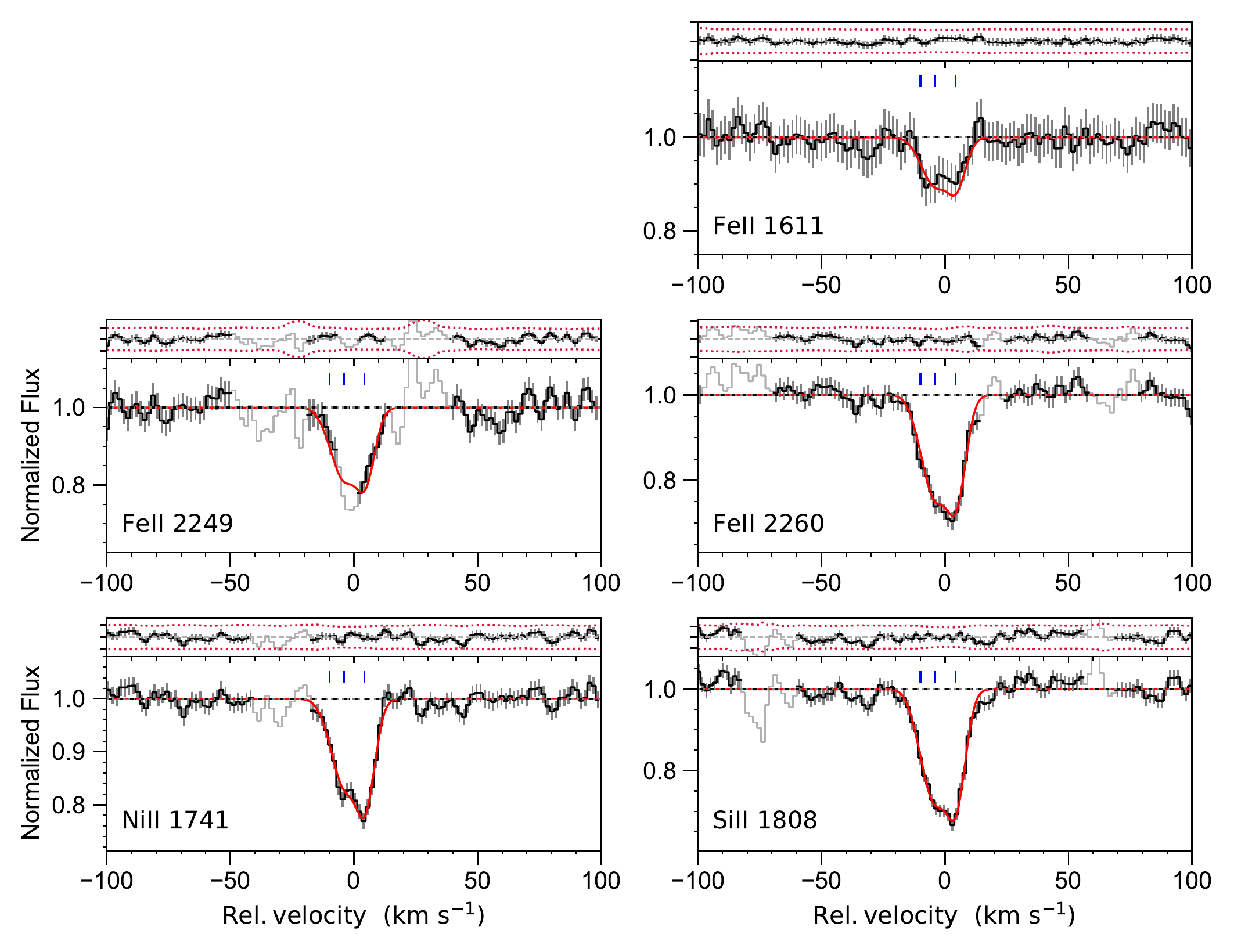}
\caption{Velocity profiles of selected low-ionization transition lines from the DLA system at $z_{\rm abs}=$ 2.85234 toward Q\,2138$-$444b.}
\end{figure}

\begin{figure}[h!]
\includegraphics[keepaspectratio=true,width=\columnwidth]{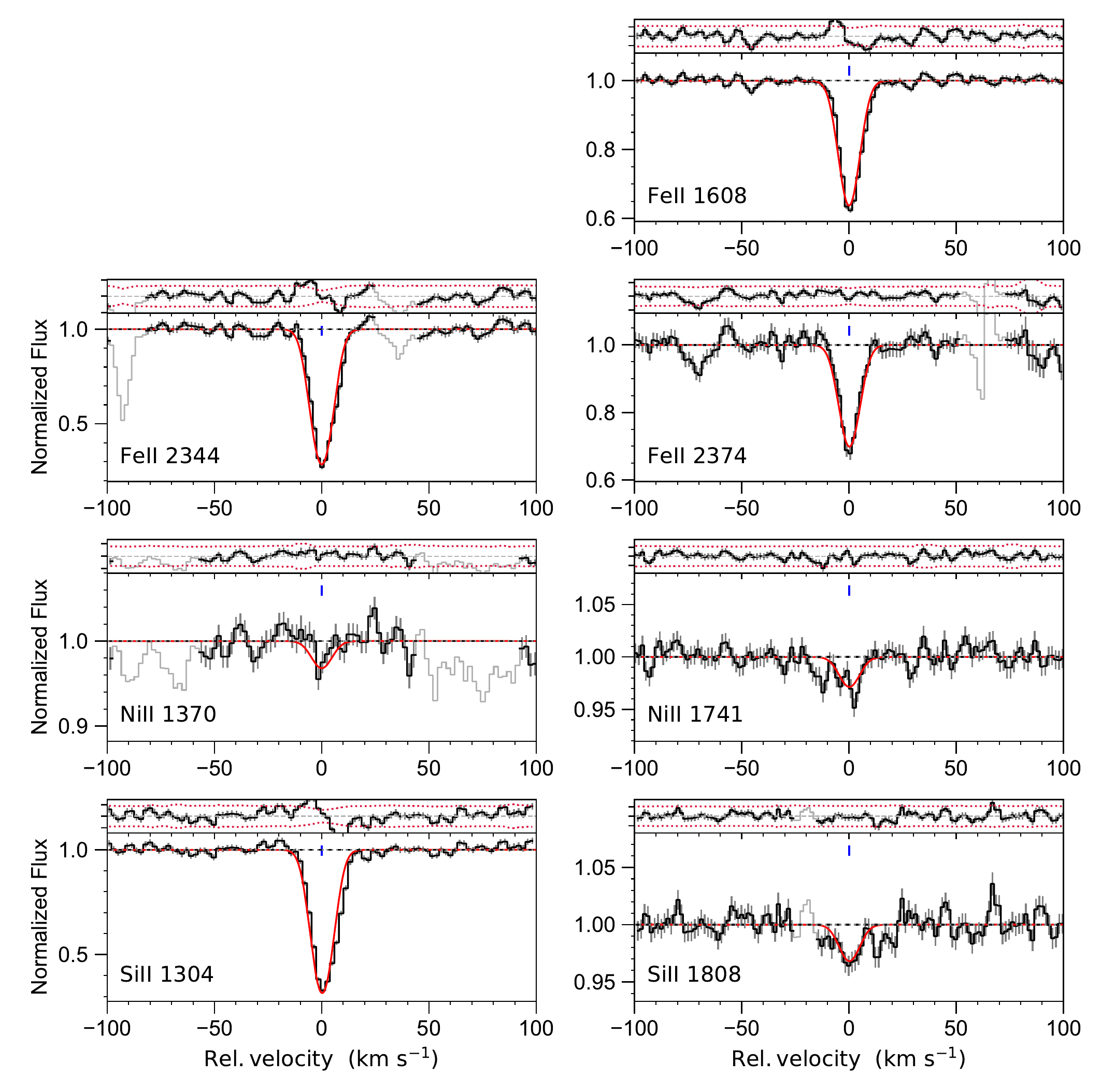}
\includegraphics[keepaspectratio=true,width=\columnwidth]{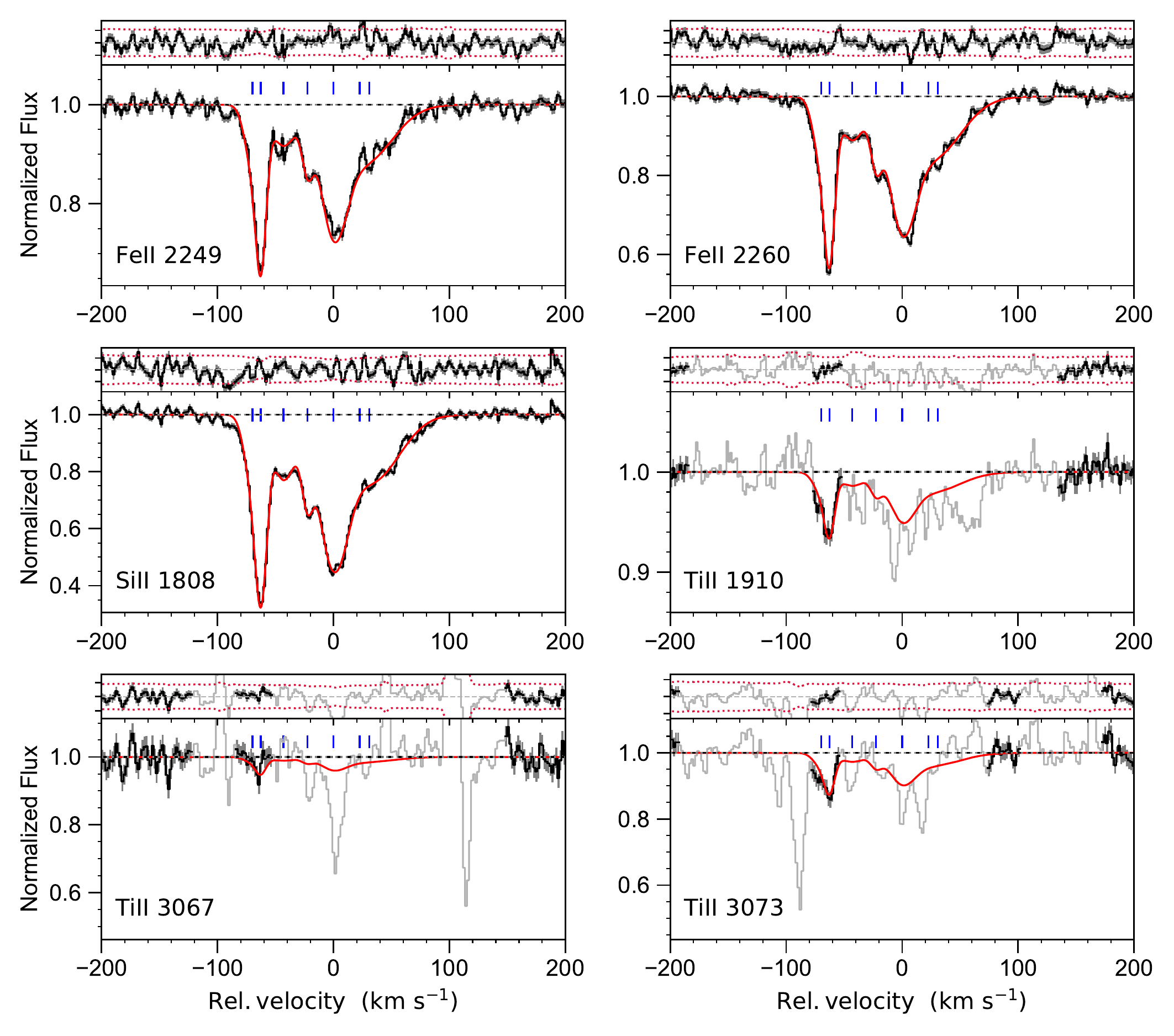}
\caption{Velocity profiles of selected low-ionization transition lines from the DLA system at $z_{\rm abs}=$ 1.92061 toward Q\,2206$-$199a.}
\end{figure}

\begin{figure}[h!]
\includegraphics[keepaspectratio=true,width=\columnwidth]{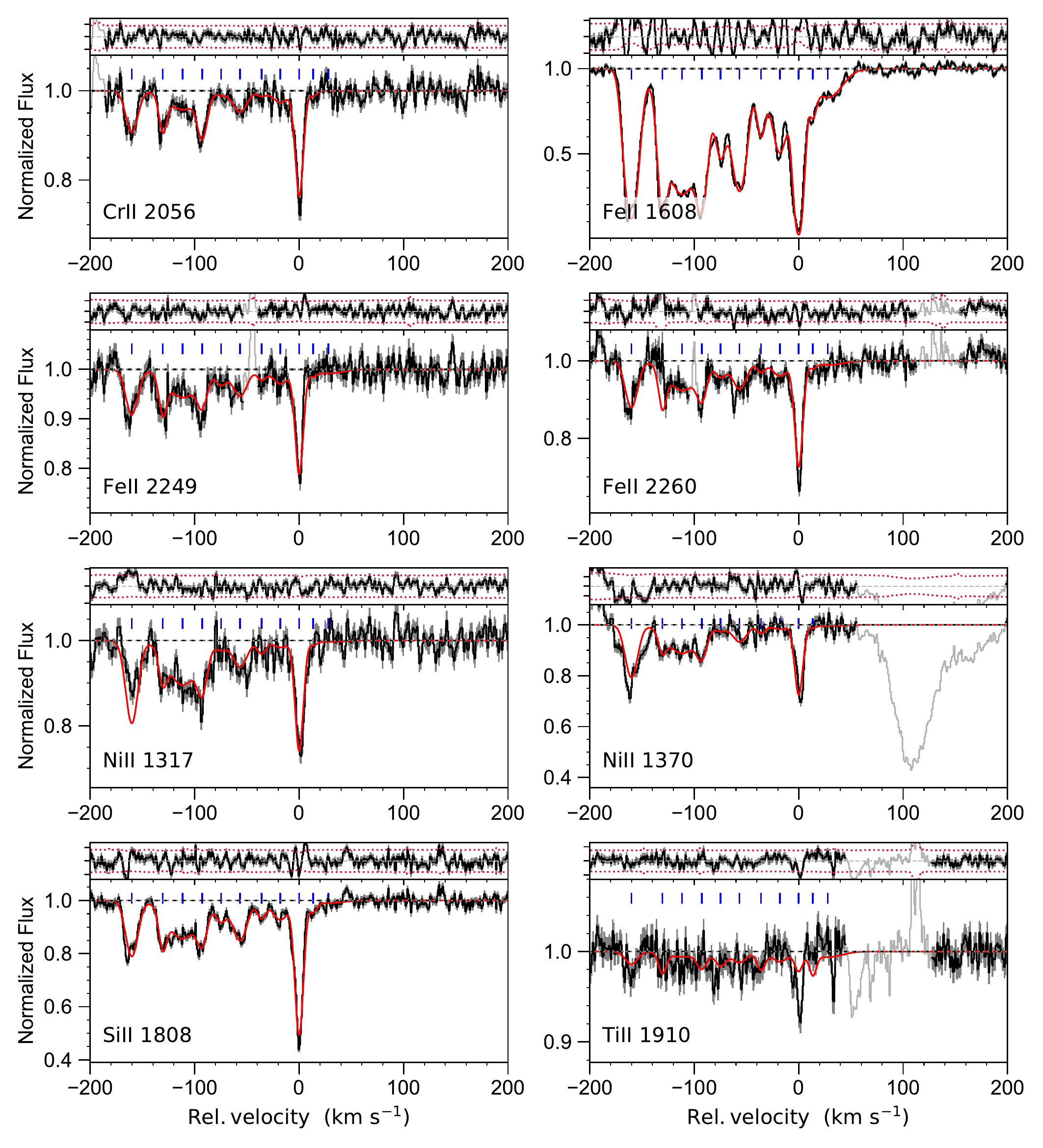}
\caption{Velocity profiles of selected low-ionization transition lines from the DLA system at $z_{\rm abs}=$ 2.33062 toward Q\,2243$-$605.}
\end{figure}

\begin{figure}[h!]
\includegraphics[keepaspectratio=true,width=\columnwidth]{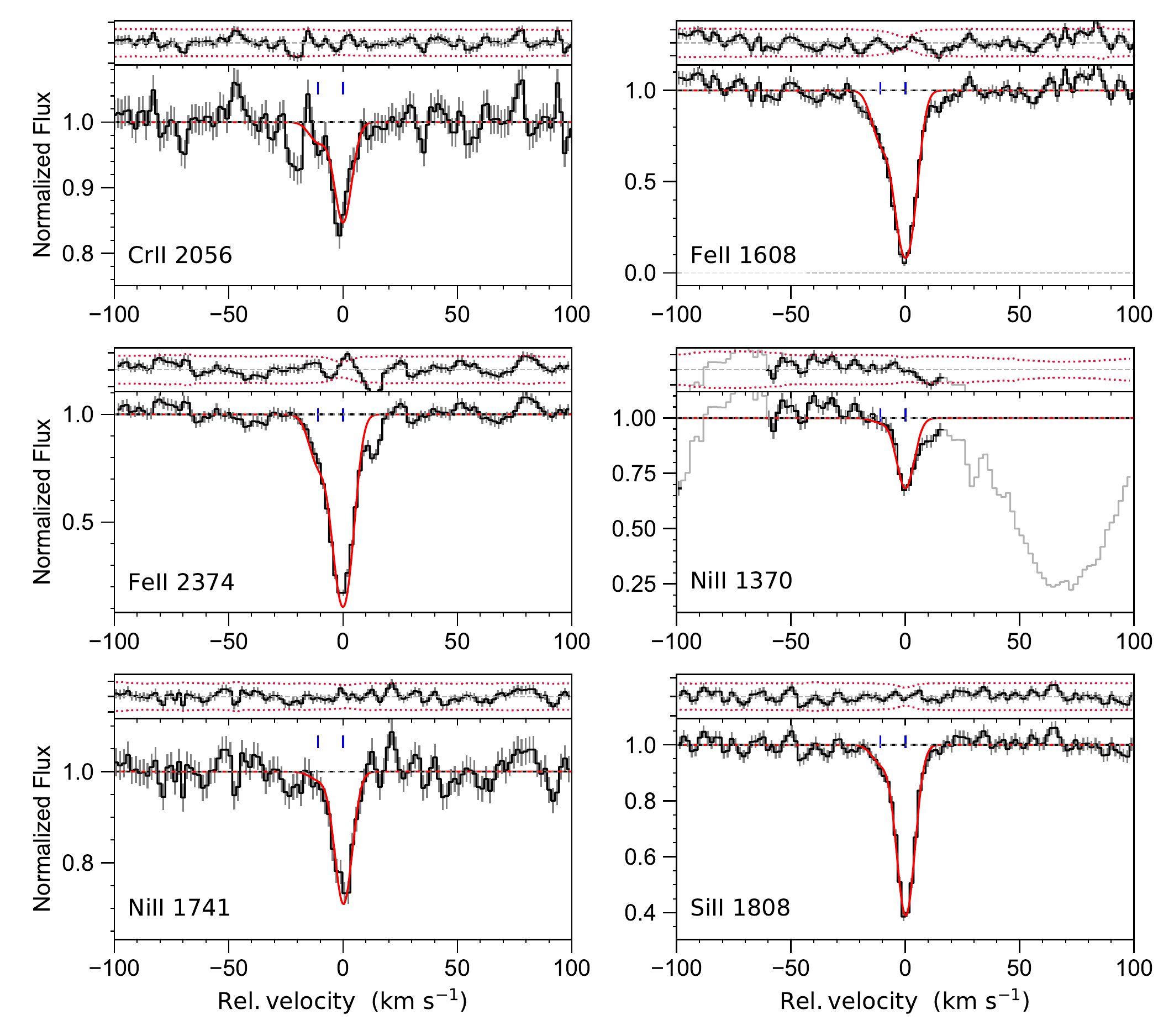}
\caption{Velocity profiles of selected low-ionization transition lines from the DLA system at $z_{\rm abs}=$ 2.28749 toward Q\,2332$-$094a.}
\end{figure}

\begin{figure}[h!]
\includegraphics[keepaspectratio=true,width=\columnwidth]{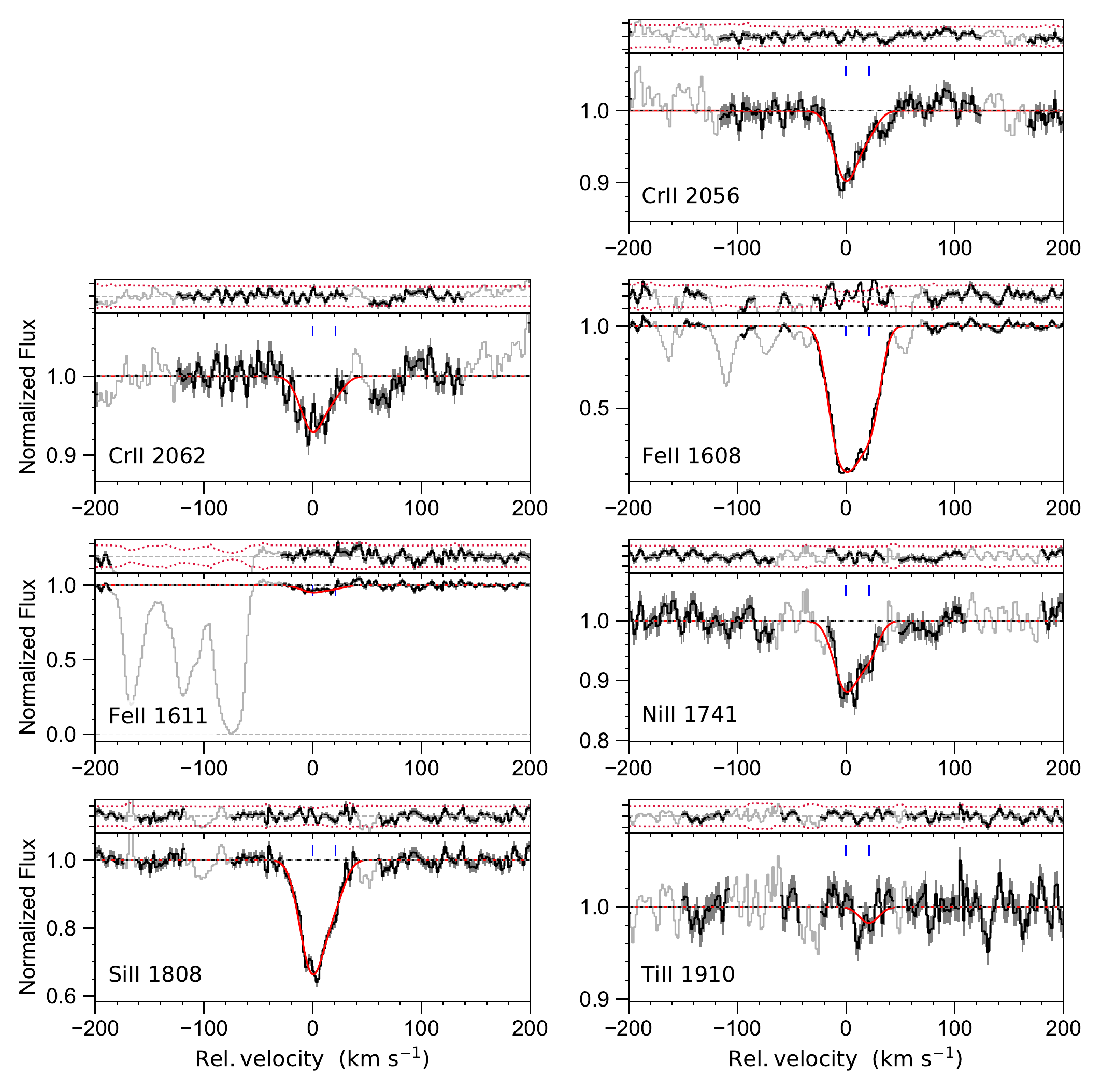}
\caption{Velocity profiles of selected low-ionization transition lines from the DLA system at $z_{\rm abs}=$ 2.43123 toward Q\,2343$+$125.}
\end{figure}

\begin{figure}[h!]
\includegraphics[keepaspectratio=true,width=\columnwidth]{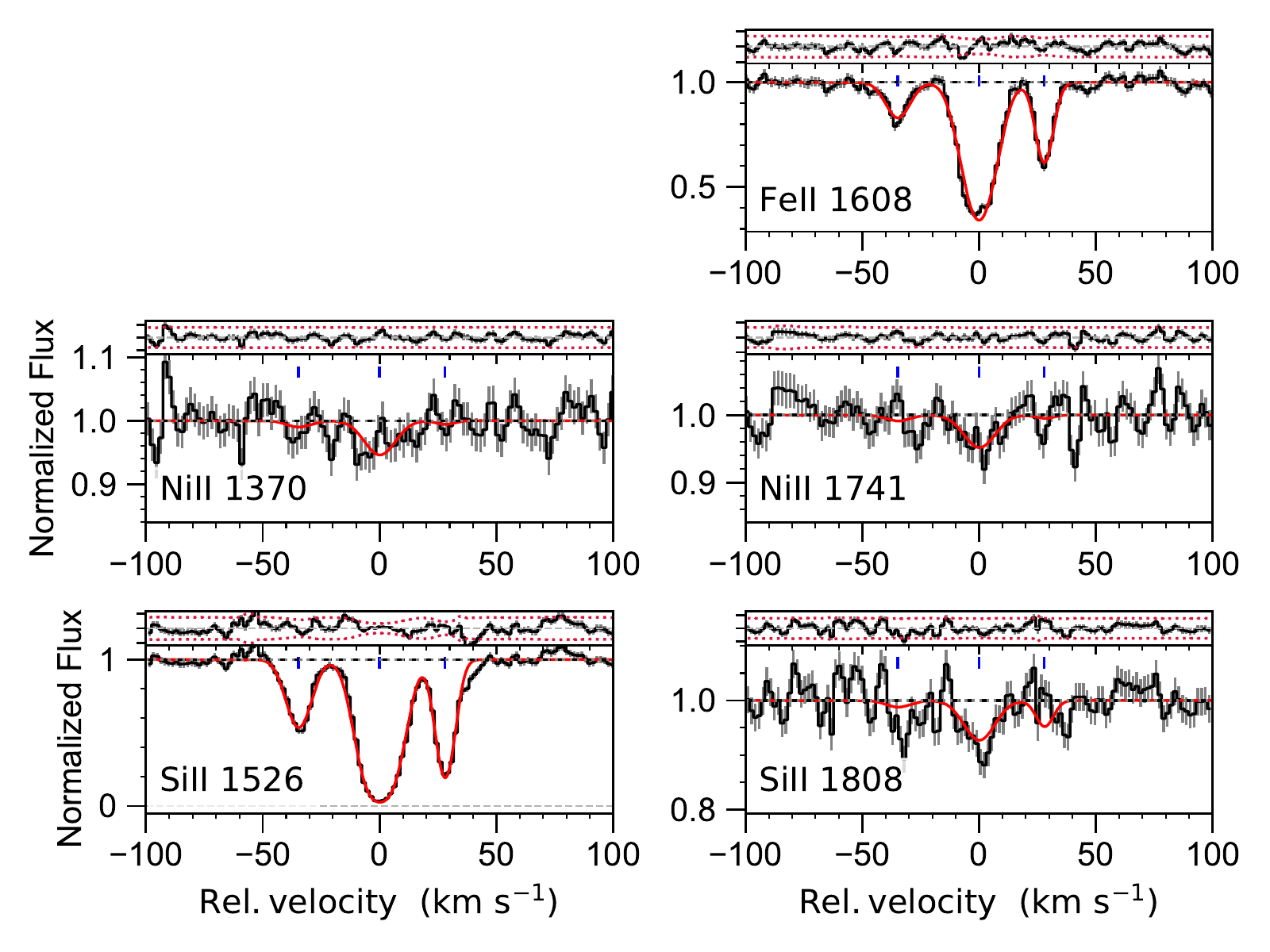}
\caption{Velocity profiles of selected low-ionization transition lines from the DLA system at $z_{\rm abs}=$ 2.0951 toward Q\,2344$+$125.}
\end{figure}

\clearpage

\section{Column densities of all the samples}
\label{appsec:columndens}
\onecolumn

\begin{landscape}
\centering
\tiny
\setlength\tabcolsep{3.5pt}
\begin{longtable}{lccccccccccccccc}
\caption{QSO-DLA column densities}
\label{dlastable}
\\
\hline
QSO & z$_{abs}$ &log N(\ion{H}{I})&log N(\ion{O}{I}) & log N(\ion{Mg}{II})& log N(\ion{Si}{II}) & log N(\ion{S}{II}) & log N(\ion{P}{II})& log N(\ion{Ti}{II})&log N(\ion{Cr}{II})& log N(\ion{Mn}{II})& log N(\ion{Fe}{II})  & log N(\ion{Co}{II})& log N(\ion{Ni}{II}) & log N(\ion{Zn}{II}) & Ref\\
\noalign{\smallskip}\hline\hline\noalign{\smallskip}
\endfirsthead
\caption{continued}\\
\hline
QSO & z$_{abs}$ &log N(\ion{H}{I})&log N(\ion{O}{I}) & log N(\ion{Mg}{II})& log N(\ion{Si}{II}) & log N(\ion{S}{II}) & log N(\ion{P}{II})& log N(\ion{Ti}{II})&log N(\ion{Cr}{II})& log N(\ion{Mn}{II})& log N(\ion{Fe}{II})  & log N(\ion{Co}{II})& log N(\ion{Ni}{II}) & log N(\ion{Zn}{II}) & Ref\\
\noalign{\smallskip}\hline\hline\noalign{\smallskip}
\endhead
\hline
\endfoot
\hline
\endlastfoot
Q0000-263      & 3.39 & 21.40$\pm$0.08  & 16.42$\pm$0.10 & ...            & 14.99$\pm$0.02 & 14.74$\pm$0.03 & 12.72$\pm$0.04           & ...                               & 13.08$\pm$0.03 & ...            & 14.87$\pm$0.03 & ...            & ...                               & 11.91$\pm$0.05  & 1\\
Q0010-002      & 2.02 & 20.95$\pm$0.10 & ...            & 15.54$\pm$0.05 & 15.24$\pm$0.03 & 15.00$\pm$0.03  & ...            & \textless{12.38} & 13.41$\pm$0.02 & ...  & 15.18$\pm$0.04 & ...            & 13.69$\pm$0.03                    & 12.09$\pm$0.04  & $\star$, 1 \\
Q0013-004      & 1.97 & 20.83$\pm$0.05 & ...  & ...            & 15.47$\pm$0.01 & 15.48$\pm$0.01 & 14.04$\pm$0.01          & ...                               & 13.17$\pm$0.03 & 13.00$\pm$0.04 & 14.91$\pm$0.01 & ...            & ...                               & 12.81$\pm$0.01 & 1 \\
Q0058-292      & 2.67 & 21.10$\pm$0.10   & ...            & 15.30$\pm$0.00  & 15.16$\pm$0.04 & 14.96$\pm$0.02 & ...            & 12.30$\pm$0.11                    & 13.11$\pm$0.02 & ...  & 14.75$\pm$0.02 & ...            & 13.55$\pm$0.08                    & 12.14$\pm$0.02 & $\star$, 1 \\
Q0100+130      & 2.31 & 21.35$\pm$0.08 & ...            & 15.51$\pm$0.06 & ...            & 15.13$\pm$0.01 & 13.08$\pm$0.02        & 12.72$\pm$0.11                    & 13.36$\pm$0.01 & 12.68$\pm$0.04 & 15.05$\pm$0.01 & ...            & ...                               & 12.34$\pm$0.01 & $\star$, 1 \\
Q0102-190a     & 2.37 & 21.00$\pm$0.08 & ... & ... & ...            & 14.30$\pm$0.02 & ...            & 12.20$\pm$0.12                    & 12.80$\pm$0.02 & 12.04$\pm$0.04 & 14.44$\pm$0.02 & ...            & 13.16$\pm$0.07                    & 11.61$\pm$0.08 & $\star$, 1 \\
Q0216+080a     & 1.77 & 20.30$\pm$0.10 & ...            & ...  & 14.83$\pm$0.02 & 14.47$\pm$0.13 & ...           & \textless{12.22} & 12.68$\pm$0.07 & ...  & 14.47$\pm$0.01 & ...            & \textless{13.08} & 11.87$\pm$0.06 & $\star$, 1 \\
\hline
\end{longtable}
\flushleft
{\footnotesize \textbf{Notes:} References:($\star$) This work; (1) \citet{DeCia2016}; (2) \citet{Berg2015}. Revised f-values from \citet{Cashman2017}, \citet{Kisielius2014S, Kisielius2015Zn} and \citet{Kurucz2017}}.
\end{landscape}

\tiny
\setlength\tabcolsep{1.8pt}
\begin{longtable}{lccccccccccc}
\caption{Milky Way column densities}
\label{mwtable}\\
\hline\hline
Star &log N(\ion{P}{II}) & log N(\ion{Mg}{II})& log N(\ion{Si}{II}) & log N(\ion{S}{II}) & log N(\ion{Ti}{II})&log N(\ion{Cr}{II})& log N(\ion{Mn}{II})& log N(\ion{Fe}{II})  & log N(\ion{Ni}{II}) & log N(\ion{Zn}{II}) & Ref\\
\hline
\hline
HD-18100    & 13.81$\pm$0.07 & 15.18$\pm$0.03            & 15.27$\pm$0.03         & ...            & ...                         & 12.89$\pm$0.04 & 12.75$\pm$0.03            & 14.67$\pm$0.04 & 13.44$\pm$0.05 & 12.62$\pm$0.05 &1\\
KSI-PER     & 14.62$\pm$0.03  & 15.92$\pm$0.04            & ...                         & ...            & 11.53$\pm$0.05         & 12.88$\pm$0.02 & 13.53$\pm$0.03            & 14.72$\pm$0.03 & 13.30$\pm$0.05 & 13.39$\pm$0.02&1 \\
23O-ORI     & 14.40$\pm$0.04 & 15.63$\pm$0.04            & 15.37$\pm$0.08         & ...            & 11.17$\pm$0.03         & 12.72$\pm$0.01 & 13.02$\pm$0.04            & 14.38$\pm$0.03 & 12.96$\pm$0.12 & 13.28$\pm$0.07 &1\\
DEL-ORI-A   & 13.42$\pm$0.20   & 15.09$\pm$0.04            & ...                         & ...            & 11.11$\pm$0.04         & 12.07$\pm$0.05 & 12.28$\pm$0.06            & 14.06$\pm$0.06 & ...                         & 12.62$\pm$0.05 &1\\
EPS-ORI     & 13.87$\pm$0.08 & 15.27$\pm$0.02            & ...                         & ...            & 11.36$\pm$0.03         & 12.43$\pm$0.05 & 12.73$\pm$0.04 & 14.21$\pm$0.05 & ...                         & 12.97$\pm$0.06 &1\\
MU-COL      & 13.70$\pm$0.02 & 15.12$\pm$0.02 & 15.09$\pm$0.02 & 15.21$\pm$0.02 & 11.46$\pm$0.02         & 12.61$\pm$0.03 & 12.42$\pm$0.02            & 14.31$\pm$0.01 & 12.94$\pm$0.05 & 12.59$\pm$0.08 &1\\
\hline
\end{longtable}
\flushleft
{\footnotesize \textbf{Notes:} References: 1=\citet{Jenkins2009};  2=\citet{DeCia2021}. Revised f-values from \citet{Cashman2017}, \citet{Kisielius2014S, Kisielius2015Zn} and \citet{Kurucz2017}.
}

\tiny
\setlength\tabcolsep{1.8pt}
\begin{longtable}{lccccccccccc}
\caption{Milky Way. Elements with limited coverage}
\label{mwtablelimited}\\
\hline\hline
Star & log N(\ion{O}{I}) &log N(\ion{Fe}{II}) & log N(\ion{Zn}{II}) & log N(\ion{Co}{II}) & log N(\ion{Ge}{II}) & log N(\ion{Kr}{I})  & log N(\ion{Cl}{II})  & log N(\ion{Cu}{II}) &log  N(\ion{Al}{II}) & log  N(\ion{C}{II})&Ref\\
\hline
\hline
HD-18100    & ...             & 14.67$\pm$0.04 & 12.62$\pm$0.05 & ...                   & ...            & ...             & ...            & ...              & ...            & 13.69$\pm$0.07 & 1 \\
KSI-PER     & 17.81$\pm$0.06 & 14.72$\pm$0.03 & 13.39$\pm$0.02 & ...                   & ...            & ...             & 14.54$\pm$0.41 & 12.21$\pm$0.07 & ...            & 14.50$\pm$0.03  & 1 \\
23O-ORI     & 17.31$\pm$0.10  & 14.38$\pm$0.03 & 13.28$\pm$0.07 & ...                   & 11.89$\pm$0.04 & ...             & 13.57$\pm$0.68 & 12.03$\pm$0.03 & ...            & 14.28$\pm$0.04 & 1 \\
DEL-ORI-A   & 16.66$\pm$0.05 & 14.06$\pm$0.06 & 12.62$\pm$0.05 & ...                   & ...            & ...             & 13.23$\pm$0.11 & ...              & ...            & 13.30$\pm$0.20   & 1 \\
EPS-ORI     & 16.97$\pm$0.05 & 14.21$\pm$0.05 & 12.97$\pm$0.06 & ...                   & ...            & 11.42$\pm$0.09 & 13.51$\pm$0.06 & ...              & ...            & 13.75$\pm$0.08 & 1 \\
MU-COL      & ...             & 14.31$\pm$0.01 & 12.59$\pm$0.08 & ...                   & ...            & ...             & ...            & ...              & ...            & 13.58$\pm$0.02 & 1 \\
\hline
\end{longtable}
\flushleft
{\footnotesize \textbf{Notes:} Milky Way column densities of the elements with limited coverage. Revised f-values from \citet{Cashman2017, Kisielius2014S, Kisielius2015Zn} and \citet{Kurucz2017}.
References: 1=\citet{Jenkins2009};  2=\citet{DeCia2021}; 3=\citet{Phillips1982}.}

\tiny
\setlength\tabcolsep{1.8pt}
\begin{longtable}{lcccccccccccc}
\caption{GRB-DLAs}
\label{grbtable}\\
\hline\hline
GRB & z$_{abs}$ & log N(\ion{H}{i}) & log N(\ion{O}{I}) & log N(\ion{Zn}{II}) & log N(\ion{S}{II}) & log N(\ion{Si}{II})$^{\rm a}$ & log N(\ion{Mg}{II})$^{\rm a}$ & log N(\ion{Mn}{II}) & log N(\ion{Cr}{II})$^{\rm a}$ & log N(\ion{Ni}{II})$^{\rm a}$ & log N(\ion{Fe}{II})  & log N(\ion{Ti}{II})$^{\rm a}$\\
\hline
\hline
090926A & 2.1069 & 21.58$\pm$0.01 & \textgreater{15.00} & 12.24$\pm$0.11  & 14.52$\pm$0.04     & 15.21$\pm$0.05     & \textgreater{14.60}   & \textless{13.00}    & \textless{12.90} & 13.40$\pm$0.11  & 14.21$\pm$0.02     & \textless{11.60} \\
120327A & 2.8143 & 22.07$\pm$0.01 & \textgreater{16.30} & 13.21$\pm$0.04  & 15.75$\pm$0.02     & \textgreater{14.50} & 16.40$\pm$0.23       & ...                  & 14.10$\pm$0.02  & 14.57$\pm$0.04  & 15.79$\pm$0.06     & 12.69$\pm$0.08  \\
120716A & 2.4874 & 21.73$\pm$0.03 & \textgreater{15.50} & 13.65$\pm$0.05  & 16.00$\pm$0.14     & \textgreater{16.20} & \textgreater{14.80}   & 13.93$\pm$0.18     & 14.38$\pm$0.22  & \textless{15.40} & 15.81$\pm$0.22     & ...               \\
120815A & 2.3582 & 22.09$\pm$0.01 & \textgreater{16.00} & 13.27$\pm$0.02  & 15.51$\pm$0.05     & \textgreater{15.40} & ...                    & 13.29$\pm$0.07     & 13.37$\pm$0.08  & 14.16$\pm$0.08  & 15.11$\pm$0.05     & \textless{12.50} \\
121024A & 2.3005 & 21.78$\pm$0.02 & \textgreater{15.00} & 13.65$\pm$0.06  & \textgreater{15.70} & 16.62$\pm$0.14     & \textgreater{17.80}   & 13.76$\pm$0.03     & 14.07$\pm$0.05  & 14.24$\pm$0.34  & 15.73$\pm$0.05     & ...               \\
140311A & 4.9550 & 22.30$\pm$0.02 & \textgreater{15.50} & 13.28$\pm$0.14  & 15.48$\pm$0.12     & \textgreater{14.70} & \textgreater{14.60}   & 13.49$\pm$0.03     & 14.11$\pm$0.08  & 14.40$\pm$0.23  & 15.78$\pm$0.15     & ...               \\
\hline
\end{longtable}
\flushleft
{\footnotesize \textbf{Notes:} Redshifts and column densities are taken from \citet{Bolmer2019}. Revised f-values from $^{\rm a}$\citet{Cashman2017} and $^{\rm b}$ \citet{Boisse2019}.}

\tiny
\setlength\tabcolsep{1.8pt}
\begin{longtable}{lcccccccccccc}
\caption{LMC column densities}
\label{lmctable}\\
\hline
\hline
Star & log N(\ion{H}{I})& log N(\ion{Cr}{II})& log N(\ion{Fe}{II}) & log N(\ion{Mg}{II}) & log N(\ion{Ni}{II}) & log N(\ion{O}{I}) & log N(\ion{P}{II}) & log N(\ion{S}{II}) & log N(\ion{Si}{II}) & log N(\ion{Zn}{II})  & log N(\ion{Ti}{II})\\
\hline
\hline
BI173    & 21.25$\pm$0.05 & 13.57$\pm$0.03 & 15.24$\pm$0.03 & 16.01$\pm$0.10  & 13.94$\pm$0.03 & ...            & 13.79$\pm$0.02  & 15.86$\pm$0.06 & 15.91$\pm$0.20  & 13.22$\pm$0.04 & ...            \\
BI184    & 21.15$\pm$0.04 & 13.61$\pm$0.06 & 15.31$\pm$0.04 & 16.05$\pm$0.05 & 14.15$\pm$0.10  & ...            & ...            & ...            & 15.78$\pm$0.16 & 13.19$\pm$0.07 & ...            \\
BI237    & 21.65$\pm$0.03 & 13.53$\pm$0.06 & 15.23$\pm$0.04 & 16.32$\pm$0.05 & 13.98$\pm$0.06 & ...            & ...            & ...            & 15.84$\pm$0.14 & 13.36$\pm$0.06 & 12.45$\pm$0.05 \\
BI253    & 21.68$\pm$0.03 & 13.72$\pm$0.04 & 15.46$\pm$0.03 & 16.38$\pm$0.09 & 14.17$\pm$0.06 & ...            & ...            & ...            & ...            & 13.44$\pm$0.03 & 12.71$\pm$0.04 \\
PGMW3120 & 21.48$\pm$0.03 & 13.60$\pm$0.06  & 15.22$\pm$0.05 & 16.05$\pm$0.12 & 14.00$\pm$0.06  & ...            & ...            & ...            & 15.93$\pm$0.26 & 13.34$\pm$0.06 & ...            \\
PGMW3223 & 21.40$\pm$0.06  & 13.51$\pm$0.06 & 15.24$\pm$0.03 & 16.12$\pm$0.11 & 13.93$\pm$0.05 & ...            & ...            & ...            & 15.84$\pm$0.17 & 13.42$\pm$0.05 & ...            \\
\hline
\end{longtable}
\flushleft
{\footnotesize \textbf{Notes:} Column densities are taken from \citet{Roman-Duval2021}, except for the $\ion{P}{II}$ for BI173, SK-70115, SK-6522, SK-675, SK-7145, SK-66172,SK-68135, which are from \citet{Tchernyshyov2015}. Revised f-values from \citet{Cashman2017}, \citet{Boisse2019} and \citet{Kurucz2017} for $\ion{P}{II}$.}

\newpage

\tiny
\setlength\tabcolsep{1.8pt}
\begin{longtable}{lcccccccccccc}
\caption{SMC column densities}
\label{smctable}\\
\hline \hline
Star & H$_{tot}$ & log N(\ion{Mg}{II})
 & log N(\ion{Si}{II})
 & log N(\ion{S}{II}) & log N(\ion{Ti}{II})
 & log N(\ion{Cr}{II})
 & log N(\ion{Mn}{II})
 & log N(\ion{Fe}{II}) & log N(\ion{Ni}{II})
 & log N(\ion{Zn}{II})
 & log N(\ion{P}{II})\\
\hline
\hline
AzV18  & 22.06$\pm$0.02 & 16.68$\pm$0.07 & ...            & ...            & 12.94$\pm$0.01 & 13.95$\pm$0.06 & 13.48$\pm$0.02 & 15.61$\pm$0.03 & 14.29$\pm$0.04 & 13.67$\pm$0.08 & ...            \\
AzV26  & 21.77$\pm$0.04 & 16.11$\pm$0.05 & 16.05$\pm$0.05 & ...            & 12.86$\pm$0.01 & 13.82$\pm$0.02 & 13.38$\pm$0.01 & 15.51$\pm$0.02 & 14.23$\pm$0.02 & 13.26$\pm$0.05 & ...            \\
AzV47  & 21.32$\pm$0.04 & 16.07$\pm$0.12 & 15.81$\pm$0.06 & 15.84$\pm$0.08 & 12.78$\pm$0.03 & 13.67$\pm$0.04 & 13.25$\pm$0.03 & 15.37$\pm$0.03 & 14.06$\pm$0.04 & 13.09$\pm$0.06 & 13.86$\pm$0.03 \\
AzV80  & 21.83$\pm$0.02 & 16.37$\pm$0.09 & ...            & ...            & 12.77$\pm$0.03 & 13.70$\pm$0.03  & 13.26$\pm$0.02 & 15.39$\pm$0.03 & 14.09$\pm$0.04 & 13.23$\pm$0.05 & ...            \\
AzV95  & 21.50$\pm$0.04  & 16.02$\pm$0.15 & 15.94$\pm$0.06 & 15.84$\pm$0.07 & 12.78$\pm$0.04 & 13.70$\pm$0.04  & 13.24$\pm$0.02 & 15.37$\pm$0.03 & 14.04$\pm$0.04 & 13.09$\pm$0.06 & 13.86$\pm$0.01 \\
AzV104 & 21.46$\pm$0.06 & 16.14$\pm$0.10  & 15.89$\pm$0.11 & 15.78$\pm$0.05 & ...            & 13.48$\pm$0.09 & ...            & 15.31$\pm$0.04 & 14.01$\pm$0.04 & 12.97$\pm$0.16 & ...            \\
\hline
\end{longtable}
\flushleft
{\footnotesize \textbf{Notes:} All column densities are taken from \citet{Jenkins2017}, except for the case of SK143, for which log N(\ion{Ti}{ii}) is taken from \citet{Welty2010} and  log N(\ion{Zn}{ii}), log N(\ion{Fe}{ii}) are taken from \citet{Tchernyshyov2015}. Revised f-values from \citet{Cashman2017}, \citet{Boisse2019} and \citet{Kurucz2017}.}

\end{appendix}

\end{document}